\documentclass[abstractpage,publicabstractpage,figures,tables,phd]{uithesis}

\usepackage{amsfonts}
\usepackage{amsmath}
\usepackage{amssymb}
\usepackage{array}
\usepackage{float}
\usepackage[toc,page,titletoc]{appendix}
\usepackage{longtable}
\usepackage{graphicx}
\usepackage[normalem]{ulem}
\usepackage{nicefrac}
\usepackage{units}
\usepackage{rotating}
\usepackage{afterpage}
\usepackage{caption}
\usepackage{mathtools}
\usepackage{mathrsfs}
\usepackage{tensor}
\usepackage{cite}
\usepackage{latexsym}
\usepackage{extarrows}
\usepackage{epsfig}
\usepackage{subfig}
\usepackage{graphics}
\usepackage{dsfont}
\usepackage{enumerate}
\usepackage{hyperref}
\usepackage[all]{hypcap}

\usepackage{etoolbox}
\apptocmd{\thebibliography}{\interlinepenalty 10000\relax}{}{}

\usepackage{enumitem}
\setlist{noitemsep,topsep=1pt, partopsep=4pt, parsep=2pt}
\setenumerate{noitemsep,topsep=1pt, partopsep=4pt, parsep=2pt}

\let\savedCaption=\caption
\renewcommand*{\caption}[2][\shortcaption]{%
  \def\shortcaption{#2}
  \normalsize\singlespace
  \savedCaption[#1]{#2}}

\makeatother 

\let\baraccent=\=

\renewcommand{\v}[1]{\ensuremath{\mathbf{#1}}}

\renewcommand{\=}[1]{\stackrel{#1}{=}}

\newcommand{\uv}[1]{\ensuremath{\mathbf{\hat{#1}}}}

\newcommand{\pd}[2]{\frac{\partial #1}{\partial #2}}

\newcommand{\ket}[1]{\left| #1 \right>} 
\newcommand{\bra}[1]{\left< #1 \right|} 
\newcommand{\braket}[2]{\left< #1 \vphantom{#2} \right| \left. #2 \vphantom{#1} \right>} 
\newcommand{\matrixel}[3]{\left< #1 \vphantom{#2#3} \right| #2 \left| #3 \vphantom{#1#2} \right>}

\newcommand{\al}[1]{\begin{align}#1\end{align}}
\newcommand{\bs}{\begin{split}}
\newcommand{\es}{\end{split}}
\newcommand{\eqr}[1]{Eq.~\ref{#1}}

\newcommand{\m}[1]{$ #1 $}
\newcommand{\LAO}{LaAlO$_3$~}
\newcommand{\BiSb}{Bi$_{1-\text{x}}$Sb$_{\text{x}}$~}

\newcommand{\BiSe}{Bi$_2$Se$_3$~}
\newcommand{\SeBi}{Bi$_2$Se$_3$}
\newcommand{\BiTe}{Bi$_2$Te$_3$~}
\newcommand{\TeBi}{Bi$_2$Te$_3$}
\newcommand{\STO}{SrTiO$_3$~}
\newcommand{\OTS}{SrTiO$_3$}
\newcommand{\LAOSTO}{LaAlO$_3$/SrTiO$_3$~}
\newcommand{\STOLAO}{LaAlO$_3$/SrTiO$_3$}
\newcommand{\kp}{$\v k \cdot \v p~$}

\newcommand{\degree}{\ensuremath{^\circ}}

\newcommand{\etal}{\emph{et al.}~}

\newcommand{\figref}[1]{Fig.~\ref{#1}}

\newcommand{ \kx	    }{\ensuremath{	k_{x}}}
\newcommand{ \ky	    }{\ensuremath{	k_{y}}}
\newcommand{ \kz	    }{\ensuremath{	k_{z}}}	

\title{Spin Dynamics of Complex Oxides, Bismuth-Antimony Alloys, and Bismuth Chalcogenides}
\author{Cuneyt Sahin}
\dept{Physics}
\setboolean{multipleSupervisors}{false}
\advisor{Professor Michael E. Flatt\'e}
\memberOne{Michael E. Flatt\'e}
\memberTwo{Thomas F. Boggess}
\memberThree{Markus Wohlgenannt}
\memberFour{Fatima Toor}
\memberFive{David R. Anderson}
\submitdate{August 2015}
\copyrightyear{2015}

\abstractfile{thesisAbstract} 
\publicabstractfile{publicAbstract}

\setboolean{figures}{true} 
\setboolean{tables}{true}

\begin{document}

\frontmatter

\chapter{Introduction and Broad Statement of the Problem}
\section{Introduction to Spintronics}

Spin is the intrinsic angular momentum of sub-atomic particles. Although the name suggests that particles are spinning objects, this phenomenon is purely quantum mechanical and originates from relativistic quantum mechanics. The spin of an electron can be oriented in two versions, either parallel to an effective magnetic field (or magnetization) or antiparallel. However, in an ensemble of many particles the total spin, or magnetization, can orient in different directions and the collective behavior of spins aligned in the same direction creates magnets. Magnets have been known for several thousand years and are used in many aspects of our lives from computer hard drives to magnetic resonance imaging (MRI) \cite{Lauterbur1973, Wang2014}. There is still a tremendous potential for the usage of spins and magnetism in the future. The one which is the subject of this study is called spintronics, and also known as spin-based electronics.

Spintronics is an emerging field of condensed matter physics, which aims to utilize spins in a system as the processing and storage units of electronic devices. The relatively new name of this technology, spintronics, may sound unfamiliar to many.  We have been using devices based on the manipulation of the collective behavior of spins, or magnetization, on a daily basis for over 15 years. One of the most prominent examples is the giant magnetoresonance (GMR) effect \cite{Baibich1988, Binasch1989}, which is the functioning principle of the computer hard drives. Magnetoresistance is a phenomenon discovered by Lord Kelvin in 1857 and can be defined as the change in the resistance of a conductor due to an applied magnetic field. On the other hand, giant magnetoresistance is the extremely large version of magnetoresistance and was discovered independently by the research groups of Albert Fert and Peter Gr\"unberg, which resulted in the Nobel Prize in Physics in 2007. A typical device in which giant magneto resistance can be observed, consists of two ferromagnetic contact and a spacer material between them. The electric current flows between these ferromagnets through the channel material with little resistance if the magnetizations of  the two ferromagnets are aligned in the same directions. Switching the magnetization of one ferromanget leads to an enormous electrical resistance. The functioning principle of the GMR effect is based on the contrast in electrical conductivities of these two configurations. This discovery had a substantial impact on the electronics industry in the last 20 years, as recent computer hard drives store information using the GMR effect \cite{Chappert2007, Fert2008}.

One of the goals of spintronics is to enhance or replace conventional metal-oxide-semiconductor field-effect transistors (MOSFET) with a spin based ones. This goal originates from the fact that as MOSFETs are produced smaller over time (Moore's law), the size of a transistor approaches the quantum limit. Reducing the size of the transistor brings also higher energy consumption and heat generation which limits the operating speed. The response to all these problems could be changing the current paradigm of charge-based transistors. In a spintronic transistor the information is carried by the spins of the electrons. As a result of this, spintronics offers higher processing speed with less power consumption as well as non-volatility. However, there are several challenges for spintronic devices  \cite{Awschalom2007} of which three basic ones related to the functioning principles can be broadly classified as:
\begin{enumerate}
\item Generating the spin-polarized current in an effective way: There are optical and electrical ways of generating a fully polarized spin current. For instance, circularly polarized light may generate a net spin current in III-V direct band gap semiconductors \cite{Parsons1969, Crooker2007} as a result of the selection rules. This method is also applicable to silicon \cite{Lampel1968}, however, it is not very effective due to the indirect band gap in silicon. On the other hand the spin Hall effect, which is a result of the spin-orbit interaction in materials, is  effective in generating such currents \cite{Kato2004}. 
\item Preserving the polarization of spins for a long time: Spin polarization is usually lost quickly due to several spin relaxation mechanisms, such as the Dyakanov-Perel mechanism \cite{Dyakonov1971, Dyakonov1972}, the Bir-Aranov-Pikus mechanism \cite{Bir1975, Aronov1983}, and the Elliott-Yafet mechanism \cite{Elliott1954,Yafet1963}. Finding and designing  materials which preserve the orientation of spins for sufficiently long times such that spin diffusion lengths are sufficient to operate spin logic elements is one of the most important challenges of spintronics. 
\item Detecting the spin polarization: Similar to the process of generating spin currents, spin polarization can be detected by electrical and optical methods such as Kerr and Faraday effects \cite{Kato2004}.
\end{enumerate}
 
\noindent There have been several proposals to replace conventional semiconductors. Several aspects of spin-related phenomena such as spin-transfer torque \cite{Brataas2012, Katine2008}, spin Hall effects \cite{Jungwirth2012} and spin caloritronics \cite{Bauer2012} are used in proposed devices. The most widely known proposal for such a device is the Datta-Das transistor \cite{Datta1990}. In this transistor, spin-polarized electrons are injected from one contact (source). Under the influence of an effective magnetic field spins precess and may reach the other contact (drain) with opposite spin depending on the gate voltage applied. This leads to no current, however if they reach with the same polarization as the drain then current can flow. Several other device proposals are available, such as magnetic bipolar transistors \cite{Flatte2003, Fabian2004}, metal-oxide-semiconductor based spin devices \cite{Tanaka2007}, spin based diodes \cite{Flatte2001a}, dynamic spin based logic units \cite{Dery2007}, although most of them have not been realized experimentally. 

In addition to these devices based on spin-polarized currents, color centers in semiconductors, especially in diamond and silicon, provide a framework for single-spin applications based on the manipulation of a single electron's spin. There are many advantages of such spin centers in diamond; firstly carbon is an abundant element, semiconducting, and strong. It is also transparent, which offers the opportunity for optical manipulation. One can change the state of the electron in such a vacancy by using visible light. Spin dynamics in nonmagnetic wide-bandgap materials has received renewed attention due to the exceptionally long spin coherence times of spin centers in diamond \cite{Balasubramanian2009} and silicon carbide \cite{Koehl2011}. Single spin centers in diamond and silicon carbide in the form of either nitrogen-vacancy centers or transition metal dopants \cite{Bocquel2013} also provide another perspective: utilization of the quantum mechanical nature of spin for quantum computation \cite{Awschalom2002}.

This study proposes two type of materials as a solution to two major problems of spintronics based on spin currents: Materials which have high capacity for generation of spin current and materials with large spin lifetimes. We report that the latter can be achieved by using complex oxide heterostructures, specifically two-dimensional electron systems at the interface of strontium titanate and lanthanum aluminate (\STOLAO), and the former is achievable by bismuth-based materials with large spin-orbit interactions, such as \BiSb alloys, \SeBi, and \BiTe topological insulators. 

The strontium titanate interface has attracted much interest since spin injection experiments in bulked doped \STO \cite{Han2013,Reyren2012} were conducted. Large Rashba coefficients \cite{Caviglia2010}, strain and growth tunability have also been reported for \LAOSTO interface that have high-density, high-mobility, two-dimensional electron gases (2DEGs) \cite{Ohtomo2004}. One important advantage of these 2DEGs is the inversion symmetry that is absent in well-explored materials such as III-V semiconductors and their heterostructures. Therefore effective pseudomagnetic fields\cite{Meier1984, Dresselhaus1955} dominate spintronic properties such as spin lifetimes  \cite{Dyakonov1972,Lau2001} as a result of the inversion asymmetry of the crystal. Thus, interfaces and heterostructures of complex oxides are expected to exhibit larger spin lifetimes compared to conventional semiconductors.

On the other hand, spin current generation is possible through the spin Hall effect, which originates from the spin-orbit interaction in a solid \cite{Engel2007, Murakami2005,Vignale2010}. The spin Hall conductivity, which is the ratio of the spin Hall current to the longitudinal  electric field,  depends on details of the electronic band structure such as the strength of the spin-orbit interaction, the Fermi energy, the direction of current relative to crystal axes and the strain\cite{Dyakonov1971,Hirsch1999,Murakami2003,Kato2004b,Sih2005,Guo2005,Yao2005,Hankiewicz2006,Sih2006,Guo2008,Lowitzer2011,Liu2012,Norman2014,Norman2014e}.
Bismuth based structures which are centrosymmetric semimetals or topological insulators, and have gigantic spin-orbit couplings, might have spin Hall angles (the ratio of the spin current to the longitudinal charge current) that are much larger than conventional semiconductors\cite{Yao2005,Valenzuela2006,Vila2007,Guo2008,Liu2012}. Also, these materials have more tunable spin Hall conductivities and longitudinal conductivities while maintaining very large spin Hall angles. For example, large spin Hall angles have been demonstrated for bismuth selenide\cite{Mellnik2014}, motivated by proposals for large spin current effects in topological insulators\cite{Burkov2010, Culcer2010, Pesin2012}. This motivates us further to investigate bismuth-antimony alloys, which are also topological insulators at certain concentration, and bismuth chalcogenides, such as \BiSe and \BiTe topological insulators which have a tremendous potential for highly tunable spin current generation transverse to the applied electric field.

The rest of this chapter serves as an introduction to these two types of material families and ends with challenges that must be faced before using these materials for future spintronic applications.

\section{Complex Perovskite Oxides}
\subsection{Introduction}
Many years of experimental and theoretical investigations resulted in semiconductor materials with high degrees of functionality that can be accurately designed, tuned and used in numerous applications. The success in the field of mainstream semiconductors such as group IV elements \cite{Adachi2009, Casey1999}, III-V compounds \cite{Madelung1964,Meier1984} and related materials \cite{Dresselhaus1996, Winkler2003, Zaitsev2001} encourages researchers to go further and explore the physics of new materials with highly correlated systems, such as complex oxides. Oxide materials with strongly correlated electrons provide many opportunities to exploit their novel features while facing new challenges. Transition metal oxides may exhibit a variety of properties depending on the details of their composition and structure. For instance, CrO$_2$ \cite{suzuki1998resistivity} and Fe$_3$O$_4$ \cite{Verwey1939} are metals if the temperature is higher than 120K. Cu$_2$O is a semiconductor \cite{de1999cu2o} whilst VO$_2$ and V$_2$O$_3$ exhibit semiconductor-metal transitions\cite{mott1974metal}. There also exist superconductors such as La(Sr)$_2$CuO$_4$ \cite{takagi1989superconductor}. Electrical and magnetic properties also shows a great diversity such as piezoelectric and ferroelectric BaTiO$_3$ \cite{kamalasanan1991structural}, ferro- and ferri magnets CrO$_2$\cite{schwarz1986cro2} and $\gamma$-Fe$_2$O$_3$ \cite{cannas1998structural} and antiferromagnet  $\alpha$ Fe$_2$O$_3$ \cite{dormann1985mossbauer}. The most attractive and specific examples of these oxides are perovskite oxides that contain a broad range of systems such as strontium titanate (SrTiO$_3$), lanthanum aluminate (LaAlO$_3$), lanthanum cobalt oxide (LaCoO$_3$) etc. They all have the general formula ABO$_3$, where A and B can be substituted by almost all of the elements in the periodic table. This is a great advantage that gives rise to heterostructures with a variety of different properties. Low-temperature superconductivity, two-dimensional systems with high mobilities at their interfaces, metal-insulator transitions and multiferroicity \cite{Ohtomo2004, Chakhalian2012, Pena2001} are only some of these versatile features.

\begin{figure}[h]
\centering
\includegraphics[width=1\textwidth]{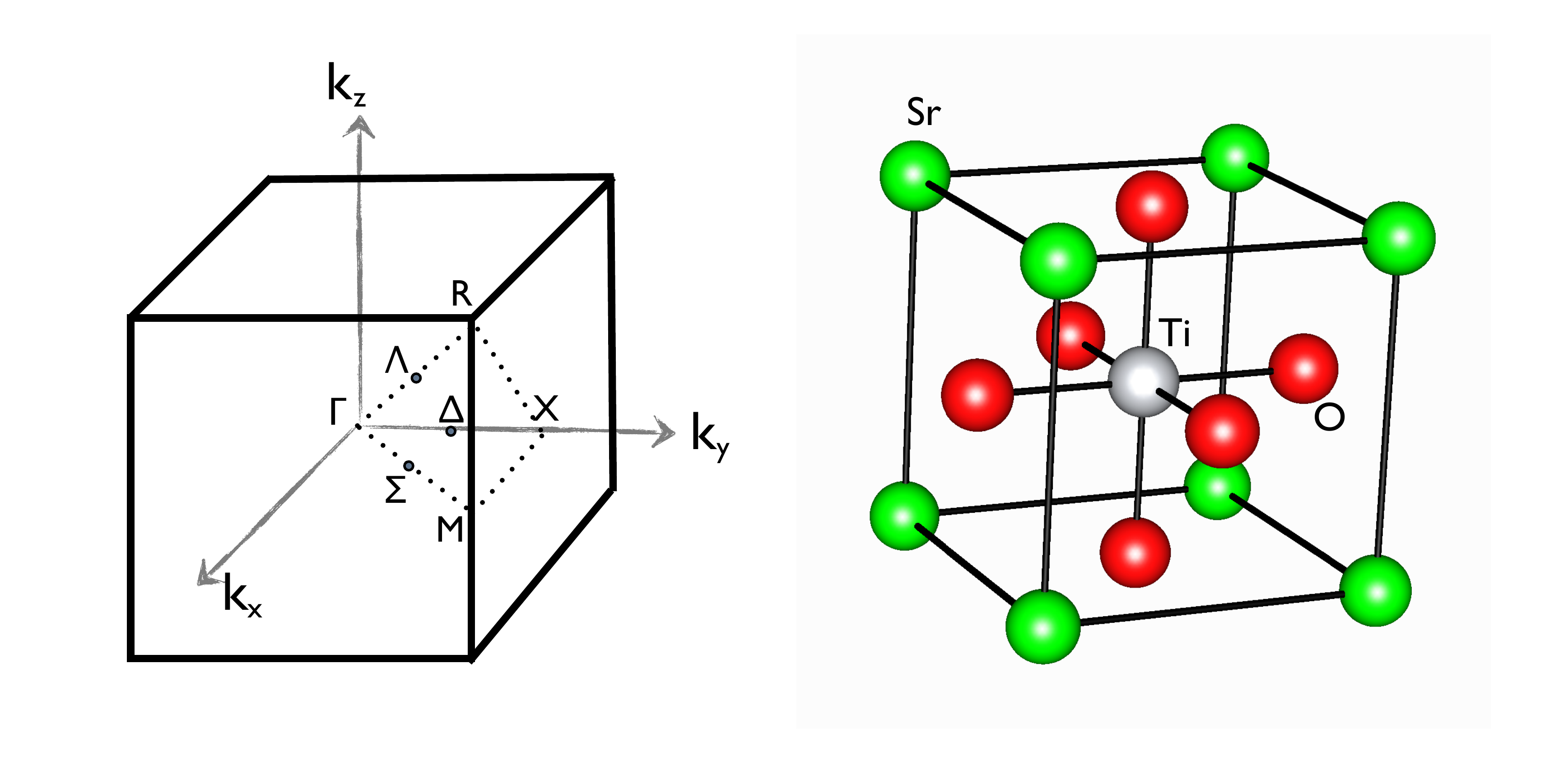}
\singlespace
\caption[Reciprocal lattice and crystal structure of a simple cubic perovskite consisting of three different elements]{\label{fig:perovskite}
Reciprocal lattice and crystal structure of a simple cubic perovskite consisting of three different elements. The general formula is ABO$_3$. As indicated in the figure atom B (titanium) is connected to six oxygen atoms (atom B) making an octahedron. Atom A (strontium) sits in the corners and usually contributes its \textit{s}-orbital electrons to oxygen's \textit{p}-orbitals.}
\end{figure}

One of the most well-known representatives of perovskite oxides is strontium titanate. \STO has cubic symmetry and space group O$_h^1$ \cite{Cardona1965}. It is a centrosymmetric crystal and therefore inversion symmetric, and time reversal symmetry exists as well. This provides doubly degenerate bands if no additional symmetry breaking occurs, e.g., by an external magnetic field. A depiction of the simple cubic perovskite crystal and its Brillouin zone can be seen in \figref{fig:perovskite}. 

\begin{figure}

\centering
\includegraphics[width=1\textwidth]{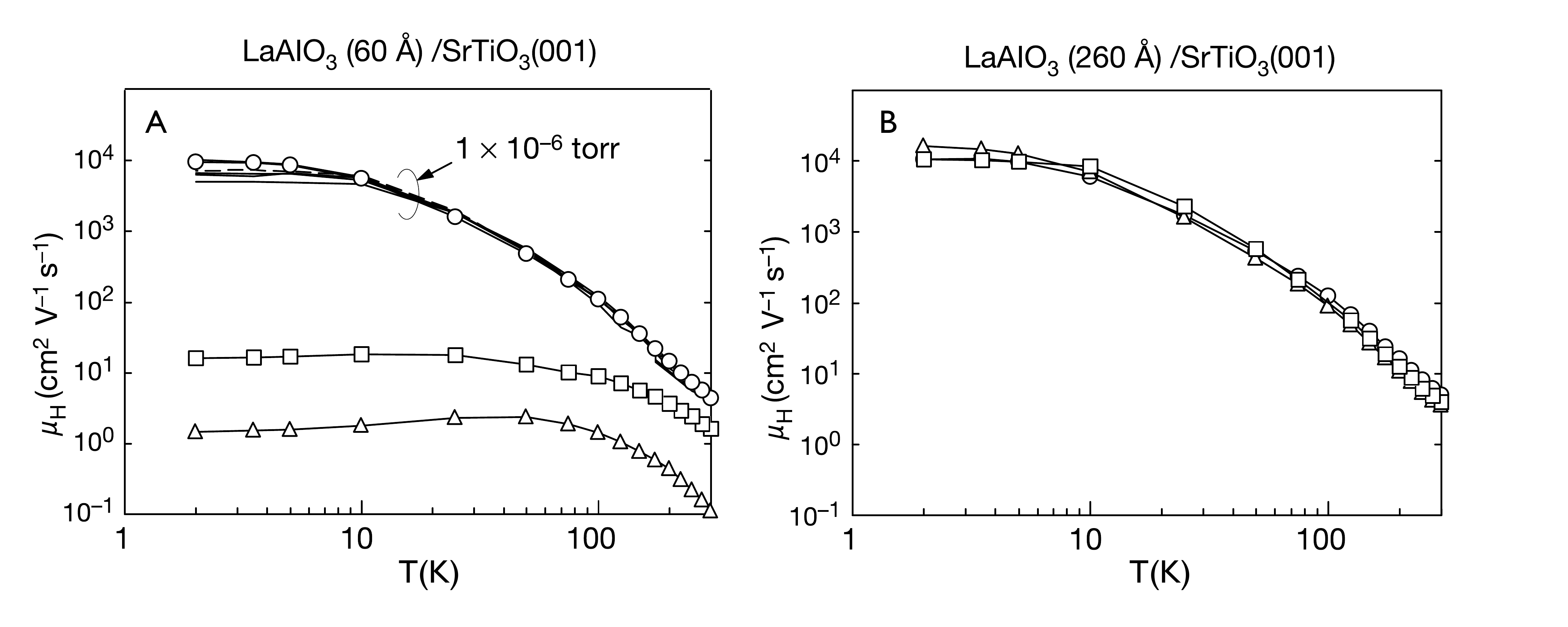}
\caption[Mobility vs. temperature of the 2DEG at the interface of \LAO  and \STO grown at different partial oxygen pressures]{\label{fig:ohtomo}
Mobility vs. temperature of the 2DEG at the interface of \LAO  and \STO grown at different partial oxygen pressures. Figure a shows the mobilities of the system for a \LAO thickness of 60 \AA ~and figure b is for 260 \AA. (From \cite{Ohtomo2004})}

\end{figure}

Another exciting feature of oxide materials is the possibility to form two-dimensional systems at their interfaces. For instance growing \LAO on top of the \STO results in the formation of a two-dimensional electron gas (2DEG) at the interface. This 2DEG has been first discovered by Ohtomo and Hwang in 2004 \cite{Ohtomo2004} by measuring the conductivity of the interface. Furthermore by Hall effect measurements they also concluded that these electron gases have high mobilities up to 10000 cm$^2$ V$^{-1}$s$^{-1}$ at low temperatures (\figref{fig:ohtomo}). This discovery aroused interest in 2 dimensional systems at oxide interfaces. Approximately three years after the discovery of \LAOSTO 2DEGs, Reyren \etal \cite{Reyren2007} conducted an experiment which showed the superconductive properties of these systems. By transport measurements, superconductivity has been observed below 200 miliKelvin for samples with 8 unit cells of \LAO deposited onto \OTS. Similarly Gariglio \etal \cite{Gariglio2009} has reported superconductivity at 200 miliKelvin. Furthermore Ben Shalom \etal \cite{BenShalom2010} showed that applying a gate voltage may tune the superconducting phase transition temperature up to 350 miliKelvin for a voltage of -50 V.

The structures of \LAO and \STO  are similar. They don't exhibit any significant lattice mismatch as their lattice constants are 3.789 \AA ~and 3.905 \AA ~for \LAO and \STO respectively \cite{Ohtomo2004}. In bulk form these two oxides are both insulators with wide band gaps of 5.6 eV \cite{Ohtomo2004} for \LAO and 3.2 eV for \STO \cite{Cardona1965}. The possibility of creating a conducting layer of two-dimensional electrons at the interface of insulating oxides is quite astonishing. This phenomenon automatically raises the question from where the carriers at the interface originate. 

\subsection{Origin of the Carriers at the Interface}
In conventional semiconductor heterojunctions such as GaAs/Al$_x$Ga$_{1-x}$As interfaces in MOSFETs, the formation of the 2DEG is due to modulation doping with band bending. \cite{YuCardona} However oxide heterojunctions suggest different ways to form the two-dimensional electronic systems. While there have been many studies trying to explain the source of the electrons at the interface of \LAOSTO, two of the possible explanations seem to be prominent and strongly supported by both theoretical and experimental evidence: the polar catastrophe and oxygen vacancies. 

\begin{figure}

\centering
\includegraphics[width=0.95\textwidth]{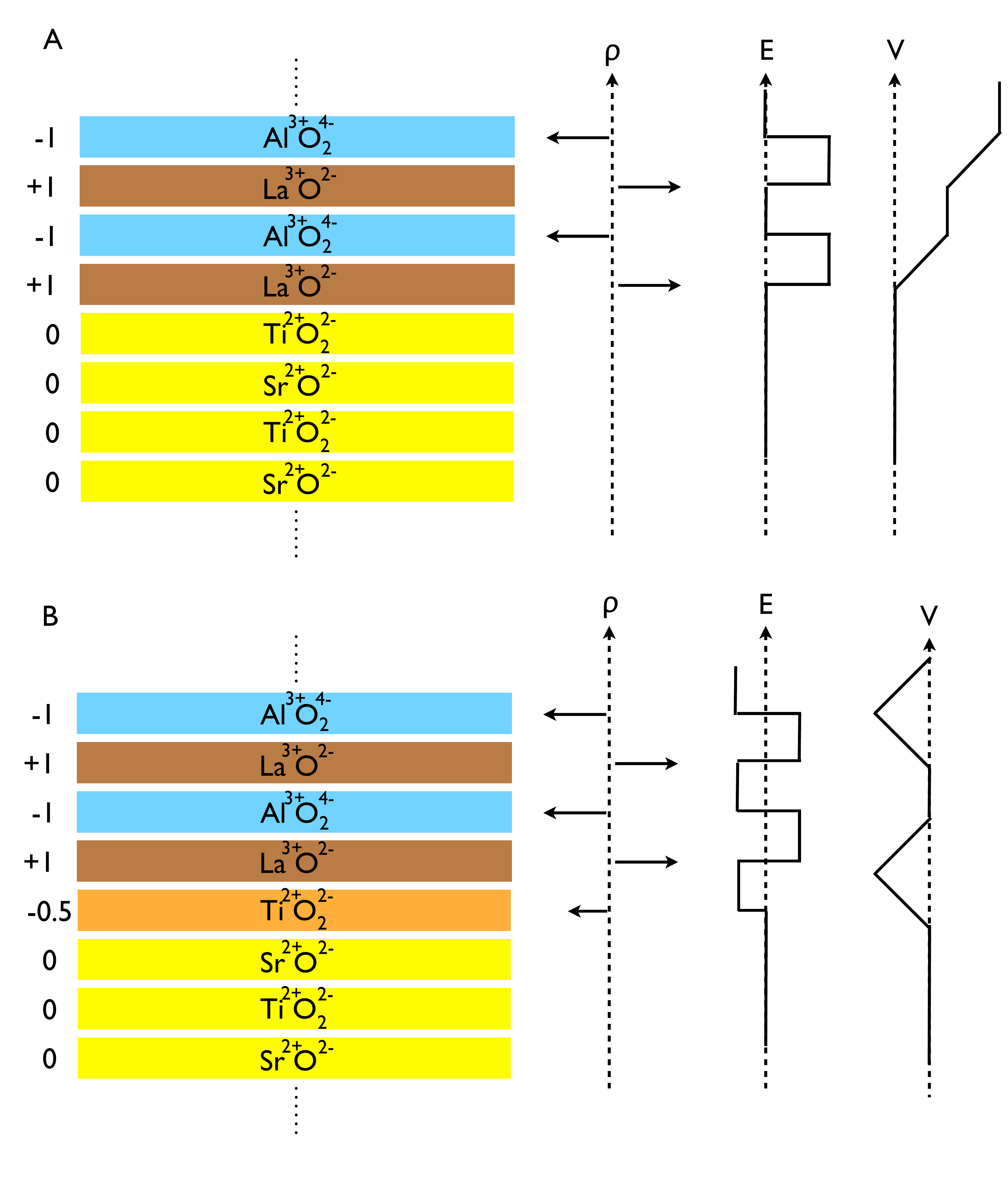}
\caption[Polar catastrophe theory at the \LAOSTO interface]{\label{fig:polarcat}
Polar catastrophe theory at the \LAOSTO interface. This theory is a result of a polar discontinuity at the interface. The polarity of the \LAO layers changes between -1 and +1 while \STO layers are not charged.\cite{Nakagawa2006} The charge distribution along the growth direction is shown right next to each layer. The electric field resulting from this charge distribution has a step function feature. The electric potential calculated from this electric field by simple integration diverges as the thickness of the \LAO layer grows.(Part A) To overcome this infinite potential 1/2 electron per unit cell is transferred from the surface of the \LAO to the interface. (orange layer in Part B) (from \cite{Nakagawa2006})
}

\end{figure}

The polar catastrophe theory is based on the fact that layers of \LAO and \STO  have different polarities. They both crystallize in Ruddlesden-Popper stacking that are alternating layers of AO and BO$_2$  \cite{Ruddlesden1958} in the general representation of perovskite oxides as ABO$_3$ (\figref{fig:polarcat}). However, the main difference is that the layers of \STO are non-polar while \LAO stackings have a polarity of -1 and +1  when they are grown in the [001] direction.  A divergent potential originates from this polar discontinuity as depicted in \figref{fig:polarcat} part A. To overcome this discontinuity, reconstruction of the charge distribution is required. The reconstruction occurs by charge transfer to the interface and prevents the potential from rising to infinity at the surface.(\figref{fig:polarcat} part B)

In their seminal paper Ohtomo and Hwang \cite{Ohtomo2004} argue that the carriers should come from the polar discontinuity effect since they eliminated most of the possible oxygen vacancies by annealing samples at high temperature and quenching to room temperature rapidly. However, they also report that the extremely high carrier densities such as $10^{17}$cm${^{-2}}$ at some samples might be the indication of the effects of oxygen vacancies which is the second possible explanation for the origin of the carriers. Popovic \etal \cite{Popovic2008} has studied the electron gas in a \LAOSTO supercell by using density functional theory with the generalized gradient approximation (GGA) and concluded that the carrier density for the intrinsic case would be around $10^{13}$cm$^{-2}$ without oxygen vacancies which supports the arguments of Ref. \cite{Ohtomo2004}.

\begin{figure}[h]
\centering

\centering
\includegraphics[width=1\textwidth]{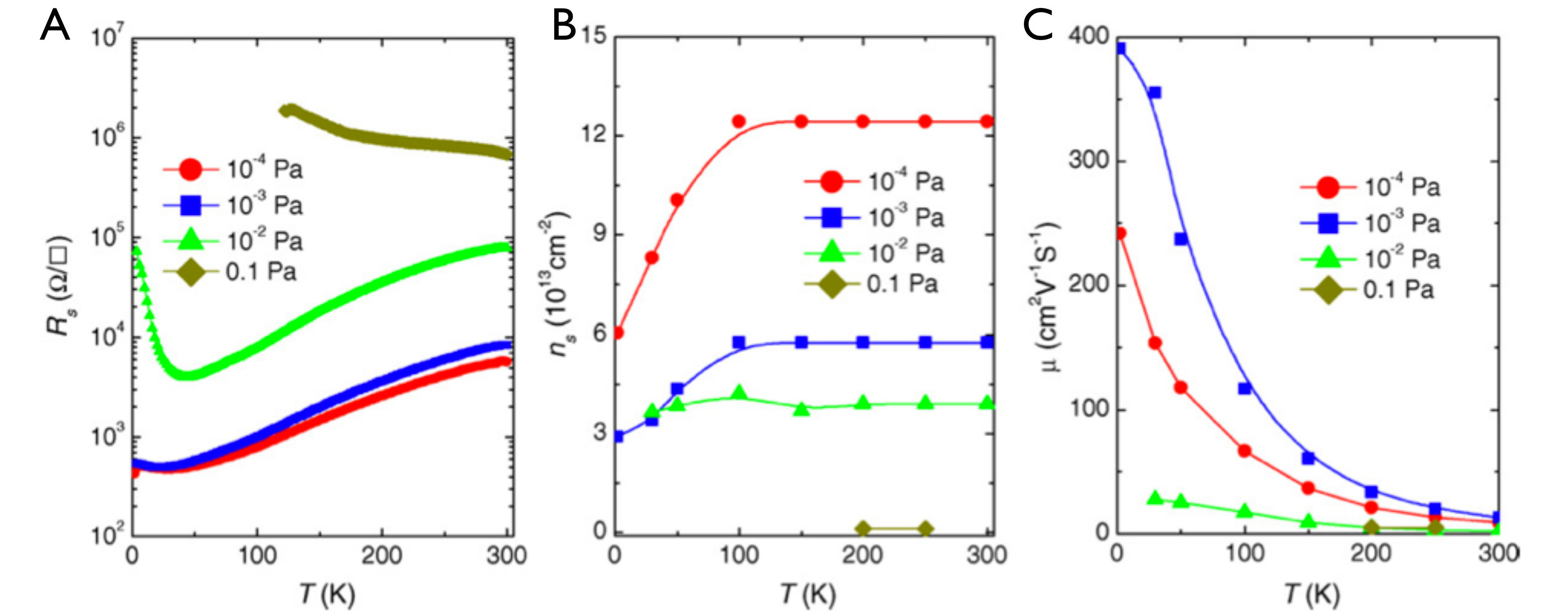}
\caption[Partial oxygen pressure dependence of the transport properties of LAO/STO 2DEGs]{\label{fig:oxygendependence}
Partial oxygen pressure dependence of the transport properties of LAO/STO 2DEGs. The first figure shows a strong dependence of the resistances of the samples on the amount of oxygen. The middle figure measures the increase in the carrier density as the oxygen pressure declines. The mobility does also strongly depend on the partial oxygen pressure in the growth chamber. (from Ref. \cite{Ariando2011})
}

\end{figure}
Oxygen has an ionic character with strong electronegativity and -2 charge in perovskite oxides. A possible lack of oxygen in the \STO or \LAO leaves two electrons free. These free electrons due to vacancies near the interface result in a metallic region at the interface. There have been many studies showing the strong dependence on different oxidation conditions during the growth process, especially in samples with very high electron densities. Ariando \etal \cite{Ariando2011} conducted transport measurements in four point van der Pauw geometry and shown that \LAOSTO samples that are grown in an environment with low partial oxygen pressures have higher carrier densities at the interface and vice versa. (\figref{fig:oxygendependence}) Furthermore Kalabukhov \etal \cite{Kalabukhov2007}  showed that the cathode luminescence (CL) of oxygen reduced \STO has the same color as the CL of \LAOSTO heterointerface while the photoluminescence measurement indicated the same wavelength of emitted light for both structures.
In a series of the experiments regarding the magnetotransport properties of \STOLAO, Herranz \etal \cite{Herranz2007} demonstrated an increasing conductivity of the interface by decreasing the partial oxygen pressure ($10^{-3}$-$10^{-6}$mbar), and hence proved a strong oxygen vacancy dependence. The conclusion of this discussion leads to an understanding that the origin of the carriers might be either oxygen vacancies or a polar catastrophe for different conditions. Additionally most of the time both mechanisms work together in the formation of an interfacial conducting layer.

\subsection{Control Mechanisms}
\begin{figure}[h]
\centering
\includegraphics[width=1\textwidth]{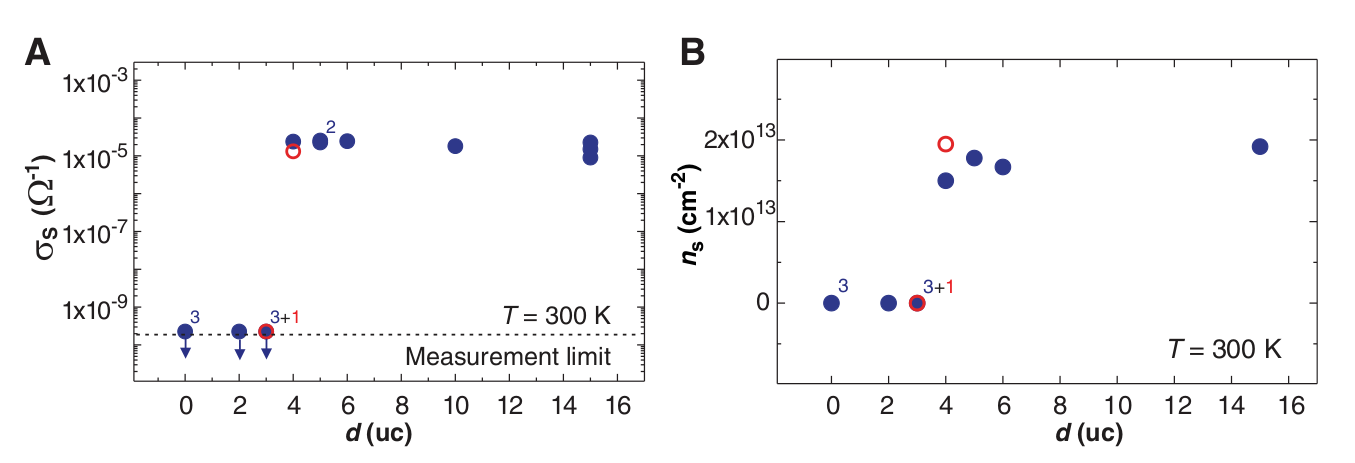}
\caption[The critical thickness for 2DEG formation at the oxide interface] {\label{fig:criticalthickness}
The critical thickness for 2DEG formation at the oxide interface. (from \cite{Thiel2006}) Figure A shows the \LAO thickness dependence of the sheet conductance. After a critical thickness of 3 unit cells, the conductance rises to a measurable level. Figure B clearly shows the formation of the 2DEG after three units cells by a carrier density measurement. Blue data is for samples grown at 770\degree C while red data is for samples grown at 815\degree C}
\end{figure}

One of the main features of the \LAOSTO interface is the critical thickness of \LAO that should be deposited to create the 2DEG. This critical thickness has been shown to be about 4 unit cells. Thiel \etal \cite{Thiel2006} demonstrated that a conducting layer of 2DEG develops after the 3rd unit cell by growing \LAOSTO layers with different \LAO thicknesses. (\figref{fig:criticalthickness})

As another control mechanism, strain has been studied by Jalan \etal \cite{Jalan2011} for La doped \STO thin films that are grown by molecular beam epitaxy (MBE). Strain is applied to the system by a three-point bending apparatus. Using Hall effect and four-terminal magnetotransport measurements, it is shown that strain of approximately -0.3\% can be used to enhance the mobility by more than 300\%  with no apparent limit. Enhancement in the mobility as a response to the strain can be understood from the change in the band structure. Strain splits the degenerate conduction bands resulting in separate light and heavy bands. The band with lighter effective mass is occupied more compared to heavier one, which causes an increase in the mobility. Another effect comes from reducing the number of domains due to strain. Domain boundaries usually scatter carriers causing lower mobilities.  Moreover, strain affects the critical thickness and causes it to increase from 4 unit cells up to 15 unit cells while reducing the carrier density with a compressional strain. Bark \etal  \cite{Bark2011} also demonstrated that tensile strain prevents the formation of the 2DEG.

\begin{figure}
    \includegraphics[width=\textwidth]{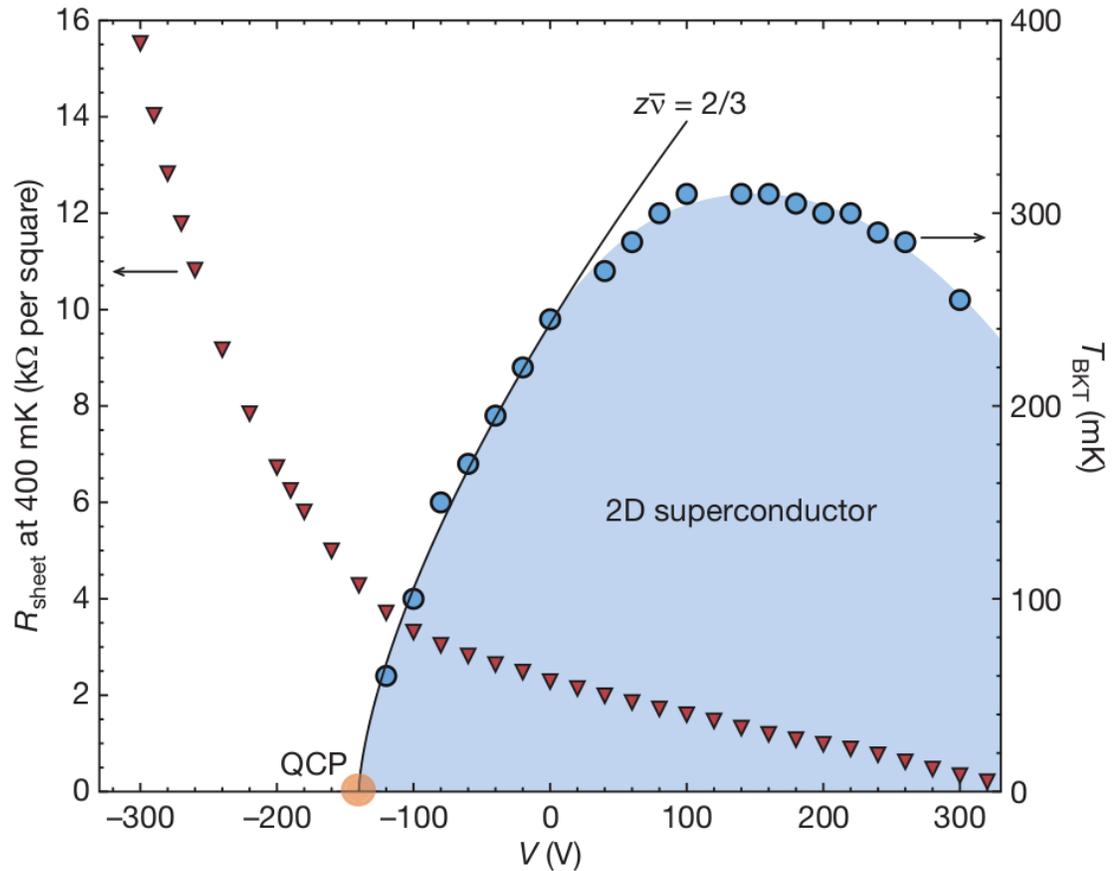}
 \caption[Phase diagram of the superconducting critical temperature]{\label{fig:phasediagram}Phase diagram of the superconducting critical temperature.
Caviglia \etal \cite{Caviglia2008} has conducted several transport measurements showing the tunability of the critical temperature with applied gate voltage. The blue region shows the superconducting phase while the 2DEG is insulating in the white sections. Gate voltage dependence of the critical temperature indicates an apparent reliance on the carrier density. After a certain positive gate voltage, the critical temperature starts to drop indicating overdoping.}
  \end{figure}

One of the most important ways to control the properties of 2DEGs is through field effects. For instance, the conductivity of the \LAOSTO interface can be altered in a broad range from insulating to conducting and even superconducting by gate fields. Such an experiment has been carried out by Caviglia \etal \cite{Caviglia2008}, and they obtained a conductivity phase diagram. (\figref{fig:phasediagram}) A quantum phase transition has been observed which allows on/off switching of the superconducting phase. The voltage dependent sheet resistivity and phase transitions from insulator to superconducting and metallic phases allow field effect tuning of the electronic properties.

Furthermore, Cen \etal \cite{Cen2008} and Xie \etal \cite{Xie2010} have reported the possibility of creating and erasing conducting islands at these interfaces by using an atomic force microscope probe as a voltage source.  A positive voltage of 4V is sufficient to change the conductivity of the interface locally. This process allows the writing and deleting of conducting regions reversibly, and also creates conductive regions with long lifetimes (up to 24 hours).

\subsection{Spin Properties}
There has been a growing interest in the spin properties of LAO/STO 2DEGs because they may be used as channel materials in future spin transistors. Two of the earliest experiments done on  spin injection in these 2DEGs were conducted by Reyren \etal \cite{Reyren2012} and Bibes \etal \cite{Bibes2012}. They confirmed spin injection from three terminal direct and inverted Hanle measurements, which is related to the change in the voltage due to spin polarization within the sample. The fundamentals of this kind of spin experiment follow. From a ferromagnetic contact (in this case cobalt)  electrons are injected with a certain polarization into the channel. The electrical Hanle effect is observed by applying an external magnetic field which is perpendicular to the magnetization of the ferromagnetic injector. The resultant magnetic field reduces the spin accumulation which can be verified by measuring negative magnetoresistance. This proves the existence 
of spin injection from the ferromagnetic contact to the 2DEG channel for temperatures below 150K. Furthermore they have shown that the spin signal can be amplified by applying a gate voltage in addition to tuning the carrier density. In a similar three terminal Hanle measurement Han \etal \cite{Han2013} demonstrated spin injection into lanthanum and niobium doped \STO and determined the spin lifetime to be 100 ps. They also mention the negative effect of scatterings at the tunnel barrier and \STO interface. In addition to this Caviglia \etal \cite{Caviglia2010} reports a large Rashba spin-orbit coupling in \LAOSTO 2DEGs and manipulation of this coupling using external electric fields. The Rashba effect is a direct result of the broken structural symmetry across the interface. This effect can be tuned up to a magnitude of 10 meV  which is comparable to the intrinsic spin-orbit coupling in the system.
\subsection{Challenges and Disputes}
The band structures are essential elements for understanding the electronic properties of materials, and a variety of different techniques have been used to calculate the electronic properties of \STO and other oxides. Each of them has advantages and drawbacks. For example Soules et al.\cite{Soules1972} calculated the electronic structure of \STO using a non-relativistic, \textit{ab initio}, self-consistent tight-binding method. Although their results are in good agreement with the ordering of bands, their calculation of the band gap is 12 eV, far larger than experimental results. Kahn and Leyendecker \cite{Kahn1964} investigated the \STO band structure by the Slater-Koster \cite{Slater1954} tight-binding model and correctly computed the band gap and effective masses. However, their approach was questioned by Simanek and Sroubek \cite{Simanek1965} for their treatment of  ionicity. Kahn and Leyendecker adjusted the ionicity of oxygens from -2 to -1.7 in order to fit the observed band gap, since with -2 charged oxygens and 3+ charged titanium and +2 charged strontium the band gap would be 17 eV. Therefore, they reduced the ionicity and thus the Madelung potential of oxygen and assumed a 15\% covalency. However, this approach doesn't take into account the spin-orbit coupling in the system. There have been several contradicting experiments about the spin-orbit coupling in the \LAOSTO system. Different experimental values for the spin splitting have been reported, including 0 meV, 18 meV, and 30 meV (Bistritzer \etal \cite{Bistritzer2011} and references therein).

In addition to these challenges there occur several phase transitions in oxides as the temperature of the system is decreased. Below the temperature 100K \STO undergoes a second-order phase transition from cubic to tetragonal structure.\cite{Lytle1964} The effect of this phase transition on the conduction bands of the \STO has been studied by Mattheiss \cite{Mattheiss1972a} and it was shown that the TiO$_6$ octahedra in \STO start to rotate around the z-axes as the temperature falls below 110K. This rotation breaks the cubic symmetry and causes a tetragonal lattice where the c/a ratio becomes 1.00056. However  this effect shows itself in the tight-binding approach only if second neighbour interactions are included.  Cao \etal \cite{Cao2000} studied the phase transition under epitaxial stress and concluded that the transition temperature can be altered and increased by 1.2 K when an epitaxial stress of 13.5 MPa is present. At temperatures lower than 50 K more complicated phase transitions can be observed.

\section{Bismuth Based Materials}
\subsection{Bismuth-Antimony Alloys: \BiSb}
\begin{figure}[h]
\centering
\includegraphics[width=1\textwidth]{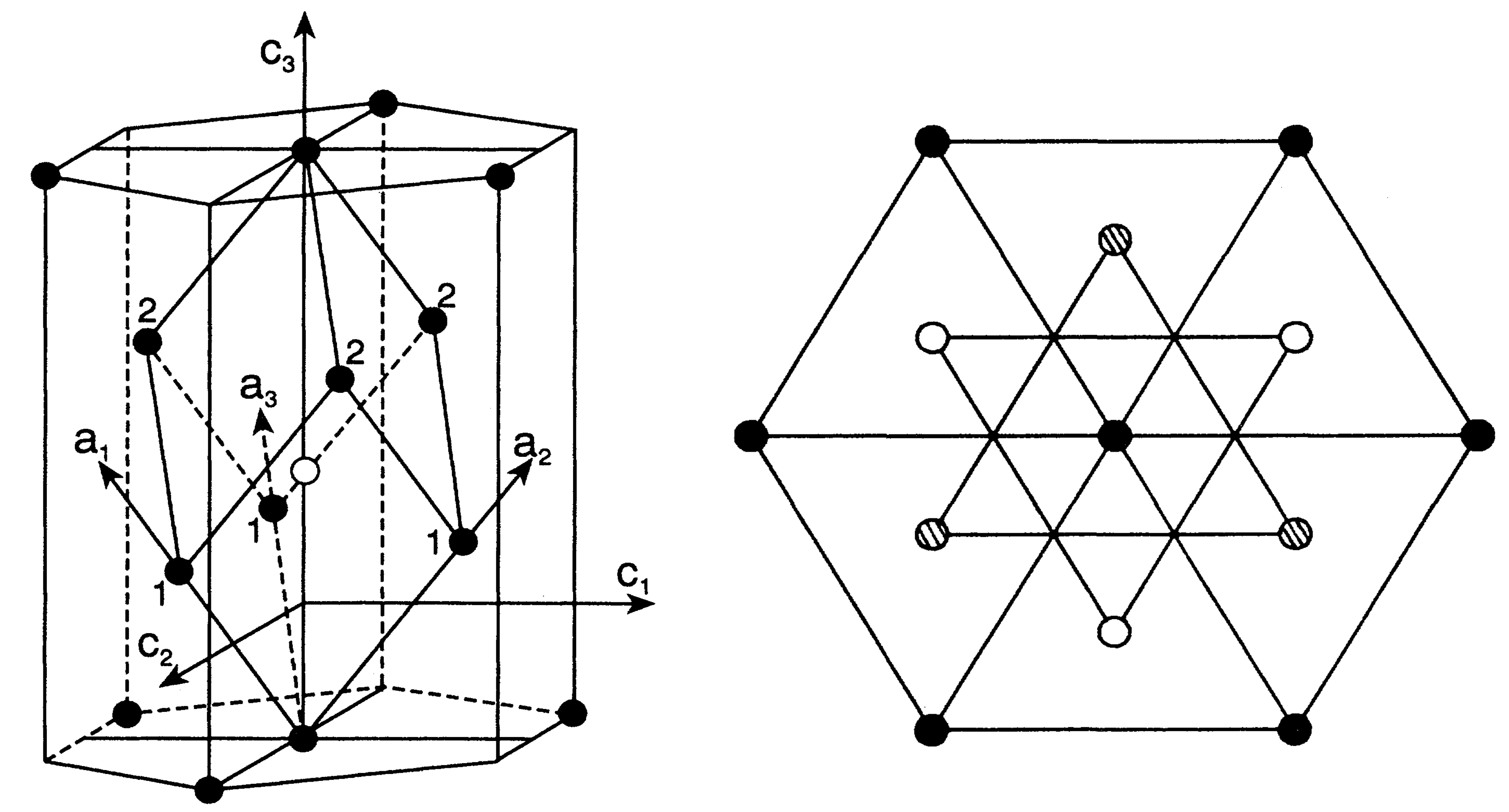}
\caption[Crystal structure of bismuth and antimony]{\label{fig:bicrystal}
 Crystal structure of bismuth and antimony. The rhombohedral unit cell of bismuth and antimony can be viewed as a deformation of a cubic unit cell along the [111] direction that creates the rhombohedral angle. (left figure) Here $\v a_1$, $\v a_2$ and $\v a_3$ are primitive lattice vectors while x,y and z-axis are denoted as c$_1$, c$_2$, and c$_3$, which are bisectrix, binary and trigonal axes. Top view of the same unit cell shows three nearest neighbors. (right figure). The first nearest neighbors are shown with empty circles; the second nearest neighbors are plotted with dashed circles. The central atom and the third nearest neighbors are shown as full black circles.(from Ref. \cite{Liu1995})
}
\end{figure}

Bismuth and antimony are both semimetals with rhombohedral crystals (also known as A7 structure) with a space group of $D_{3d}^5$ (R$\bar 3$m) and a point group $D_{3d}$. ($\bar 3$m) \cite{Ast2003} as shown in \figref{fig:bicrystal}. There are two atoms per unit cell which are separated by a vector $\v d =(0,0,2\mu )c$, where $\mu$ is the internal displacement parameter and c is the lattice constant of the hexagonal unit cell that is conventionally used. The central atom shown as an empty circle in \figref{fig:bicrystal}, and has three nearest neighbors located in the plane above at $\v a_1 -\v d$, $\v a_2 -\v d$, and $\v a_3 -\v d$ as well as three second nearest neighbors located in the plane below at $\v a_1 +\v a_2 =\v d$, $\v a_1 +\v a_2 -\v d$, and $\v a_2+\v a_3 -\v d$. Here $\v a_1$, $\v a_2$ and $\v a_3$ are primitive lattice vectors. There are also 6 third-nearest neighbor which are shown in \figref{fig:bicrystal} (right figure). The nearest and second nearest neighbor distances are almost identical each other, that leads to inclusion of the third nearest neighbor interactions into the electronic band structures.

The valence band is higher in energy than the conduction band by 40 meV in bismuth and by 180 meV in antimony which results in an indirect negative band gap and free electrons and holes \cite{Liu1995}. In group V semimetals, many unconventional electronic properties originate from their unique band structures. For instance, the electron effective masses are small, 0.06 m$_0$\cite{Isaacson1969, Smith1964} on x and y-axes for bismuth and 0.091 m$_0$ \cite{Datars1964, Issi1979} along [111] for antimony. Small effective masses and small conduction and valence band overlap make them ideal semimetals for quantum confinement studies \cite{Ogrin1966, Huber2007}. In fact, some significant experiments in condensed matter physics were done and first explored on bismuth such as the first experimental study of the Fermi surface in metals \cite{Shoenberg1939}, the Nernst-Ettingshausen effect \cite{Nernst1886}, the Shubnikov-de Haas effect \cite{Schubnikov1930}, and the de Haas-van Alphen effect \cite{deHaas1930}. Additionally, bismuth and antimony, as well as \BiSb alloys, were extensively studied in terms of their thermoelectric properties (which is the utilization of electrical energy for extracting heat for cooling or vice verse) \cite{Hicks1996}. \BiSb alloys can be semimetallic or semiconducting depending on the alloy concentration, and for small concentrations of antimony, \BiSb can be used as an n-type thermoelectric compound at room temperature \cite{Yim1972}. 

From the perspective of spintronics, one of the most important features of these group V semimetals is their enormous spin-orbit couplings, which are 1.5 eV and 0.6 eV for bismuth and antimony \cite{Gonze1990}. and a lot larger compared to spin-orbit splittings in conventional semiconductors, such as silicon 0.044 eV, carbon 0.06 eV, GaAs 0.34 eV and Ge 0.3 eV \cite{Madelung1986}. Most of the spin properties of materials are directly linked to the strength of the spin-orbit interaction in the system. Therefore, bismuth and antimony are of particular interest. Many properties, such as spin Hall conductivity, are expected to be expressed robustly in these materials. Bismuth also is shown to have an enormous spin-diffusion length, 70 $\mu$m, which can be increased up to 230 $\mu$m by alloying with Pd \cite{Lee2009}. Recent experiments using spin pumping techniques on amorphous bismuth indicated a large spin Hall response \cite{Emoto2014,Hou2012} as well as in earlier experiment on bismuth wires \cite{Fan2008}. In these experiments, the spin Hall signal depends mainly on temperature and signal is lost at room temperature. A similar experiment on the inverse spin Hall effect conducted in bismuth-permalloy films, confirmed a strong spin response of bismuth related materials. However, there hasn't been a reliable experiment on the spin Hall conductivity of bulk bismuth, antimony, or \BiSb alloys. Bismuth and antimony alloys also exhibit a semimetal-semiconductor transition at certain concentrations of antimony resulting in the opening of a small gap at 10\% of antimony \cite{Cho1999}. Both \BiSb alloys and bismuth thin films form topologically protected edge and surface states as confirmed by several experiments \cite{Wada2011, Hsieh2008, Hsieh2009, Bihlmayer2010}. Spin split surface states were also observed \cite{Hirahara2007}.

\subsection{Bismuth Chalcogenides: \BiSe and \BiTe}\label{sec:bischal}
\begin{figure}[h]
\centering
    \includegraphics[width=1\linewidth]{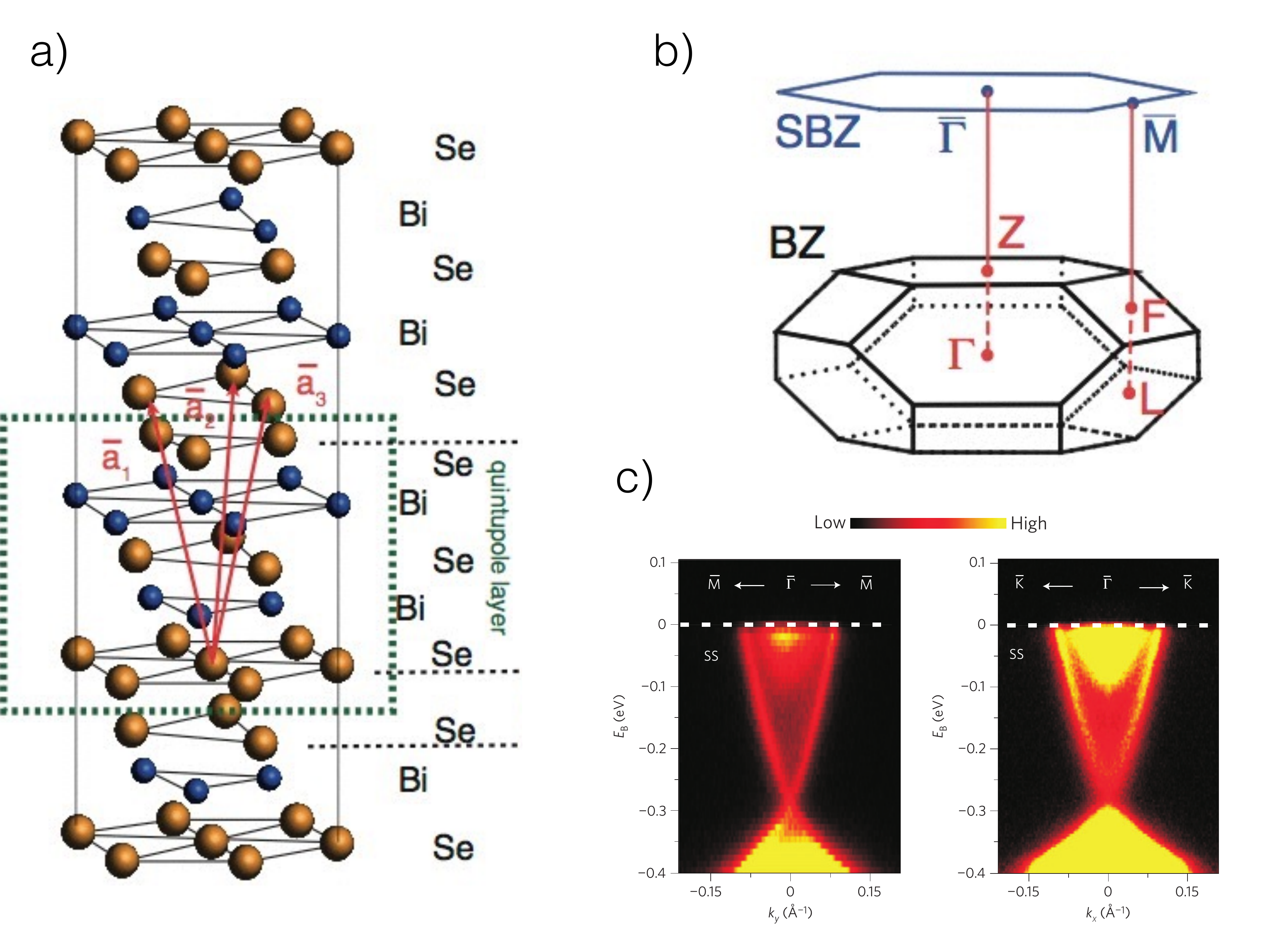} 
  \caption[Crystal structure of \BiSe with quintuple layers and lattice vectors]{Crystal structure of \BiSe with quintuple layers and lattice vectors. a) One of the quintuple layer is demonstrated by the green box. b) Three-dimensional bulk (BZ) and two-dimensional surface (SBZ) Brillouin zone (from Ref.~\cite{Bianchi2012}, and c) ARPES (angle-resolved photoemission spectroscopy) measurement of Dirac cones of the surface electronic band structure in two directions, from $\Gamma$ to K and from $\Gamma$ to M. (from Ref.~\cite{Xia2009})}
   \label{fig:bise}
\end{figure}

Bismuth chalcogenides that include compounds in the form of Bi$_2$A$_3$, where A is the chalcogen atom, have a crystal structure similar to bismuth and antimony. This results in 5-fold layered structures, namely quintuple layers (QL) as shown in \figref{fig:bise} part a). The crystal has a rhombohedral unit cell with D$_{3d}$, R$\bar 3$m, point and space groups respectively. The unit cell is converted to a conventional hexagonal unit cell by a transformation of axes as in the case of bismuth and antimony. The lattice vectors and axes are similar to the group V structure: the x-axis is the binary axis with twofold rotation symmetry, the y-axis corresponds to a bisectrix axis that is on the reflection plane, and the z-axis is along the trigonal axis that has three-fold rotation symmetry and is usually the growth direction. However, the structure of bismuth chalcogenides differs from group V semimetals by the fact that one unit cell contains five layers of atoms in the order Se$_2$-Bi-Se$_1$-Bi-Se$_2$ along c axis. The quintuple layers are connected by rather weak van der Waals forces while layers within the quintuple layer are covalently bonded with different bond lengths between Se$_2$-Bi and Se$_1$-Bi. Any alloy of the form of Bi$_2$Te$_{3-x}$Se$_x$ has the same crystal structure \cite{Nakajima1963}. Bismuth chalcogenides have a larger band gap than \BiSb alloys; \BiSe has a bulk band gap of 0.35 eV \cite{Pejova2004} and \BiTe has a band gap of 0.16 eV \cite{Black1957}.

Bismuth chalcogenides have been studied for their extraordinary thermoelectric properties for a long time, since they exhibit (along with their alloys) the largest figure of merit at room temperature \cite{Disalvo1999}. The figure of merit can be defined as $ZT=S^2 \sigma T/k$, where $S$ is the Seebeck coefficient, $\sigma$ is the electrical conductivity, T is the absolute temperature and $k$ is the thermal conductivity, and signifies how efficient a material is for thermoelectric purposes. However, they have attracted tremendous interest from the scientific community recently for a different reason. In 2011 King \etal \cite{King2011} discovered that bismuth selenide based two-dimensional electron gases can generate a Rashba effect that is orders of magnitude stronger than in other semiconductors, such that the Rashba parameter is around 1.3 eV\AA ~even at room temperature. Recently injection and detection of spin-polarized currents by spin potentiometric measurements in the bismuth chalcogenide alloys (Bi$_2$Te$_2$Se) has been demonstrated \cite{Tian2015}. Tunable, large Rashba coefficients, as well as the possibility of creating and detecting spin current in these materials provide an excellent opportunity for spintronic applications along with the topologically protected surface states, which will be discussed next.

\subsection{Topological Insulators based on \BiSb, \BiSe and \BiTe}
The topological insulator is a new phase of matter in which materials possess a band gap and are insulating in the bulk while the material surface (or edge in two dimensions) is conducting and contains topologically protected surface (edge) states \cite{Kane2005,Zhang2009}. Topologically invariant quantities are protected under smooth transformations of the manifold to which they belong. This is called the topological order and this strong order can only be broken  by a metallic transition by closing the bulk band gap \cite{Moore2007, Murakami2007}.  As a result of the strong spin-orbit interaction and time reversal symmetry, the surface states possess a spin texture, such that their spin is locked to the momentum of electrons (spin-momentum locking). These helical states result in the observation of the quantized spin Hall effect \cite{Qi2010, Roy2009}. Topologically non-trivial states are immune to backscattering by impurities or imperfections and are of particular interest for many electronic and spintronic studies \cite{Zhao2011, Kobayashi2011a}. Another feature of topological insulators is that energies of the surface states describe a Dirac cone with linear dispersion similar to graphene. This feature can be detected by an experimental tool that, angle-resolved photoemission spectroscopy (ARPES) (\figref{fig:bise} part C).

Studies concerning topologically insulating materials were pioneered by several groups in the most recent decade. \BiSb alloy is the first material where three-dimensional topologically protected surface states were observed \cite{Hsieh2008} and theoretically predicted \cite{Fu2007}. Furthermore, the \BiSb alloy at x=9\% antimony concentration exhibits an enormous mobility (up to 85000 cm$^2$/V$\cdot$s). Spin texture of the surface states of topological insulators have also been verified by scanning tunneling microscope techniques for \BiSb \cite{Roushan2009} and \BiTe \cite{Zhang2009a}. The bulk band gap in which surface states live is small for \BiSb alloys, however other bismuth related materials such as bismuth chalcogenides offer the opportunity of obtaining such states within a larger band gap \cite{Xia2009, Zhang2009}. Furthermore, recently it has been shown that \BiSe topological insulators have robust spin Hall conductivities ranging from 500 to 1000 $(\hbar/e)\Omega^{-1}cm^{-1}$ at room temperature \cite{Mellnik2014}, thus confirming the significance of these materials with robust spin-orbit couplings and topologically protected surface states for spintronic studies.

\section{Summary}
In conclusion, future spintronics devices require materials that possess a broad range of flexibility with tunable spin properties. 2DEGs at the \LAOSTO interfaces serve as promising candidates for both electronic applications and spin-based devices. Their highly correlated electrons with large mobilities and tunable Rashba couplings indicate that these 2DEGs may be an ideal channel material for spin transistors such as suggested by Datta and Das \cite{Datta1990}. However, there have not been many theoretical studies concerning the spin properties of this interface. The spin lifetimes and spin diffusion lengths are not known and the spin current responses to the charge currents have not been investigated. This research serves as a comprehensive introduction to spin calculations of complex oxide interfaces, specifically for the 2DEG at the \LAOSTO interface.

On the other hand, bismuth-based materials such as \BiSb alloys, \BiSe and \BiTe topological insulators with enormous intrinsic spin-orbit couplings and topologically protected surface states are one of the primary candidates for generating pure spin currents through the spin Hall effect. There has not been any theoretical or experimental study of most of these materials. This study calculates that bismuth-antimony alloys and bismuth chalcogenides exhibit giant spin Hall conductivities compared to conventional semiconductors, and confirms that they have spin properties tunable by gate voltage or doping. 

In the next chapter, Chapter 2, I will describe the tight-binding method which is used to obtain the Hamiltonian of systems for which spin calculations are carried out. Chapter 2 will be followed by the derivation and calculation of the effective spin-orbit interaction from the eigenstates and eigenvalues of this Hamiltonian in Chapter 3. This chapter also serves as a basis for the computation of the spin lifetimes in complex oxides from the Elliott-Yafet relaxation mechanism and scatterings by impurities in the same section. Chapter 4 shows the calculation of another significant spin property, the intrinsic spin Hall conductivity, from the Kubo formula for 2DEGs at \LAOSTO interfaces, bismuth-antimony alloys, and bismuth chalcogenides. The concluding section of this work, Chapter 5, summarizes the results of the spin dynamics calculations and remarks on directions to which this study could be extended in the future. The appendix includes the tight-binding Hamiltonians that are constructed in this study for future reference.

\chapter{Tight-Binding Method}

\section{Introduction}
There exist several techniques to calculate the electronic band structures of solids, such as uding the \kp Hamiltonian which is based on momentum matrix elements and the symmetry properties of crystals \cite{Kane1956, Kane1959}, or pseudopotential method which is based on empirical parameters and uses the orthogonalized plane waves \cite{Phillips1959, Andersen1973}, or ab-initio, first principles methods based on density functional theory \cite{Gross2013 ,Levy1979}. The tight-binding method (TBM), which is also commonly known as the linear combination of atomic orbitals (LCAO), has been widely used in the last 60 years due to its ability to describe physical and chemical properties of materials accurately with a small number of interpolation parameters. The computational cost of tight-binding calculations is extremely small compared to methods based on density functional theory. Furthermore, TBM can produce electronic bands for the whole Brillouin zone while \kp theory focuses on bands near the zone center \cite{Hamaguchi2001}. 

TBM is based on the assumption that electrons are tightly bound to atoms. Hence, it is quite fast and reliable to calculate the wave functions and energies of valence band electrons. This technique can also be improved to give a reasonable description of the wavefunctions and energies of conduction band electrons as well. Atomic orbitals are not the only option as a basis. One can choose different bases such as Hartree-Fock atomic functions \cite{Chaney1971}, Gaussian type or Slater-Koster type atomic orbitals \cite{Slater1954}, or basis sets of the representations of the point group of the crystal. (such as at the $\Gamma$ point)\cite{Granovskii1973}. Although there are several ways of constructing TB Hamiltonian, one of the most widely used methods is the Slater-Koster parametrization. In their seminal paper, Slater and Koster \cite{Slater1954} explained how to construct such a Hamiltonian using Bloch sums of normalized and orthogonal L\"{o}wdin orbitals. This method takes a set of atomic orbitals, $\psi_n$, which have symmetry properties of s, p$_x$, p$_y$, p$_z$, d$_{xy}$, d$_{yz}$, d$_{zx}$, d$_{x^2-y^2}$, or d$_{3z^2-r^2}$ and constructs Bloch sums from these atomic orbitals with the periodic boundary conditions:
\al{
u_n (\v k , \v r)=\frac{1}{\sqrt N}\sum_{\v R_i}e^{i(\v k \cdot \v R_i)}\psi_n(\v r -\v R_i)
}
where N is the number of all unit cells in the crystal, n is the band index, and $\v R_i$ is the position of the atom within the unit cells. Then the matrix elements of the TB Hamiltonian between different Bloch sums are:
\al{
\matrixel{u_n(\v k , \v r)}{H}{u_m(\v k , \v r)}=\frac{1}{N}\sum_{\v R_i,\v R_j}e^{i\v k\cdot(\v R_j- \v R_i)}\int \psi_n^*(\v r -\v R_i)H_{tb}\psi_m(\v r -\v R_j)
}
The factor of 1/N cancels when the sum is done over all the unit cells. Furthermore, the position of one of the atoms in the unit cell can be taken as the origin, which makes $\v R_i$ vanish. The matrix elements become:
\al{\label{overlap}
\matrixel{u_n}{H}{u_m}=\sum_{\v R_j}e^{i\v k\cdot \v R_j}\int \psi_n^*(\v r)H_{tb}\psi_m(\v r -\v R_j)
}

This process reduces the integral to a simpler 2-center integral. The integral in Eq.~\ref{overlap} is called an overlap integral. At this point, one should decide how many nearest neighbors are relevant for the purpose of the research. Slater and Koster \cite{Slater1954} tabulated all of the possible overlap integrals between \textit{s},\textit{p}, and \textit{d} orbitals in terms of the $\sigma,\pi \text{~and},\delta$ bonds with appropriate directional cosines in their seminal paper. The basic problem of the TB method is to find all the matrix elements in a chosen basis and the number of nearest neighbors to construct the Hamiltonian. The values of overlap integrals are determined with the help of symmetry analysis and by fitting them to either experimental observations or other theoretical calculations of certain properties of the materials such as band gaps, effective masses, and g-factors.

\section{A Short Note on the Choice of the Tight-Binding Basis}
It is important to determine type and size of basis that will be used for the tight-binding Hamiltonian since every basis has its advantages and drawbacks. In theory, one could add many near neighbors to the tight-binding Hamiltonian. Adding more neighbors may improve the band structure and results in correct energies throughout the Brillouin zone. As a second option one may add more atomic orbitals, such as \textit{d} and \textit{f}-orbitals. In one of the earliest band calculations, Chadi and Cohen \cite{Chadi1975} used a basis with only s and p orbitals, namely the $sp^3$ basis to calculate the valence band structures of diamond and zinc-blende crystals. This simple basis is very successful in explaining the valence band structure of most semiconductors but lacks an accurate description of the conduction bands. As an alternative many studies \cite{Newman1984, Nestoklon2006}  tried to add second neighbor overlap integrals for Si$_{1-x}$Ge$_x$ alloys. By adding second neighbor interaction, they could fit energy values at the symmetry point L, as well as other major symmetry points. However this Hamiltonian results in an incorrect effective mass of the conduction band at the zone center.

To get accurate conduction bands with correct energies and effective masses is essential for optical calculations, spin relaxation times and the spin Hall effect. Although adding more neighbors may improve the accuracy of the tight-binding calculations for valence bands, it does not provide a better picture of the conduction bands. Furthermore, it may even cause extra complexities, such as too many parameters to fit. It should be noted that the tight-binding model uses atomic orbitals to express Bloch functions that are eventually used for wavefunctions. Valence bands have similar characteristics to bound atomic orbitals. Hence, TBM is a pretty good approach for these bands. However, in the case of conduction bands, localized atomic wavefunctions don't provide an adequate picture. To solve this issue Vogl et al.\cite{Vogl1983} included a s$^*$ orbital to their sp$^3$ basis and obtained the sp$^3s^*$ basis. This new 'fictitious' and "excited" s$^*$ band is used to fit the energy values of the conduction band and results in a better depiction of the conduction band.  
The success of the sp$3s^*$ basis for conduction band energies led scientists to use this for many calculations of the most zinc-blende type crystals. (mostly III-V semiconductors) However, this turned out to be ineffective for diamond type crystals and also inaccurate for a few zinc-blende compounds. \cite{Klimeck2000a,Oh2005} On the other hand, Grosso and Piermarocchi \cite{Grosso1995}  successfully computed spin splitting and the band structure with a sp$3s^*$ basis including second nearest neighbors. However, this attempt failed to express the energies at X and L points correctly. 

Jancu et al. \cite{Jancu1998} added \textit{d}-orbitals to these calculations and constructed sp$^3$d$^5$s$^*$ basis. This basis provides accurate band structures, and also allows calculations of magnetic properties of group IV and III-V crystals because of the existence of \textit{d} orbitals.

\section{Complex Oxides}
\subsection{Tight-Binding Hamiltonian for Strontium Titanate}
As explained in the previous chapter \STO has a perovskite crystal structure. A titanium atom sits in the center, and six oxygen atoms around the titanium are located at ($\pm\frac{1}{2}$a,0,0),(0,$\pm\frac{1}{2}$a,0), and (0,0,$\pm\frac{1}{2}$a) while a is the size of the unit cell. (\figref{fig:perovskite}) The overlap integrals between atomic orbitals are listed in the Slater-Koster tables \cite{Slater1954} in terms of direction cosines. For instance the interaction between a $d_{xy}$ orbital  and $p_x$ orbitals is of the form:
\al{
E_{x,xy}=\sqrt 3l^2m(pd\sigma)+m(1-2l^2)(pd\pi),
}
where l, m, (and n) are the directional cosines along the x, y, (and z) directions. From the locations of oxygen atoms it can be seen that the matrix elements of the Hamiltonian are not zero only when m$\neq 0$, that is for p$_x$ orbitals located at (0,$\pm\frac{1}{2}$a,0) and, in this case, the overlap integral is in the form of $E_{x,xy}=pd\pi$. The rest of the procedure follows  Eq.~\ref{overlap}:
\al{\label{eq:TBmatrixelement}
\matrixel{d_{xy}}{H}{x_2}=\sum_{\v R_j}e^{i\v k\cdot \v R_j}(pd\pi)=(e^{\frac{a}{2}i\ky}-e^{-\frac{a}{2}i\ky})pd\pi=2isin(\frac{a}{2}\ky)(pd\pi),
}
where $\v R_j$ runs over nearest neighbor oxygen atoms. One can construct each element of the Hamiltonian by this procedure. For a 14x14 Hamiltonian as in the case of \STO, this may take too much time. However, by using cyclical permutations and the Hermitian property of the Hamiltonian, this matrix can be constructed relatively quickly. The unknown values of overlap integrals (such as pd$\pi$) are then fitted by using experimental results, such as the band gaps, effective masses, and energy values at certain high symmetry points of the crystal.

The tight-binding parameters of the Hamiltonian for \STO were studied by Kahn and Leyendecker \cite{Kahn1964}. They considered the interactions between d orbitals of titanium and p orbitals of oxygen atoms as the backbone of the Hamiltonian while s orbitals of oxygen and strontium were omitted. Energies of neglected orbitals are either far below or above the band gap. Therefore, they don't play a significant role in the electronic structure or transport properties. Three oxygen ions with three  2\textit{p} orbitals and one titanium ion with five 3\textit{d} orbitals constitute the 14x14 Hamiltonian matrix in the basis of atomic wave functions. The magnitude of the overlap integrals and Madelung energies, etermined in Ref. \cite{Kahn1964}, are tabulated below:
\begin{table}[h]
\caption[TB Parameters]{TB Parameters}
\begin{center}
\begin{tabular}{|cc|cc|}
\hline
M$_{Ti}$  & -6.8 eV   &pd$\sigma$ & 2.1 eV \\
M$_{O}$  & -10.5 eV & pd$\pi$ & 0.84 eV \\
d$_{el}$ & 0.62 eV  &pp$\sigma$ & -0.16 eV \\
p$_{el}$ & 0.48 eV   & pp$\pi$ & 0.062 eV  \\
 \hline
\end{tabular}
\end{center}
\end{table}

\noindent M$_{Ti}$ and M$_{O}$ are the ionization potential and Madelung energy of Ti and O, d$_{el}$ and p$_{el}$ are electrostatic splitting of d and p orbitals respectively. The remaining four values are overlap integrals between various orbitals. I have constructed a 14x14 tight-binding Hamiltonian using  the parameters above which can be found in Appendix \ref{app:strontiumhamiltonian}.

\subsection{Strain Hamiltonian and Interface Effects}
For epitaxially grown strontium titanate films, it is highly possible to observe an effective strain which can have a large influence on the energy levels of the conduction bands. In the case of strontium titanate based two-dimensional systems we observe an interfacial quantum confinement effect which has same consequences as the strain.  The conduction bands that are not on the xy plane ( growth is in the z-direction) are shifted towards higher energies as a result of the strain (in bulk) or quantum confinement at the interface (in 2-dimensions)

The first three conduction bands of \STO with xy, yz, and zx symmetry transform as x, y, and z orbitals. From Janotti \etal \cite{Janotti2011}, we see that the effect of the strain on the conduction bands can be considered as same as the effect of strain on the valence band of zinc-blende crystals. Referring to the work by van de Walle \cite{Walle1989} one can obtain relations between deformation potentials and conduction band splittings as a function of strain in the system for the [001] and [111] direction. On the other hand, the valence bands of \STO should behave exactly the same way since they possess the same symmetry as the valence bands of zinc-blende crystals.

For a general strain in the tight-binding Hamiltonian, we observe two changes. The first one is the change in the lengths between atoms. They may be longer or shorter depending on the stress. The second change is the angle of the vector which connects two atoms. For instance, this angle is 90 degrees between titanium and oxygen atoms for the simple cubic \STO before strain. After strain, it changes the location of the atoms. Therefore, both the strength of the overlap integrals and the directional cosines must be altered as a result of strain. 

\noindent \textit{A general form of the strain}: If we consider u being the displacement vector of a point due to strain, and then strain can be expressed up to first order as:
\al{
\epsilon_{ij}=\frac{1}{2}\left (\pd{u_i}{x_j}+\pd{u_j}{x_i}\right )
}
In elastic theory, strain is connected to the stress by the stiffness tensor:
\al{
\epsilon_{ij}=\sum_{kl} S_{ijkl} \sigma_{kl}
}
Diagonal elements of  $\sigma_{kl}$ are called normal stress while non-diagonal elements constitute shear stress. For both stress and strain tensors the equality of $\sigma_{ij}=\sigma_{ji}$ holds. A vector (such as a primitive lattice vector) under the effect of general strain transforms as:
\al{
\v a' = (1+\epsilon \v a)
}

In the case of perovskite oxides, the locations of the three oxygen atoms transform for a general strain $\epsilon_{xy}$:
\al{
\v a_1 '= \frac{a}{2}(1+\epsilon_{xx}, \epsilon_{xy},\epsilon_{xz})\\
\v a_2 '= \frac{a}{2}(1+\epsilon_{yx}, \epsilon_{yy},\epsilon_{yz})\\
\v a_3 '= \frac{a}{2}(1+\epsilon_{zx}, \epsilon_{zy},\epsilon_{zz})
}
while titanium atom's position remains unchanged at the center. Additionally, the volume of a crystal unit cell with new primitive vectors is written as:
\al{
\Omega'=\Omega_0 (1+\epsilon_{xx}+\epsilon_{yy}+\epsilon_{zz})=\Omega_0 (1+Tr(\epsilon))
}
The distance between titanium and oxygen atoms changes from a/2 to 
\al{
d_i'=\frac{a}{2}\sqrt{(1+\epsilon_{ii})^2+\epsilon_{ij}^2+\epsilon_{ik}^2}
}
where i=x, y, or z. If we assume that $\epsilon_{ij}$ is small then the distance becomes: $a/2(1+\epsilon_{ii})$. One must modify the overlap integrals and directional cosines, according to the new distances and angles. The strength of the overlap integrals should also be adjusted by using Harrison's scaling law \cite{Froyen1979}, which states that the strength of the interaction is related to the bond length with an inverse square rule (d$^{-2}$ rule).

As a result of this modification of directional cosines and bond lengths, the matrix element in Eq.~\ref{eq:TBmatrixelement} will transform under a general strain as:
\al{
\matrixel{d_{xy}}{H}{x_2}=&\left [ \frac{\sqrt{3}\epsilon_{yx}^2(1+\epsilon_{yy}))}{(\epsilon_{yx}^2+(1+\epsilon_{yy})^2+\epsilon_{yz}^2)^{3/2}} pd\sigma ' 
+\frac{(1+\epsilon_{yy}) (1-2\epsilon_{yx}^2)}{(\epsilon_{yx}^2+(1+\epsilon_{yy})^2+\epsilon_{yz}^2)^{3/2}}pd\pi ' \right] \\
& \times 2i sin \left(\frac{a}{2} (k_x \epsilon_{xy}+k_y (\epsilon_{yy}+1)+k_z \epsilon_{yz})\right)
}
This expression can be greatly simplified by assuming that elements of the strain tensor are small, in other words $\epsilon_{ij}^2\approx 0$,  $\epsilon_{ij}^3\approx 0$, and $(1+2\epsilon_{yy})^{3/2}\approx 1+3\epsilon_{yy}$ This simplification results in a matrix element:
\al{
\matrixel{d_{xy}}{H}{x_2}=\frac{2 i (1+\epsilon_{yy}) \sin \left(\frac{a}{2} (k_x \epsilon_{xy}+k_y(1+\epsilon_{yy})+k_z\epsilon_{yz})\right)}{1+3 \epsilon_{yy}}pd\pi '
}
where $pd\pi ' = pd\pi(\frac{1}{1+2\epsilon})$ is the overlap integral which is scaled according to Harrison's rule at small strain. We also observe that this matrix element converges to Eq.~\ref{eq:TBmatrixelement} as the strain approaches zero. Other elements of the Hamiltonian can be studied and computed in a similar fashion. The full Hamiltonian with all the matrix elements can be found in Appendix \ref{sec:strainhamiltonian}. With this, the eigenvalues and eigenvectors of the Hamiltonian of strained complex oxides can be calculated efficiently. There are numerous directions of stress and strain widely used in the literature and experiments. Here I list stress and strain tensors in 3 different directions.\\
\noindent Stress and strain along [100]:
\al{
\sigma=
 \begin{pmatrix}
  P & 0 & 0  \\
  0 & 0 & 0  \\
  0  & 0 & 0   
 \end{pmatrix}
 ~~~~~
 \epsilon=
 P\begin{pmatrix}
  s_{11}& 0 & 0  \\
  0 & s_{12}& 0  \\
  0  & 0 & s_{12}   
 \end{pmatrix}
 }

\noindent Stress and strain along [110]:
\al{\sigma=
P \begin{pmatrix}
  1/2 & 1/2 & 0  \\
  1/2 & 1/2 & 0  \\
  0  & 0 & 0   
 \end{pmatrix} 
  ~~~~~
  \epsilon=
P \begin{pmatrix}
  s_{11}+s_{12} & s_{44}/2 & 0  \\
  s_{44}/2 & s_{11}+s_{12} & 0  \\
  0  & 0 & 2s_{12}   
 \end{pmatrix} 
}

\noindent Stress and strain along [111]:
\al{\sigma=
\frac{P}{3} \begin{pmatrix}
  1 & 1 & 1  \\
  1 & 1 & 1  \\
  1  & 1 & 1   
 \end{pmatrix} 
   ~~~~~
   \epsilon=
P \begin{pmatrix}
  s_{11}+2s_{12} & s_{44}/2 & s_{44}/2  \\
  s_{44}/2 & s_{11}+2s_{12} & s_{44}/2  \\
  s_{44}/2  & s_{44}/2 & s_{11}+2s_{12}   
 \end{pmatrix} 
}
\begin{figure}
\vspace{3\li}
\centering
\includegraphics[width=1\linewidth]{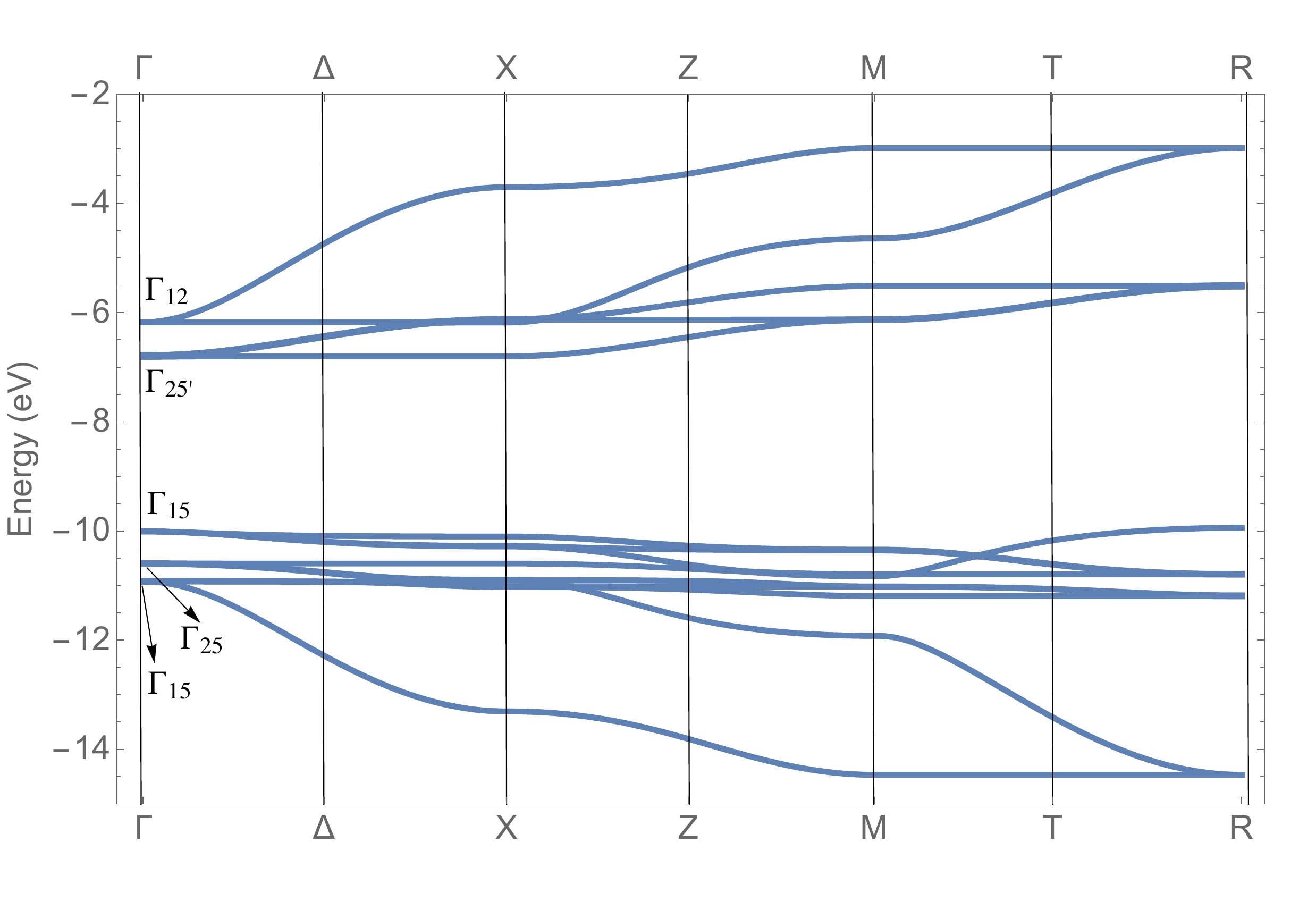}
\caption[Electronic band structure of strontium titanate]{Electronic band structure of strontium titanate. The band structure clearly shows that the minimum of the conduction band is located at the zone center ($\Gamma$ point). The valence bands at zone center consist of \textit{p}-orbitals of oxygen while they come from \textit{d}-orbitals of titanium for conduction bands. Both valence and conduction band pairs are separated from each other by the crystal potential, electrostatic potential, band gap and as well as a small splitting due to spin-orbit coupling. (30 meV at the conduction band minimum) The influence of the spin-orbit coupling is less than the band gap which is about 3.22 eV \cite{Cardona1965}. The maximum of the valence band is located at point R, which is very close to $\Gamma$ but slightly higher in energy. Besides the conduction band is flat along the [100] direction  ($\Gamma$-X), which gives rise to enormous effective masses in that direction.
}\label{fig:stobands}
\vspace{3\li}
\end{figure}

\subsection{Electronic Band Structures of \STO}
\begin{figure}[h]
\centering
\includegraphics[width=1\textwidth]{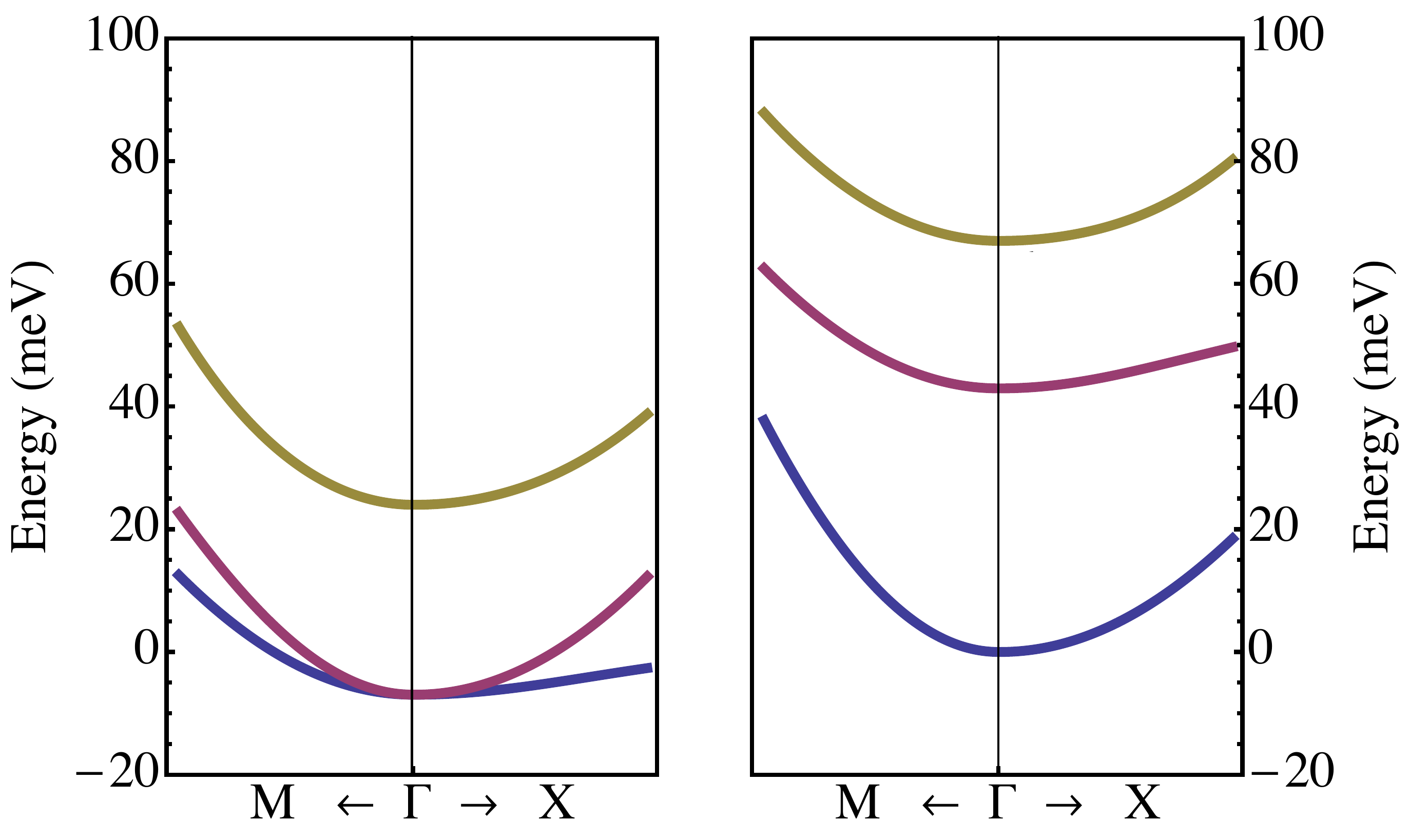}
\caption[Electronic band structure of 2DEG at the \LAOSTO interface]{Electronic band structure of 2DEG at the \LAOSTO interface. Doubly degenerate conduction bands of STO$_3$ are calculated by the Slater-Koster tight-binding method. Spin-orbit couplings are computed from atomic spectra tables of Moore \cite{Moore1,Moore2} and the splitting due to spin-orbit coupling is taken to be 30 meV. Notice the four-fold degeneracy of the lowest conduction band at the $\Gamma$ point consisting mostly of $d_{xy}$ and $d_{yz}$ atomic orbitals while the upper conduction band has mostly $d_{zx}$ character. Due to spin-orbit coupling in the system these states are mixed with each other. (left figure) A compressive uniaxial stress of 50 meV removes the four-fold degeneracy and results in three doubly degenerate conduction bands. (right figure) Here X=$\pi$/a(1,0,0) and M=$\pi$/a(1,1,0) where a is the lattice constant. }\label{fig:2degbands} 
\end{figure}

The electronic band structure is shown for the full Brillouin zone in \figref{fig:stobands}. Effects of the spin-orbit coupling and strain (or confinement at the interface) on the conduction bands can be seen in \figref{fig:2degbands}. For \STO the electronic states near the conduction band minimum at the Brillouin zone center mostly consist of Ti \textit{d}-orbitals. The lowest conduction band constitutes of 5-fold degenerate d-orbitals. The crystal potential splits these conduction bands into sixfold t$_{2g}$ bands (d$_{xy}$, d$_{yz}$, d$_{zx}$) and fourfold (higher-energy) e$_g$ bands (d$_{x^2-y^2}$, d$_{3z^2 -r^2}$ which are not shown in the figure); spin-orbit coupling results in a further splitting ($\approx$ 30 meV) of the lower t$_{2g}$ bands into fourfold and twofold bands, as shown in \figref{fig:2degbands}(a).  We consider strained STO, in which the compressive strain breaks the fourfold degeneracy at the $\Gamma$-point and results in well-resolved, doubly degenerate subbands in the plane perpendicular to the growth direction, as shown in Fig.~\ref{fig:2degbands}(b) for a splitting of $\sim 50$~meV. The same energy splitting is produced by an interface and leads to the electronic structure of the LAO/STO 2DEG\cite{Salluzzo2009} The spin-orbit couplings, absent in Ref.~\cite{Kahn1964}, are computed from atomic spectra tables\cite{Moore1,Moore2} by using the Land\'e interval rule. The 30 meV spin-orbit splitting from the atomic energies is in agreement with first principle calculations\cite{Marel2011}.
The constant energy surfaces of the lowest conduction band do not show any elliptical or spherical symmetry. However, by playing with the spin-orbit coupling and the amount of the strain in the system one can achieve a spherically symmetric lowest conduction band on the k$_x$-k$_y$ plane for a spin splitting of 30 meV and a strain splitting of 107 meV.

\section{Bismuth-based Materials}
\subsection{Hamiltonian and Band Structures of \BiSb Alloys}
\begin{figure}[h]
\centering
    \includegraphics[width=1\linewidth]{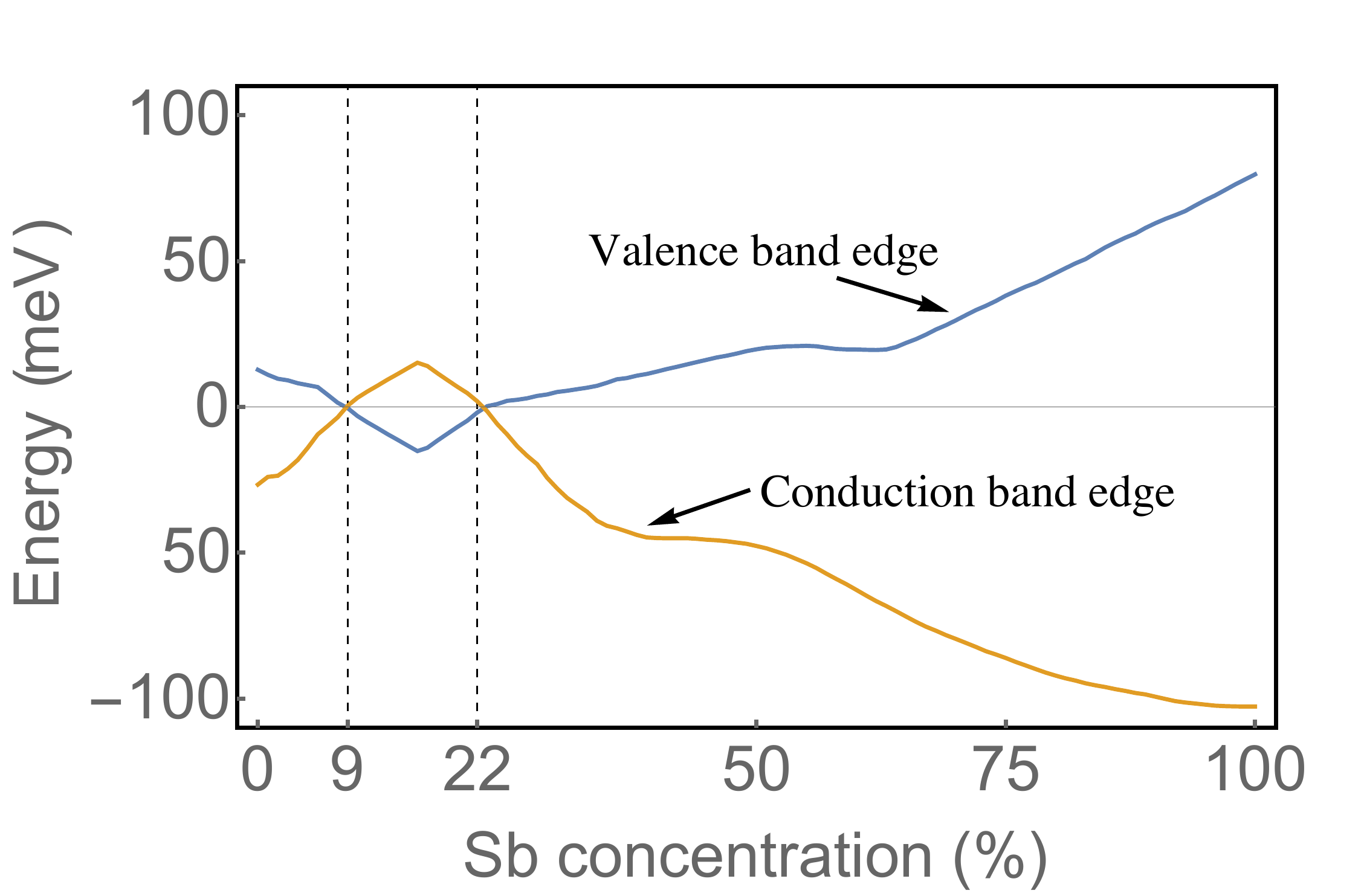} 
  \caption[Valence and conduction band edges as a function of Sb concentration]{Valence and conduction band edges as a function of Sb concentration. This plot reveals the characteristics of the alloy and locations where transitions between semiconducting and semimetallic states occur. A similar plot of the band edges calculated by an alternative virtual crystal approximation can be found in Ref. \cite{Teo2008} up to 20\% of antimony. This approximation is further discussed in Section \ref{sec:alternativeVCA}.}
   \label{fig:sbdependence2}
\end{figure}

 Bismuth and antimony are rhombohedral crystals (also known as A7 structure) with the space group of $D_{3d}^5$ (R$\bar 3$m) and point group $D_{3d}$. ($\bar 3$m) \cite{Ast2003}. Both materials are ideal semimetals for quantum confinement studies \cite{Huber2007} with enormous spin-orbit couplings, which are 1.5 eV and 0.6 eV respectively\cite{Gonze1990}. There have been many attempts to calculate band structures of this materials, such as early tight-binding models \cite{Mase1958}, pseudopotential approaches \cite{Golin1968} or simple 2-band models \cite{Brown1963,Cohen1961}, however these are overly simplified models and each lacks one important property, either g-factors, effective masses, or optical properties for \BiSb alloys. Liu and Allen \cite{Liu1995} developed and parameterized a tight-binding model with a $sp^3$ basis and a conventional hexagonal unit cell which contains two atoms. This Hamiltonian includes up to the third-nearest-neighbor interactions which is sufficient to mimic the characteristics of the electronic band structure and effective masses around the Fermi energy. First and second nearest neighbor atoms are very close to each other; as a result of that it is essential to include third nearest neighbor overlap integrals to the Hamiltonian. 
 \begin{figure}[t]
\centering
 \includegraphics[width=.96\linewidth]{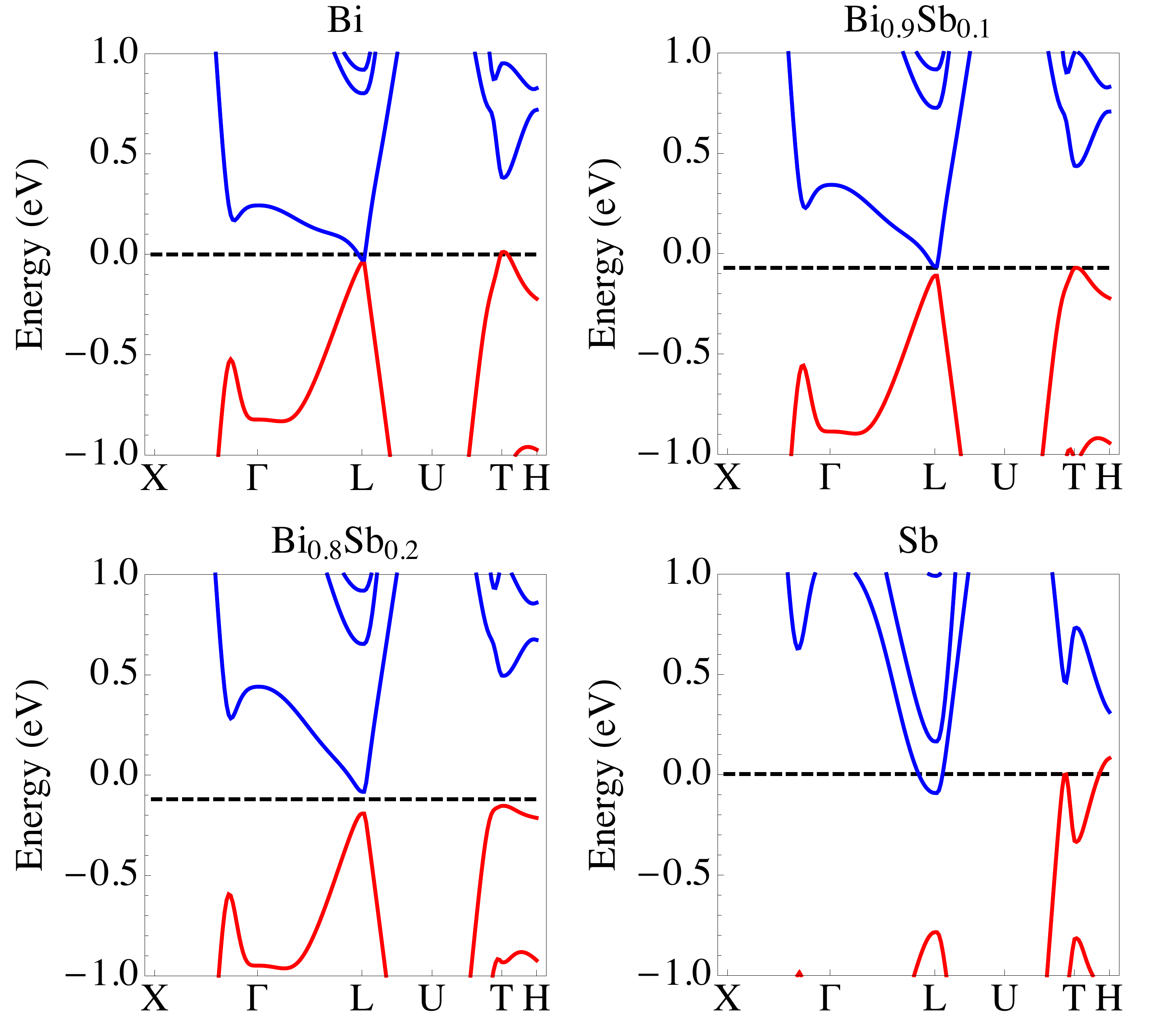} 
  \caption[Electronic band structure of \BiSb alloys]{Electronic band structure of \BiSb alloys. The band structure for a) pure bismuth and b) Bi$_{0.9}$Sb$_{0.1}$ with disappearing band overlap c) semiconducting Bi$_{0.8}$Sb$_{0.2}$ and d) pure antimony. Dashed lines indicate the location of the Fermi levels. Red and blue curves represent valence and conduction bands respectively.}
   \label{fig:bands}
\end{figure}
The semimetal behavior of Bi and Sb comes from slightly overlapping conduction and valence bands resulting in electron and hole pockets. The minimum of the conduction band is located at L point for both materials while the top of the valence band is at T for bismuth and H for antimony. The overlap between L (00$\frac{1}{2}$) and T ($\frac{1}{2}$ $\frac{1}{2}$ $\frac{1}{2}$) is 40 meV in Bi and between L and H (around T) is 180 meV in Sb \cite{Xu1993, Issi1979}. To calculate properties of bismuth and antimony alloys, we used the virtual crystal approximation (VCA) which is based on averaging tight-binding overlap parameters as a function of the alloy concentration. For an alloy of Bi$_{1-x}$Sb$_{x}$ this technique requires modifying overlap integrals (\textit{e.g.} $sp\sigma$) such that: \cite{Bellaiche2000}
\al{\label{eq:vca}
V_{Bi_{1-x}Sb_x}(sp\sigma)=(1-x)V_{\text{Bi}}(sp\sigma)+xV_{\text{Sb}}(sp\sigma)
}
The virtual crystal approximations allow us to observe that a semimetal-semiconductor (SMSC) transition occurs if bismuth is alloyed with antimony. There exists a certain range for the amount of antimony where valence and conduction bands are separated with a small direct band-gap. The electronic band structure around the Fermi energy is in Fig.~\ref{fig:bands} and the energies of the valence and conduction band edges with the Fermi levels are in Fig.~\ref{fig:sbdependence2} part a) for different antimony concentrations. For pure bismuth and antimony the Fermi levels are at 0 eV. Alloying bismuth with antimony causes the bands and Fermi energies shift to lower energies. At around 9\% of Sb the band overlap disappears, and we observe the SMSC transition. As the antimony concentration is increased, the valence bands move faster than the conduction bands, and therefore we find an opening of a gap. A maximum gap of 28 meV occurs for Bi$_{0.83}$Sb$_{0.17}$. Up to 22\%, the alloy is still a semiconductor with a decreasing indirect band gap. At 22\% of antimony another SMSC transition occurs, and the alloy becomes a semimetal and stays a semimetal with increasingly overlapping conduction and valence bands, which agrees with previous experiments(\cite{Cho1999} and references within).

 \subsection{Hamiltonian and Band Structures of \BiSe and \BiTe}
We briefly review the crystal structure of bismuth chalcogenides in Section~\ref{sec:bischal} with quintuple layers. Construction of this Hamiltonian requires an extra step since there are two types of selenium atoms in one QL. The first type is surrounded by 6 bismuth atoms; three in the upper layer and another three in the lower layer, while the second type of Se is has three selenium and three bismuth atoms at nearest neighbor distances.  The distances between Bi and Se$_1$ and between Bi and Se$_2$ are different. Furthermore, bonding between layers within the quintuple layer is stronger than bonding between QL layers. This is because between QLs, there are van der Waals bonds while the bonding is covalent between layers inside the QL.
 
We use the parameters provided by \cite{Kobayashi2011a} and construct a sp$^3$ basis TB model with two nearest neighbor interactions and spin-orbit coupling. This Hamiltonian is a 40x40 matrix with 51 TB parameters. This parametrization also assigns different tight-binding parameters between two orbitals of atom 1 and atom 2 that are located in a different order. For instance, an sp$\sigma$ overlap integral, which is hopping between an s-orbital at atom 1 and a p orbital at atom 2, is different than the ps$\sigma$ parameter which is the same interaction with interchanged orbitals. Exchanging one orbital at one atom with another orbital at another atom brings a minus sign only when the sum of the parities of two orbitals are odd. Otherwise interchanging s and d orbitals in parameters, such as ds$\sigma$ to sd$\sigma$ has no effect on the parameter.

\begin{figure}[h]
\begin{minipage}[t]{0.49\linewidth}
\includegraphics[width=1\textwidth]{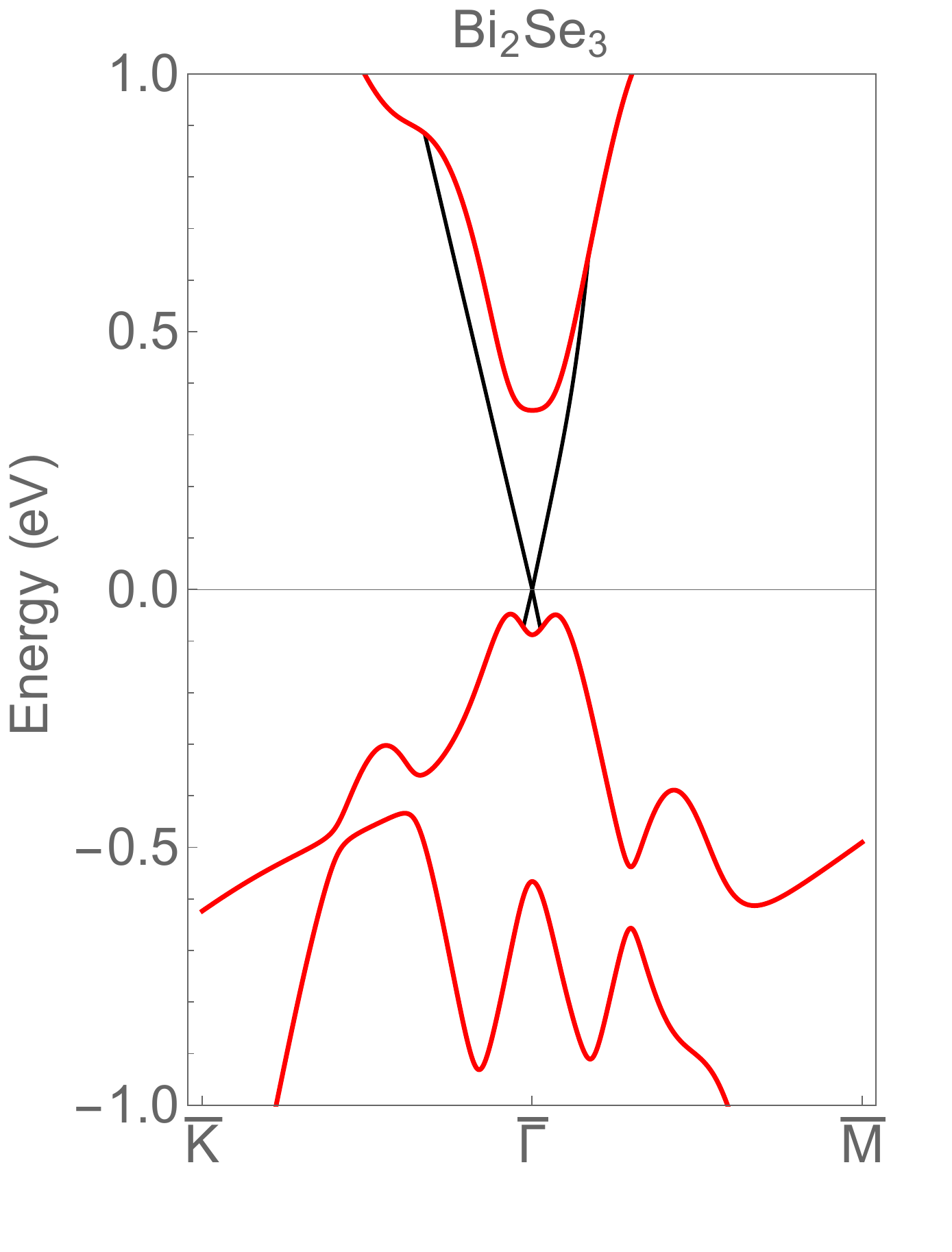}
  \end{minipage} 
  \begin{minipage}[t]{0.49\linewidth}
\includegraphics[width=1\textwidth]{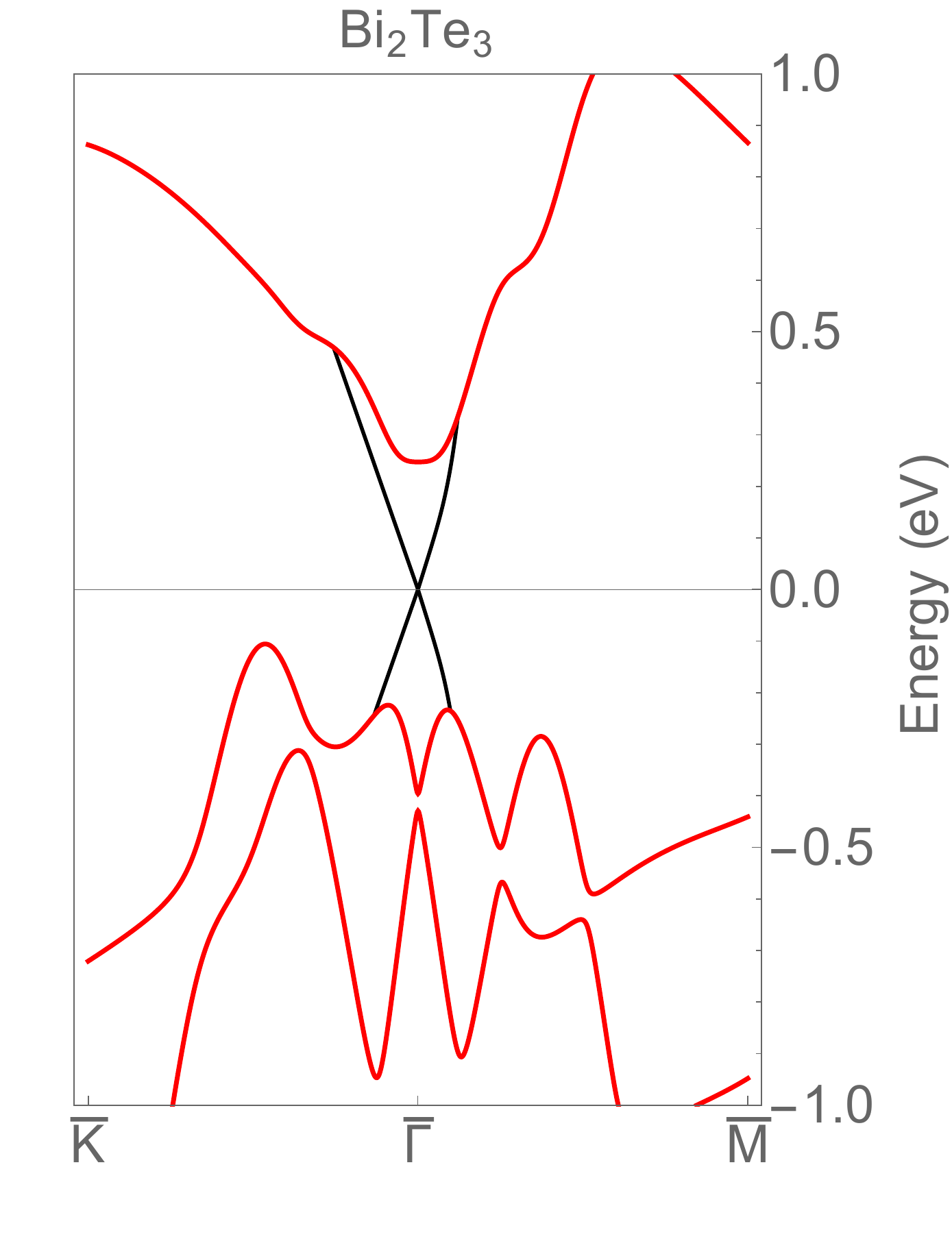}
\end{minipage}
\caption[The bulk and the surface bands of \BiSe and \BiTe crystals]{The bulk and the surface bands of \BiSe and \BiTe crystals. Electronic band structures of \BiSe and \BiTe are plotted around Fermi energy. Dirac cones are located at 0 eV. Here red figures represent bulk valence and conduction bands while black figures are for surface bands. Parameters for both bulk and surface band structures are taken from Ref.~ \cite{Kobayashi2011a}.}
\label{fig:bisetebands}
\end{figure}

By plotting band structures, we observe that neither conduction nor valence band edges are located at any of high symmetry points for both \BiSe and \BiTe as shown in \figref{fig:bisetebands}. In addition to the tight-binding model of the bulk bands, the Hamiltonian for the surface states can be modeled as:
\al{\label{surfaceHam}
H=v( \hat p_x \sigma_y -\hat p_y \sigma_x) +\frac{\lambda}{2}(\hat p_+^3 + \hat p_-^3)\sigma_z
}
where $v$ is velocity, momentum operators $ \hat p_\pm$ are defined as $\hat p_\pm=\hat p_x \pm \hat p_y$. As a result of the crystal structure the Dirac cones of bismuth chalcogenides are not perfectly spherical, rather they are hexagonally warped. The parameter $\lambda$ is the warping coefficient. This term is stronger in \BiTe compared to \BiSe usually. Here we plot both bulk bands calculated from the tight-binding Hamiltonian we constructed and the surface Hamiltonian from Eq.~\ref{surfaceHam}. Parameters for $v$ and $\lambda$ are  2.55 eV$\cdot$\AA~ and 250 eV$\cdot$\AA$^3$~ respectively for \BiTe, while they are 3.55 eV$\cdot$\AA~ and 128 eV$\cdot$\AA$^3$~for the \BiSe crystal. We observe that warping is stronger in \BiTe. Electronic band structures are obtained using parameters mentioned before as shown in \figref{fig:bisetebands}.

\section{Spin-Orbit Hamiltonian}
The basis of a NxN tight-binding Hamiltonian doubles its size once the spin-orbit coupling is introduced to the system. Then the total Hamiltonian takes this form:
\[
H_{\text{Total}}= 
  \begin{matrix}\\\mbox{}\end{matrix}
  \begin{pmatrix} H_{\text{TB}} & 0 \\ 0 & H_{\text{TB}} \end{pmatrix} +H_{\text{SO}}
\]
where H$_{\text{SO}}$ is the spin-orbit Hamiltonian which can be calculated from $H_{so}=\lambda \v L \cdot \v S$ in the Russell-Saunders coupling scheme. Here $\v L$ is the linear momentum operator, $\v S$ is the spin operator, and $\lambda $ is the strength of the renormalized atomic spin-orbit coupling. This last value is different for $p$ and $d$ orbitals, $\lambda_p$ and $\lambda_d$, while it is zero for $s$ orbitals. Moreover, the matrix elements of orbitals of different atoms are also zero. The non-zero elements of $H_{so}$ can be calculated using the $\ket{m_l,m_s}$ basis. For instance, in this basis a $p_x$ orbital with spin up is written as $\ket{p_x,\uparrow}=\frac{1}{\sqrt {2}}\left( \ket{1,\frac{1}{2}}+\ket{-1,\frac{1}{2}}\right)$, while a $p_z$ orbital with spin down is $\ket{p_z,\downarrow}=\ket{0,-\frac{1}{2}}$. The matrix element of the $H_{so}$ between $p_x$ and $p_z$ orbitals can be calculated such as:
\al{
\matrixel{p_x,\uparrow}{H_{so}}{p_z,\downarrow}&=\matrixel{p_x,\uparrow}{\lambda_p \v L \cdot \v S}{p_z,\downarrow}\\&=\frac{\lambda_p}{2}\matrixel{p_x,\uparrow}{ L^+S^- +L^-S^++2L^zS^z}{p_z,\downarrow}\\&=
\frac{\lambda_p}{2\sqrt{2}}\matrixel{-1,\frac{1}{2}|+\langle 1,\frac{1}{2}}{ L^+S^- +L^-S^++2L^zS^z}{0,-\frac{1}{2}}\\
&=\frac{\lambda_p\sqrt{2}}{2\sqrt{2}}\braket{-1,\frac{1}{2}|+\langle 1,\frac{1}{2}}{-1, \frac{1}{2}}\\
&=\frac{\lambda_p}{2}
}
where $L^{\pm}$ and $S^\pm$ are raising and lowering operators for linear momentum and spin. The rest of the matrix elements can be obtained in a similar fashion.

\noindent Here I list the spin-orbit Hamiltonian for atomic $p$ and $d$ orbitals which has been recently published by Jones and Albers \cite{Jones2009} for \textit{p}, \textit{d} and \textit{f} orbitals:\\
\[
H^p_{\text{so}}= \frac{\lambda_p}{2}
  \begin{matrix}\\\mbox{}\end{matrix}
  \begin{pmatrix} 0 & -i & 0 & 0 & 0 & 1 \\ 
          i & 0 & 0 & 0 & 0 & -i \\
          0 & 0 & 0 & -1 & i & 0 \\
          0 & 0 & -1 & 0 & i & 0 \\
          0 & 0 & -i & -i & 0 & 0 \\
          1 & i & 0 & 0 & 0 & 0 \\
   \end{pmatrix}
\]

\[
H^d_{\text{so}}= \frac{\lambda_d}{2}
  \begin{matrix}\\\mbox{}\end{matrix}
  \begin{pmatrix} 0 & 0 & 0 & 2i & 0 & 0 & 1 & -i & 0 & 0\\ 
          0 & 0 & i & 0 & 0 & -1 & 0 & 0 & -i & -i\sqrt 3\\
          0 & -i & 0 & 0 & 0 & i & 0 & 0 & -1 & \sqrt 3\\
          -2i & 0 & 0 & 0 & 0 & 0 & i & 1 & 0 & 0\\
          0 & 0 & 0 & 0 & 0 & 0 & i\sqrt 3 & -\sqrt 3 & 0 & 0\\
          0 & -1 & -i & 0 & 0 & 0 & 0 & 0 & -2i & 0\\
          1 & 0 & 0 & -i & -i\sqrt 3 & 0 & 0 & -i & 0 & 0\\
          i & 0 & 0 & 1 & -\sqrt 3 & 0 & i & 0 & 0 & 0\\
          0 & i & -1 & 0 & 0 & 2i & 0 & 0 & 0 & 0\\
          0 & i\sqrt 3 & \sqrt 3 & 0 & 0 & 0 & 0 & 0 & 0 & 0\\
   \end{pmatrix}
\]\\
Bases of spin-orbit Hamiltonians are $p_x\uparrow$, $p_x\downarrow$, $p_y\uparrow$, $p_y\downarrow$, $p_z\uparrow$, $p_z\downarrow$ for p orbitals and $d_{xy}\uparrow$, $d_{xy}\downarrow$, $d_{yz}\uparrow$, $d_{yz}\downarrow$, $d_{zx}\uparrow$, $d_{zx}\downarrow$, $d_{x^2-y^2}\uparrow$, $d_{x^2-y^2}\downarrow$,$d_{3z^2-r^2}\uparrow$, $d_{3z^2-r^2}\downarrow$ for d orbitals.

Values of coefficients such as $\lambda_p$ and $\lambda_d$, spin-orbit couplings are related to atomic spin-orbit couplings, $\xi_{i}$, and can be obtained  using atomic spectra, which were tabulated by Moore \cite{Moore1, Moore2, Moore3}. The atomic spin-orbit coupling depends on the particular configuration of the p or d electrons. \cite{Dunn1961} Usually $\lambda$ being a state parameter is related to $\xi$ which is a one electron parameter through total spin S:$\approx \pm \frac{\xi}{2S}$ . S is positive if the atomic valence shell is less than half filled, negative if more than half filled, and 0 if half or fully filled. \cite{ColeJr1970}. The value of $\xi$ depends on the energy difference in the atomic spectra \cite{Fisk1968a}:
The method of getting the SOC from spectra using the Land\'e interval rule can be written for a term with a ground state configuration such as $^{2S+1}X_J$ where S is total spin, 2S+1 is the multiplicity, J is total angular momentum and X is named after L which is orbital angular momentum (S for 0, P for 1, D for 2 etc.) \cite{Fisk1968a} 
\al{\label{eq:Lande}
  \xi(SLJ)=\frac{E(J)-E(J-1)}{J}
}

As an example, we take the number of electrons in the transition metal ions from Chanier et al. \cite{Chanier} and calculate the spin-orbit splittings using atomic energy tables. The total number of electrons in several TM (TM) ions in diamond have been calculated using density functional theory and tabulated by Ref. \cite{Chanier}. (Table~\ref{tab:atomicSO})
\begin{table}[h]\label{tab:atomicSO}
\caption[Atomic Spin-Orbit Couplings of Several Transition Metals in a Diamond]{Atomic Spin-Orbit Couplings of Several Transition Metals in Diamond}
\begin{center}
\begin{tabular}{| l |c| c| c| c |}
\hline
&&&&\\
TM           &G. State   & Electron \#  & d-orbital  (cm$^{-1}$ / meV
)
 & p-orbital  (cm$^{-1}$ / meV
 )  \\
 \hline
 &&&&\\

Sc     &$^2 D_{3/2} $  &19.3  & 79.0 / 9.8  & 315.8 / 39.2\\
Ti     &$^3 F_{2}   $     &20.1  & 60.4 / 7.5  &84.7 / 10.5\\
V      &$^4 F_{3/2} $   &21.2  & 55.9 / 6.9  &95.8 / 11.9\\
Cr     &$^5 D_{0} $      &22.3  & 58.4 / 7.2  &85.8 / 10.6\\  
Fe     &$^6 D_{9/2} $        &24  &  -1002/-124      &-491/-61 \\
Fe     &$^5 D_{4} $        &25  &  -943/-117      &-1227/-152 \\
Co     &$^3 F_{4}  $     &25.9  &-226.6 / -28.1 &-158.4 / -19.6\\
Ni     &$^2 D_{5/2} $  &26.9  &-602.8 / -74.7 &-352.2 / -43.7\\
Cu     &$^1 S_{0} $     &28.1  &0 / 0     &-841.3 / -104.3  \\
Zn     &$^2 S_{1/2} $  &29.0  & 20.3 / 2.5 &582.5 / 72.2  \\
\hline
\end{tabular}
\end{center}
\end{table}

However, this is not the whole picture. The atomic spin-orbit couplings may be quite different than splitting in crystal band structures due to spin-orbit coupling. The relation between atomic hyperfine splitting and crystal splitting has attracted considerable interest. It was thought that the partial ionicity of elements in III-V compounds could be used to relate these two splittings. Braunstein \cite{Braunstein1962} stressed that if it has been assumed that an electron spends 35 percent of time in the III atom and the rest in the V atom, so the crystal splitting can be found by multiplying the atomic splitting by a normalization factor about 29/20, which also works for germanium. Unfortunately, this turned out to be a mere coincidence since it didn't work in other III-V compounds. A normalization constant is required for two reasons. First, the top of the valence bands in III-V materials, which consists of p-like j=3/2 and j=1/2 states does, in fact, include higher order atomic d-like orbitals. Second, the Wannier functions have a tendency of extending more than the typical size of the Wigner-Seitz cell. This fact causes a volume effect \cite{Chadi1977}. To sum up, by defining the normalization constant as $C_N$, the atomic hyperfine splitting due to spin-orbit coupling is related to splitting in a crystal through:
\al{
\Delta_0=\frac{E(J)-E(J-1)}{J} \times (2S)\times \frac{2L+1}{2} \times C_N
}
The normalization constant is reported in the literature for most crystals. $C_N$ is equal to 1 for carbon, and 1.5 for germanium and gallium arsenide. One must be cautious using the Land\' e interval rule since it assumes that spin-orbit coupling is in the form of $\v L \cdot \v S$. If the calculated spin-orbit splitting deviates from experimental values, this is an indication that residual spin interactions are also important, such as spin-spin interactions, and Russel-Sounders coupling approximation is not valid.\cite{Blume1964, Blume1963, Blume1962}.

\section{Conclusions}
In this chapter, we explained the formulation of the tight-binding Hamiltonians for several materials. The tight-binding Hamiltonian is proved to be significantly efficient in calculating correct eigenenergies and eigenvectors, as long as the parametrization of the Hamiltonian is carried out correctly. We have also investigated the effects of strain on \STO based systems. The electronic band structures of \BiSb are plotted using the virtual crystal approximation, and we observed the semimetal-semiconductor phase transitions as a function of alloy concentration. \BiSe and \BiTe Hamiltonians are constructed, and band structures are plotted for both bulk and surface states. The chapter ends with an explanation of how to add the spin-orbit Hamiltonian into the TB Hamiltonian and extract atomic spin-orbit coupling values from atomic spectra tables. All of the Hamiltonians which are constructed for this chapter and used in the next chapters are tabulated in the Appendix. 

\chapter{Spin-Orbit Interaction and Spin Relaxation Times}
\section{Derivation of the Spin-Orbit Interaction Tensor}
The relativistic Hamiltonian for a free particle is given by the 4-component Dirac equation. By eliminating anti-particle wavefunctions this equation can be reduced to the 2 component Pauli equation \cite{Crepieux2001} and it takes this form: (in the order of 1/c$^2$)
\al{\label{eq:Pauliequation}
\hat{H} =\frac{p^2}{2m} +V -\frac{p^4}{8m^3c^2}+\frac{\hbar}{4m^2c^2}(\v{\sigma}\times \nabla V)\cdot \v p +\frac{\hbar^2}{8m^2c^2}\delta V
}
where the first two terms constitute the non-relativistic Hamiltonian. The third term is the relativistic mass correction; the fourth term is the spin-orbit coupling and finally the last one is the Darwin term. The relativistic mass correction is quite small, and the Darwin term can be neglected. Hence, the Hamiltonian which is relevant for this research has this form:
\al{
\hat{H} =\frac{p^2}{2m} +V +\frac{\hbar}{4m^2c^2}(\v{\sigma}\times \nabla V)\cdot \v p
}
The term beginning with $\frac{\hbar}{4m^2c^2}$ is often called the spin-orbit coupling. However in the case of crystalline structures and periodic potentials this spin-orbit coupling increases by several orders of magnitude depending upon the constituting atoms. Furthermore, effective masses that are far larger or smaller than the bare mass of the electron causes the effective spin-orbit coupling in the system to differ for each of the electronic bands. Therefore, the relativistic effect of the potential that is experienced by the electron in a band depends merely on the effective spin-orbit interaction constant with that band.

In order to calculate the effective spin-orbit interaction due to an external potential at the conduction band of semiconductors we first calculate the wavefunctions at the minimum of the conduction band. That is the point about which conduction electrons are located most of the time. In systems with time-reversal symmetry and spatial inversion symmetry the electronic states are at least two-fold degenerate at each crystal momentum $\v k$.  The wavefunctions are in the form of the Bloch states:
\al{ 
\psi_{n \mathbf k \alpha}(\mathbf r, \sigma)=e^{i\mathbf k \cdot \mathbf r} u_{n \mathbf k \alpha}(\mathbf r, \sigma)}
where n is the band index (c for conduction band), k is the wave vector, and pseudospin index $\alpha$ labels the two degenerate states at each $\v k$.  Here $u_{n \mathbf k \alpha}$, and, therefore, $\psi_{n \mathbf k \alpha}$  have the same periodicity as the crystal lattice. 

In principle, it is possible to choose the wavefunctions of any k-point in the Brillouin zone as a basis and expand all other wavefunctions around this point using the Bloch function of this basis. The conduction band edge is located at the zone center ($\Gamma$) point for SrTiO$_3$, while for group IV semiconductors there exist multiple valleys. (e.g. six valleys for Si and diamond are located at 85\% and 75\% of the line between $\Gamma$ and $X$ respectively, while four valleys for germanium are located at $L$ points.) Since the parameter space of the Hamiltonian has a curvature, expanding states around a point becomes a parallel transport problem and requires a new derivative to be defined, namely the covariant derivative:
\al{
D_\mu=\partial_{\mu}+i  A_\mu =0
}
where $\v A$ is a Hermitian matrix and is called the \textit{connection} (also known as the Berry connection \cite{Berry1984}) :
\al{ \label{Berryconnection}
\mathbf A_{c \alpha, n \beta}\equiv i \braket{u_{n \beta}}{\pd{u_{c \tilde{\mathbf k} \alpha}}{\tilde {\mathbf k}}}_{\tilde{\mathbf k} =0} \text{~~~~~and~~~~~} \mathbf A_{c \alpha , n \alpha}=\mathbf A^* _{n \beta , c \alpha}}
The wavefunctions in the vicinity (denoted as $\tilde {\v k}$) of the minimum point (indicated as $\v k_i$) have this form in first order:
\al{ \label{expansion}
\psi_{c \tilde{\mathbf k}\alpha}(\mathbf r, \sigma)=e^{i\mathbf k_{i} \cdot \mathbf r}e^{i\tilde{\mathbf k} \cdot \mathbf r}  \Bigg [u_{c  \alpha}(\mathbf r, \sigma) -i\sum_{{\beta},{n}}\tilde{\mathbf k}\cdot \mathbf A_{c\alpha ,n\beta}u_{n  \beta}(\mathbf r, \sigma) \Bigg]}
where $u_{c  \alpha}$ ($u_{n  \beta}$) is the periodic part of the conduction band (any other band) wavefunction. It is difficult to calculate the derivative of the wave function with respect to $\v k$. However, it is possible to write \eqr{Berryconnection} in terms of derivatives and eigenfunctions of the Hamiltonian. Starting with Schr\"odinger's equation:
\al{
\hat H_{\tilde{\v k}}  \ket {u_{n \tilde{\v k} \alpha}}=E_{n \tilde{\mathbf k}} \ket {u_{n \tilde{\v k} \alpha}}
}
Taking the derivative of both sides:
\al{
(\nabla_{\tilde{\mathbf k}}H_{\tilde{\mathbf k}})\ket {u_{n \tilde{\mathbf k} \alpha}} + H_{\tilde{\mathbf k}} \nabla_{\tilde{\mathbf k}}\ket {u_{n \tilde{\mathbf k} \alpha}}=(\nabla_{\tilde{\mathbf k}}E_{n \tilde{\mathbf k}})\ket {u_{n \tilde{\mathbf k} \alpha}} + E_{n \tilde{\mathbf k}} \nabla_{\tilde{\mathbf k}}\ket {u_{n \tilde{\mathbf k} \alpha}}
}
Multiplying the previous equation by $\bra{u_{m \tilde{\v k}\beta}}$, orthogonal to $\ket{u_{n \tilde{\mathbf k} \alpha}}$, we get:
\al{
 \matrixel{u_{m \tilde{\mathbf k} \beta}}{\nabla_{\tilde{\mathbf k}}H_{\tilde{\mathbf k}}}{u_{n \tilde{\mathbf k} \alpha}} &+  E_{m \tilde{\mathbf k}}\matrixel{u_{m \tilde{\mathbf k} \beta}}{\nabla_{\tilde{\mathbf k}}}{u_{n \tilde{\mathbf k} \alpha}} \\
 &= \matrixel{u_{m \tilde{\mathbf k} \beta}}{\nabla_{\tilde{\mathbf k}}E_{n \tilde{\mathbf k}}}{u_{n \tilde{\mathbf k} \alpha}}+E_{n \tilde{\mathbf k}}\matrixel{u_{m \tilde{\mathbf k} \beta}}{\nabla_{\tilde{\mathbf k}}}{u_{n \tilde{\mathbf k} \alpha}}
}
The first term on the right side of the equation is zero unless $n=m$ and $\alpha=\beta$ due to the orthogonality of states. Therefore, we end up with an equation:
 \al{
\braket{u_{m \tilde{\mathbf k} \beta}}{\pd{u_{n \tilde{\mathbf k} \alpha}}{\tilde {\mathbf k}}} =\frac{\matrixel{u_{m \tilde{\mathbf k} \beta}}{\nabla_{\tilde{\mathbf k}}H_{\tilde{\mathbf k}}}{u_{n \tilde{\mathbf k} \alpha}}}{E_{n\tilde{\mathbf k}}-{E_{m\tilde{\mathbf k}}}}}
where the connection $\mathbf A_{n\alpha, m\beta}=i\braket{u_{m \tilde{\mathbf k} \beta}}{\pd{u_{n \tilde{\mathbf k} \alpha}}{\tilde {\mathbf k}}}$.

\section{The Effective External Potential in the Conduction Band}
As the wavefunction of the lowest conduction band is expanded around its minimum energy states we now can write the matrix elements of an external potential $V(\mathbf r)$ between conduction band states $\psi_{c \tilde{\mathbf k} \alpha}$ and $\psi_{c \tilde{\mathbf k '} \alpha '}$:
\al{ 
 V_{\tilde{\mathbf k}'\alpha',\tilde{\mathbf k}\alpha}=\sum_{\sigma}\int d\mathbf r \psi_{c\tilde{\mathbf k}'\alpha'}(\mathbf r, \sigma)V(\mathbf r)\psi_{c\tilde{\mathbf k}\alpha}(\mathbf r, \sigma)}
by substituting \eqr{expansion} into the previous equation we get:
\al{
\bs
 V_{\tilde{\mathbf k}'\alpha',\tilde{\mathbf k}\alpha}=\sum_{\sigma}&\int d\mathbf r e^{-i\tilde{\mathbf k}' \cdot \mathbf r} \Bigg[ u^*_{c\alpha'}(\mathbf r, \sigma) + i\sum_{\beta',n'}\tilde{\mathbf k}'\cdot \mathbf A^*_{c\alpha',n'\beta'}u^*_{n'\beta'}(\mathbf r, \sigma)
 \Bigg] \\
 & V(\mathbf r) \Bigg[ u_{c\alpha}(\mathbf r, \sigma) - i\sum_{\beta,n}\tilde{\mathbf k}\cdot \mathbf A_{c\alpha,n\beta}u_{n\beta}(\mathbf r, \sigma) \Bigg]e^{i\tilde{\mathbf k}\cdot \mathbf r}
\es}
As a result of the orthogonality properties of the $u_{n \mathbf k}(\mathbf r)$ and by summing over the unit cells we conclude that:
\al{ 
 V_{\tilde{\mathbf k}' \alpha',\tilde{\mathbf k}\alpha}=\int d\mathbf r e^{-i \tilde{\mathbf k'}\cdot \mathbf r} \Bigg[ \delta_{\alpha' \alpha} V(\mathbf r) +\sum_{\beta, n} (\tilde{\mathbf k'}\cdot \mathbf A^*_{c \alpha', n\beta}) V(\mathbf r)(\tilde{\mathbf k}\cdot \mathbf A_{c \alpha, n\beta})\Bigg]e^{i \tilde{\mathbf k} \cdot \mathbf r}
}
Two Berry connections can be multiplied to get a simpler form such that:
\al{ 
 \sum_{\beta , n} \mathbf A^*_{c \alpha',n\beta} \mathbf A_{c\alpha, n\beta} =\Bigg[-i \braket {\pd{u_{c\tilde{\mathbf k}\alpha'}}{\tilde{\mathbf k}}}{{u_{n\beta}}}i \braket {{u_{n\beta}}}{\pd{u_{c\tilde{\mathbf k}\alpha}}{\tilde{\mathbf k}}}\Bigg]_{\tilde{\mathbf k} =0}}
\al{ 
=\braket {\pd{u_{c\tilde{\mathbf k}\alpha'}}{\tilde{\mathbf k}}}{\pd{u_{c\tilde{\mathbf k}\alpha}}{\tilde{\mathbf k}}}_{\tilde{\mathbf k} =0}}
since $ \sum_{\beta , n}\ket{u_{n\beta}}\bra{u_{n\beta}}$ is the identity matrix. We conclude:
\al{ 
 V_{\tilde{\mathbf k'}\alpha',\tilde{\mathbf k}\alpha}=\int d\mathbf r e^{-i\tilde{\mathbf k'}\cdot \mathbf r}\Bigg[\delta_{\alpha' \alpha}V(\mathbf r) +\sum_{ij} B^{ij}_{\alpha \alpha'}\nabla_i V(\mathbf r)\nabla_j\Bigg]e^{i\tilde{\mathbf k}\cdot \mathbf r}
}
where
\al{ 
 B^{ij}_{\alpha' \alpha}\equiv \braket {\pd{u_{c\tilde{\mathbf k}\alpha'}}{\tilde{k_i}}}{\pd{u_{c\tilde{\mathbf k}\alpha}}{\tilde{k_j}}}_{\tilde{\mathbf k}=0}
}
It is possible to expand any 2x2 matrix as an operator in (pseudo)spin space by using Pauli matrices and $B^{ij} _{\alpha \alpha'}$  becomes:
\al{
 B^{ij}_{\alpha \alpha'}=B_{ij0}\delta_{\alpha \alpha'}+\sum_k B_{ijk}[\sigma_k]_{\alpha' \alpha}
}
where
\al{ 
 B_{ijk}=\frac{1}{2}\sum_{\alpha \alpha'} B^{ij}_{\alpha \alpha'}[\sigma_k]_{\alpha \alpha'} 
}
\al{ 
B_{ij0}=\frac{1}{2}\sum_{\alpha} B^{ij}_{\alpha \alpha}
}
This reduces the effective interaction near the valley minimum to:
\al{ 
 \tilde{V}(\uv {r},\uv {\sigma})=\Bigg[ V(\uv{r}) + \sum_{ij} B_{ij0}\nabla_i V(\uv{r}) \nabla_j\Bigg] +\sum_{ijk}B_{ijk}\nabla_i V(\uv{r})\nabla_j \uv{\sigma_k }
}
$B_{jik}$ are Hermitian with respect to the first two indices, i.e. $B_{ijk}=B^*_{ijk}$. We also conclude that $B_{ij0}$ is real, and hence symmetric in the first two indices:
\al{
B_{ij0}=\lambda _{ij}=\lambda _{ji}
}
while $B_{ijk}$ is imaginary and  antisymmetric:
\al{ 
B_{ijk}=i\lambda _{ijk}=-i\lambda _{jik}
}
The real quantities $\lambda _{ij}$ and $\lambda _{ijk}$ can be expressed as:
\al{ 
\lambda_{ij}=\frac{1}{2}Re\sum_{\alpha} \braket {\pd{u_{c \tilde{\mathbf k} \alpha}}{\tilde{k_i}}} {\pd{u_{c \tilde{\mathbf k} \alpha}}{\tilde{k_j}}}_{\tilde{\mathbf k}=0} 
}
\al{ \label{eq:soifirst}
\lambda_{ijk}=\frac{1}{2}Im\sum_{\alpha \alpha'} [\sigma _k]_{\alpha \alpha'} \braket {\pd{u_{c \tilde{\mathbf k} \alpha'}}{\tilde{k_i}}} {\pd{u_{c \tilde{\mathbf k} \alpha}}{\tilde{k_j}}}_{\tilde{\mathbf k}=0}  
}
This quantity can also be expressed as:
\al{ 
 \lambda_{ijk}=\frac{1}{2}Im\sum_{\alpha \alpha' \beta} [\sigma _k]_{\alpha \alpha'} \braket {\pd{u_{c \tilde{\mathbf k} \alpha'}}{\tilde{k_i}}} {u_{c\tilde{\mathbf k}\beta} }_{\tilde{\mathbf k}=0}  \braket{u_{c\tilde{\mathbf k} \beta}} {\pd{u_{c \tilde{\mathbf k} \alpha}}{\tilde{k_j}}}_{\tilde{\mathbf k}=0} 
}
The intra band contributions come into play when $\alpha =\alpha' =\beta$. Consider first the matrix element with k=z: ($\sigma_z$)
\al{
 \lambda_{ijz}^{intra}=\frac{1}{2}Im\sum_{\alpha} [\sigma _z]_{\alpha \alpha} \braket {\pd{u_{c \tilde{\mathbf k} \alpha}}{\tilde{k_i}}} {u_{c\tilde{\mathbf k}\alpha} }_{\tilde{\mathbf k}=0}  \braket{u_{c\tilde{\mathbf k} \alpha}} {\pd{u_{c \tilde{\mathbf k} \alpha}}{\tilde{k_j}}}_{\tilde{\mathbf k}=0} 
}
Since $\braket {\pd{u_{c \tilde{\mathbf k} \alpha}}{\tilde{k_i}}} {u_{c\tilde{\mathbf k}\alpha} }_{\tilde{\mathbf k}=0}$ and $\braket{u_{c\tilde{\mathbf k} \alpha}} {\pd{u_{c \tilde{\mathbf k} \alpha}}{\tilde{k_j}}}_{\tilde{\mathbf k}=0} $ are purely imaginary, while $[\sigma _z]_{\alpha \alpha}$ is real, this product doesn't have any imaginary component. Thus we conclude that:
\al{
 \lambda_{ijz}^{intra}=0
}
Other matrix elements with k=x ($\sigma_x$) and k=y ($\sigma_y$) can be figured out and shown to be vanishing as well from the fact that $u_{c\tilde{\mathbf k} \alpha}$ and $u_{c\tilde{\mathbf k} -\alpha}$ are degenerate at all $\tilde{ \mathbf k}$, and we are free to make a change of basis, independent of $\tilde{ \mathbf k}$, which maps $\sigma_x$ or $\sigma_y$ to $\sigma_z$. While this transformation does not change the value of $\lambda_{ijk}^{intra}$ we can conclude for all components:
\al{
 \lambda_{ijk}^{intra}=0}
for all k. 
Therefore, the final expression for this quantity is expressed as 
\al{\label{eq:so-tensor}
\lambda_{ijk}=\frac{1}{2}Im\sum_{\alpha \alpha'}[\sigma_k]_{\alpha \alpha'} \sum_{n\neq c, \beta} \frac{\matrixel{u_{c \tilde{\mathbf k} \alpha'}}{\nabla_{\tilde{k}_i} \unit H_{\tilde{\mathbf k}}} {u_{n \tilde{\mathbf k} \beta}} \matrixel{u_{n \tilde{\mathbf k} \beta}}{\nabla_{\tilde{k}_j} \unit H_{\tilde{\mathbf k}}}{u_{c \tilde{\mathbf k} \alpha}}}{(E_{c \tilde{\mathbf k}} -E_{n \tilde{\mathbf k}})^2} \Bigg |_{\tilde{\mathbf k}=0} 
} 
We call this quantity the effective \textit{spin-orbit interaction} tensor which gives the strength of the effective potential's spin part. (\eqr{eq:effpot}). This strength is also important to calculate the spin lifetimes of conduction band electrons since it is the part of the potential which can couple to the spin of the electrons. Thus, the final form of the external effective potential becomes:
\al{ \label{eq:effpot}
 \tilde{V}(\uv{r},\unit{\sigma}) = \Bigg[V(\uv{r}) +\sum_{ij}\lambda_{ij}\nabla_i V(\uv{r}) \nabla_j\Bigg] + i\sum_{ijk}\lambda_{ijk}\nabla_i V(\uv{r}) \nabla_j \unit{\sigma}_k
}

\section{Application to III-V Semiconductors}
\subsection{Spin-Orbit Interaction Tensor Using \kp Theory}
For direct band III-V compounds it is possible to determine the wave functions of conduction and valence bands around the band gap. For these materials, the minimum of the conduction band and the maximum of the valence band are located at the zone center. (${\bf k_i}=\Gamma$ point). Within the 8-band model, the wavefunctions of the first conduction band and three highest valence bands are well-known. The method called $\v k \cdot \v p$ is based on perturbation theory and therefore it is limited to near the point where the perturbation is carried out, which is usually the Brillouin zone center. The Hamiltonian of \kp is:\cite{YuCardona}
\al{ \label{eq:kdotp}
\Bigg ( \frac{p^2}{2m}+\frac{\hbar \v k \cdot \v p}{m} +\frac{\hbar^2k^2}{2m} +V \Bigg )u_{n\v k}=E_{n\v k}u_{n\v k}
}
where m is the electron's free mass, V is the crystal potential, $u_{n \v k}$ is the periodic part of the Bloch function and $E_{n \v k}$ is the energy. At zone center (the $\Gamma $ point) the wavefunction possesses the full symmetry of the crystal. Zinc-blende III-V compounds have $T_d$ symmetry and therefore one can write the wave functions at the k=0 point using irreducible representations and basis functions of the $T_d$ group. A schematic cartoon of the band structure of 8 band $\v k \cdot \v p$ and the irreducible representations of the bands are shown below. In the eight-band model, the eigenstates of $H_{\mathbf k=0}$ correspond to the conduction band spin up and down states as well as heavy, light and split-off holes with spin up and down. The Hamiltonian for this set of basis states 

\begin{equation}
H_{\mathbf k} =\left[\begin{array}{cccccccc}
E_g&0&\frac{i\hbar P}{\sqrt{2}m}{k_+}&0
&\frac{i\sqrt{2}\hbar P}{\sqrt{3}m}k_z &\frac{i\hbar P}{\sqrt{6}m}k_-
&\frac{i\hbar P}{\sqrt{3}m}k_z &\frac{i\hbar P}{\sqrt{3}m}k_-\\
0&E_g&0&\frac{i\hbar P}{\sqrt{2}m}{k_-}
&\frac{i\hbar P}{\sqrt{6}m}k_+&\frac{i\sqrt{2}\hbar P}{\sqrt{3}m}k_z
&\frac{i\hbar P}{\sqrt{3}m}k_+&\frac{i\hbar P}{\sqrt{3}m}k_z\\
-\frac{i\hbar P}{\sqrt{2}m}{k_-}&0&0&0&0&0&0&0\\
0&-\frac{i\hbar P}{\sqrt{2}m}{k_+}&0&0&0&0&0&0\\
-\frac{i\sqrt{2}\hbar P}{\sqrt{3}m}k_z&-\frac{i\hbar P}{\sqrt{6}m}k_-&0&0&0&0&0&0\\
-\frac{i\hbar P}{\sqrt{6}m}k_+&-\frac{i\sqrt{2}\hbar P}{\sqrt{3}m}k_z&0&0&0&0&0&0\\
-\frac{i\hbar P}{\sqrt{3}m}k_z &-\frac{i\hbar P}{\sqrt{3}m}k_-&0&0&0&0&-\Delta&0\\
-\frac{i\hbar P}{\sqrt{3}m}k_+&-\frac{i\hbar P}{\sqrt{3}m}k_z&0&0&0&0&0&-\Delta\\

\end{array}
\right]
\end{equation}
where $k_+ = k_x+ik_y$ and $k_- = k_x - ik_y$, E$_g$ is the band gap, $\Delta$ the spin-orbit splitting in the valence bands, and $P$ the magnitude of the momentum matrix element between the conduction and valence bands\cite{Cardona1988}. $P$ is the magnitude of the momentum matrix element between the conduction band and valence bands in the atomic units listed in Ref.~\cite{Cardona1988}. 
\al{ 
P=-i\frac{2}{3} \matrixel{S}{p_z}{Z^v}
}
Furthermore $P=-i\frac{2}{3} \matrixel{S}{p_x}{X^v}=-i\frac{2}{3} \matrixel{S}{p_y}{Y^v}=-i\frac{2}{3} \matrixel{S}{p_z}{Z^v}$. The other matrix elements are zero by symmetry. Spin-orbit interaction (\eqr{eq:so-tensor}) within this \kp  model for the ${\bf k_i}=0$ point yields only 6 non-zero  elements with the symmetry:
\al{ 
\lambda_{ijk}=\lambda\epsilon_{ijk}
}
\begin{figure}[t]
\centering
\includegraphics[width=1\textwidth]{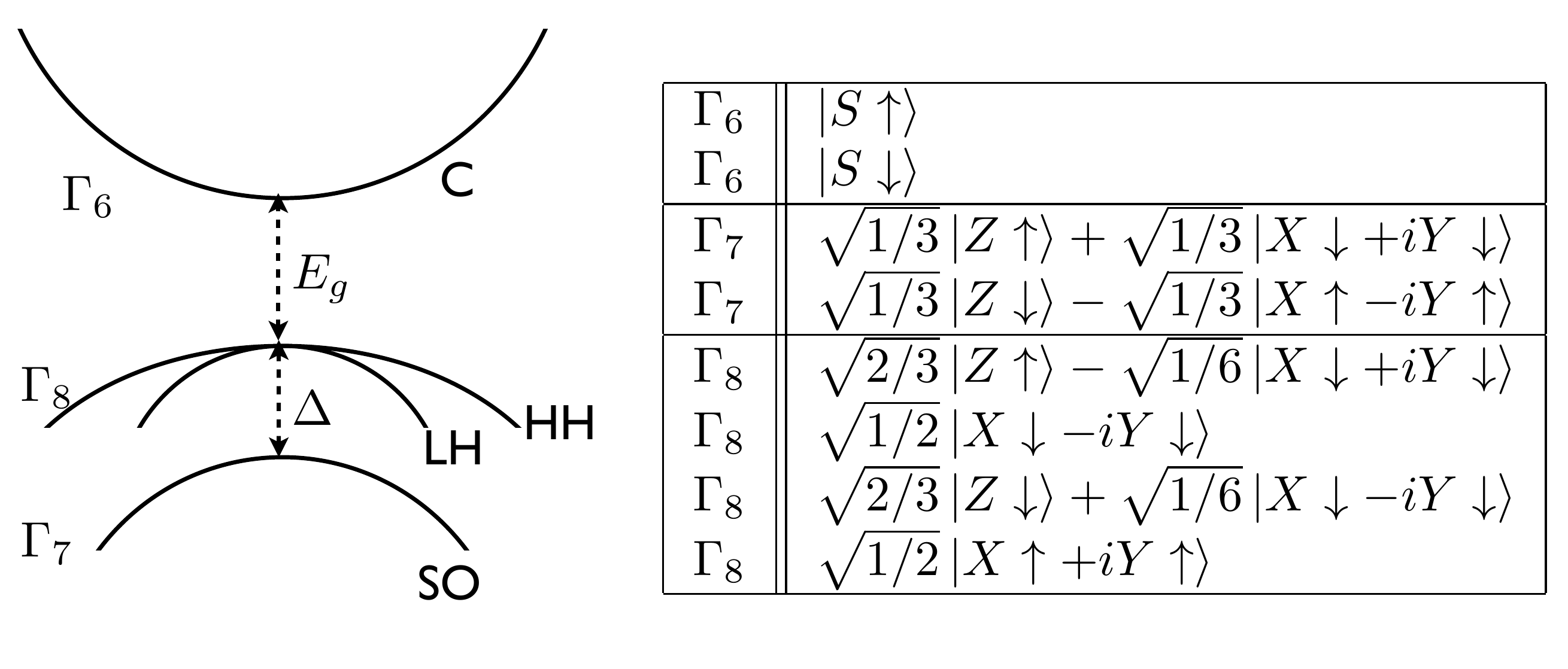}
\caption[8-band \kp bands and wavefunctions]{8-band \kp bands and wavefunctions. Here C denotes conduction band, and LH, HH, and SO are for heavy-hole, light-hole and split-off bands, respectively. The conduction band is spherically symmetric and represented by $\Gamma_6$ (also commonly as $\Gamma_1$), while HH and LH are basis functions of $\Gamma_8$ representation, and split-off band has $\Gamma_7$ symmetry. Here $E_g$ is band gap, and $\Delta$ is the spin-orbit splitting of the valence bands. (from \cite{Flatt1996})}
\label{fig:kpbands}
\end{figure}
and the analytic expression 
\al{ 
\lambda=\frac{\hbar^2 P^2}{3m^{*^2}}\left(\frac{1}{E_g^{2}}-\frac{1}{(E_g+\Delta)^{2}}\right) 
}
where E$_g$ is the band gap and $\Delta$ is the spin-orbit splitting in the valence bands. We have performed 8-band \kp calculations on several III-V compounds and compared these results with ones from other techniques in Table~\ref{tab:SOIcomparison}.

\subsection{Spin-Orbit Interaction Tensor Using the Tight-Binding Hamiltonian of III-V Semiconductors}
We have also constructed an spds$^*$ tight-binding Hamiltonian by taking parameters from Ref. \cite{Jancu1998} and computing the effective spin-orbit interaction, Eq.~(\ref{eq:so-tensor}) for GaAs, InP, GaSb and InSb. A comparison of this analytic \kp expression and the $\lambda$ computed from an spds$^*$ tight-binding Hamiltonian is shown in Table~\ref{tab:SOIcomparison} for GaAs, InP, GaSb, and InSb. We can conclude from these results that there exists an excellent agreement between \kp and tight-binding models. Further details of the spds$^*$ tight-binding Hamiltonian are given in Appendix~\ref{appendixzincblende}.

When calculating tight-binding results we see that the effective mass is crucial to be able to get similar results. This can be seen from the expression of the effective mass by $\v k \cdot \v p$ theory which depends on the matrix element of the momentum operator between the conduction and valence bands. 
\al{\frac{1}{m^*}=\frac{1}{m}+\frac{2}{m^2k^2}\sum_{n' \neq n}\frac{| \langle u_{n0}|\v k \cdot \v p|u_{n'0} \rangle|^2}{E_{n0}-E_{n'0}} \approx \frac{1}{m}+\frac{2P^2}{mE_0}}
We concluded that a tight-binding parameterization which excludes correct masses results in an incorrect spin-orbit interaction.

\subsection{Spin-Orbit Interaction Tensor from Relaxation Time Comparison}
It is also possible to relate this spin-orbit interaction to spin relaxation times and compare it to well-known analytical expressions of spin relaxation time and momentum relaxation time for the Elliott-Yafet relaxation mechanism. These relaxation times are calculated from the 8x8 Kane Hamiltonian of Chapter 3 of the Ref.\cite{Meier1984} by Pikus and Titkov:
\al{
&\frac{1}{\tau_s}=2\frac{2\pi}{\hbar}\int \frac{d\Omega_k}{4\pi}\sum_{k'}|H_{m'k',mk}|^2\delta({E_{k'}-E_k})\\
&\frac{1}{\tau_p}=\frac{2\pi}{\hbar} \sum_{k'}|H^{int}_{m'k',mk}|^2(1-cos\theta) \delta({E_{k'}-E_k})
}
where $H^{int}_{m'k',mk}$ and $H_{m'k',mk}$ are interaction Hamiltonians for momentum and spin scatterings respectively. The relation between these operators is given by:
\al{
H_{m'k',mk}=H^{int}_{m'k',mk}\frac{i\hbar^2 (\sigma [k'k])\eta(1-\frac{1}{2}\eta)}{3m^*E_g(1-\frac{1}{3}\eta)}
}
and for elastic scattering (k$\approx$ k') one can substitute $\frac{\hbar^2}{m^*}[k'k]$ as $2E_ksin\theta$ in the previous equation. Then the ratio of spin and momentum relaxation times is:
\al{
\frac{\tau_p}{\tau_s}=\frac{32}{81} \left(\frac{1}{E_g}\right )^2\eta^2 \left (\frac{1-\eta /2}{1-\eta /3}\right )^2 E_k^2
}
where $\eta=\Delta /(E_g+\Delta)$. 
This ratio depends on the square of the energy $E_k$ and everything in front of it is constant. I will call it $C_1$:
\al{\label{eq:C1}
C_1=\frac{32}{81} \left(\frac{1}{E_g}\right )^2\eta^2 \left (\frac{1-\eta /2}{1-\eta /3}\right )^2 
}
On the other hand, spin and momentum relaxation times can be also related to each other by using the effective spin-orbit interaction tensor $\lambda$ of \eqr{eq:so-tensor}:
\al{
&\frac{1}{\tau_s}=\frac{2\pi}{\hbar}\sum_{k'}|iV_0\sum_i\lambda(\v q\times \v k)_i[\sigma_i]_{\alpha'\alpha}|^2\delta({E_{k'}-E_k})\\
&\frac{1}{\tau_p}=\frac{2\pi}{\hbar} \sum_{k'}|V_0\delta_{\alpha'\alpha}|^2(1-cos\theta) \delta({E_{k'}-E_k})
}
where $\v q=\v k' -\v k$, $\lambda$ is the spin-orbit interaction tensor which is spherically symmetric for III-V compounds. First we should note that $\v q\times \v k=(\v k' -\v k)\times \v k =\v k'\times \v k =|k||k'|sin\theta$ in magnitude. If we assume that the scattering is elastic (no large momentum transfer) as in Ref.\cite{Meier1984}, then this is equal to $k^2 sin\theta =\frac{2m^*}{\hbar^2}E_ksin\theta$.
So the ratio of spin and momentum relaxation times becomes:
\al{
\frac{\tau_p}{\tau_s}=\lambda^2\frac{8{m^*}^2}{3\hbar^4} E_k^2
}
This ratio has the same form as in \eqr{eq:C1}. Let's call the coefficient in front of the previous equation $C_2$:
\al{
C_2=\lambda^2\frac{8{m^*}^2}{3\hbar^4} 
}
By comparing $C_1$ and $C_2$ we conclude that spin lifetimes have the same functional form. If
\al{\label{eq:lambdacomparison}
\lambda=\frac{\hbar^2}{2m^*}\frac{4}{3\sqrt 3}\frac{1}{E_g} \eta \left (\frac{1-\eta /2}{1-\eta /3}\right ),
}
then the two expressions agree. 
\begin{table}[h]
\caption[Spin-orbit Interaction Comparison]{\label{tab:SOIcomparison}
Spin-orbit Interaction Comparison}
\begin{center}
\begin{tabular}{|l||cccr|}
Method&
GaAs&
InP&
GaSb&
InSb\\
\hline
$\v k \cdot \v p$ & 4.4 & 1.7  & 32.5 & 544.1\\
Tight-binding & 4.6 & 1.8 & 34.6 & 583.8\\
$\lambda$ from \eqr{eq:lambdacomparison} & 5.1 & 1.7 & 39.7 &630.9 \\
\hline
\end{tabular}
\end{center}
\end{table}
We report in Table \ref{tab:SOIcomparison} the implied value of $\lambda$ from Eq.~(\ref{eq:lambdacomparison}), indicating an excellent agreement between our formalism and previously obtained results for spin lifetimes in III-V semiconductors. All of the values are in units of \AA$^2$ and material parameters are taken from Ref. \cite{Madelung1986} ($\lambda_{ijk}=\lambda\epsilon_{ijk}$)  As the ratio of the spin-orbit splitting to the band gap decreases, the results of $\lambda$ starts to differ from each other. This can be guessed easily since this approximation takes spin-orbit splitting as a perturbation. When the splitting becomes large, then all techniques fail to predict the value of the spin-orbit interaction for the same reason.

\section{Application to \STO}
\begin{figure}
    \includegraphics[width=\textwidth]{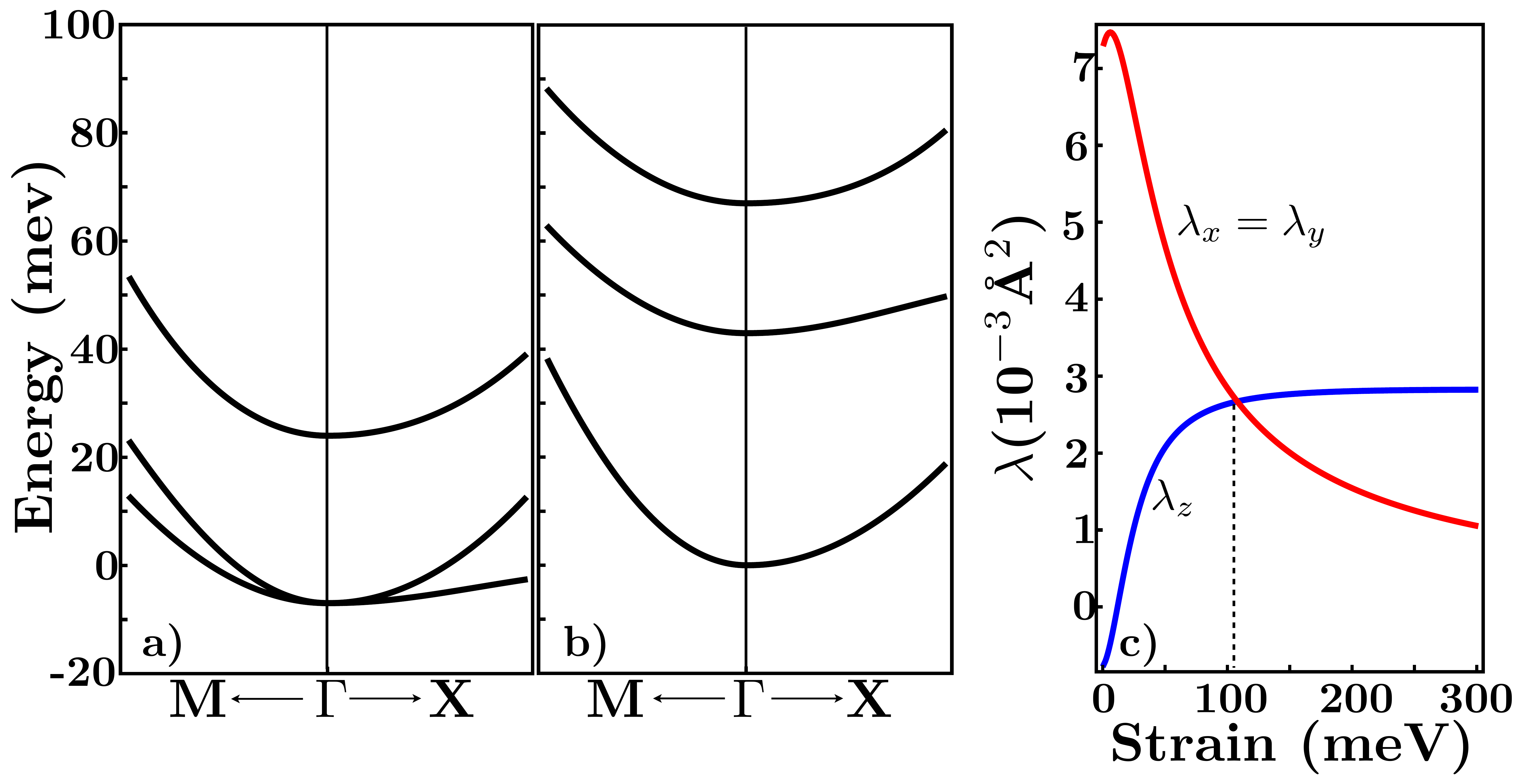}
    \caption[Strain dependence of the spin-orbit interaction]{Strain dependence of the spin-orbit interaction. Magnitude of spin-orbit interaction $\lambda$ as a function of the conduction band splitting at $\Gamma$ due to strain or confinement. The spin-orbit interaction elements $\lambda_x$ and $\lambda_y$ declines with increasing strain while $\lambda_z$ increases. However this increase doesn't continue, and the value of this element is saturated around $2.7 \times 10^{-3}\text{\AA}^2 $. The dashed line corresponds to a strain of 107 meV. This formulation is not applicable to the case of no strain because of the fact that the first and second lowest conduction bands would become four-fold degenerate at the $\Gamma$ point for no strain.} \label{fig:soistraindependence}
\end{figure}

For \STO, there exists only one momentum corresponding to the conduction band minimum, and the electronic states near this minimum at the Brillouin zone center mostly consist of Ti \textit{d}-orbitals. The crystal potential splits these conduction bands into sixfold t$_{2g}$ bands (d$_{xy}$, d$_{yz}$, d$_{zx}$) and fourfold (higher-energy) e$_g$ bands (d$_{x^2-y^2}$, d$_{3z^2 -r^2}$); spin-orbit coupling results in a further splitting ($\approx$ 30 meV) of the lower t$_{2g}$ bands into fourfold and twofold bands, as shown in Fig.~\ref{fig:soistraindependence}(a).  We consider strained STO, in which the compressive strain breaks the fourfold degeneracy at the $\Gamma$-point and results in well-resolved, doubly degenerate subbands in the plane perpendicular to the growth direction, as shown in Fig.~\ref{fig:soistraindependence}(b) for a splitting of $\sim 50$~meV. The same energy splitting is produced by an interface and leads to the electronic structure of the LAO/STO 2DEG\cite{Salluzzo2009}.  Starting from the tight-binding band structure of SrTiO$_3$, we have calculated the spin-orbit coupling tensor  \m{\lambda_{ijk}} from Eq.~(\ref{eq:so-tensor}).   There are only six non-zero elements 
at the minimum of the conduction band (\m{\Gamma} point): 
 \al{ \label{eq:lambdasto}
\bs
      &\lambda_{xyz}=-\lambda_{yxz} \equiv \lambda_z\\
      & \lambda_{xzy}=-\lambda_{zxy} \equiv \lambda_y\\
      & \lambda_{zyx}=- \lambda_{yzx}\equiv \lambda_x\,,
 \es
 }
where $z$ is the direction of the uniaxial strain and growth. Note that $\lambda_x=\lambda_y$, as expected from the symmetry of the Hamiltonian. One can write this tensor compactly as $\lambda=\epsilon_{ij}\lambda_k$.   The numerical values of the $\lambda$'s are $\lambda_x=\lambda_y= 0.0047$~\AA $^2$ and $\lambda_z= 0.0021$~\AA$^2$ for a splitting of 50 meV due to strain (or equivalently quantum confinement in \LAOSTO 2DEGs.) Compared to the values obtained for III-V compounds (Table \ref{tab:SOIcomparison}) these values are quite small, which is due to the fact that the atomic spin-orbit coupling of \STO is at least one order of magnitude less than the III-V semiconductors. 

We also studied how the spin-orbit interaction depends on the strain splitting. From \figref{fig:soistraindependence}(c) it can be concluded that the strain significantly changes the strength of the effective spin-orbit interaction. Increasing strain has a negative effect on the $\lambda_x$ and $\lambda_y$ components while $\lambda_z$ component increases. A strain of 107 meV causes the lowest conduction band to become isotropic in 3 directions. As a result of that the spin-orbit interaction becomes uniform as in the case of III-V semiconductors with spherical ($\Gamma_1$-like) conduction bands.

\section{Spin Lifetime Calculations}
The phenomena of spin relaxation describes the evolution of an ensemble of spins with an inequilibrium polarization to an equilibrium configuration where the ensemble is unpolarized. Although formal description of spin dynamics is determined by Bloch equations \cite{Fabian2004, Zutic2004} an easier definition of the spin lifetime is possible in the relaxation time approximation.\cite{Overhauser1953} Consider a total of N electrons where the number of spin up electrons is N$_+$, spin down electrons is N$_-$ and net polarization is D=N$_+$-N$_-$. If the equilibrium polarization is D$_0$ then  the time it is required to go from D to D$_0$ can be written as
\al{
\frac{dD}{dt}=\frac{D_0-D}{\tau_s}
}
where $\tau_s$ is a spin relaxation time. (also known as T$_1$ time).  The underlying reasons for spins to flip one way or another are scattering processes such as impurity and phonon scatterings, and there are three primary relaxation mechanisms that contribute to the relaxation processes:  
\begin{enumerate}
     \item Dyakanov-Perel mechanism \cite{Dyakonov1971, Dyakonov1972}, which can be dominant especially for systems lacking inversion symmetry, such as III-V semiconductors. This mechanism manifests itself through an emergent spin-orbit coupling and $\v k$-dependent effective magnetic field due to the non-centrosymmetric nature of the crystal. The phase of the spin ensemble is lost as a result of random scatterings and precession of the spin around different effective magnetic fields at each $\v k$. 
     \item Bir-Aranov-Pikus mechanism \cite{Bir1975, Aronov1983}, which accounts for electron-hole exchange scatterings and can play a major role in p-type systems.
     \item Elliott-Yafet mechanism \cite{Elliott1954,Yafet1963} is largely responsible for spin relaxation in inversion symmetric materials, such as \STO and group IV elements. This mechanism is explained in a more detailed fashion in the next section.
\end{enumerate}
In addition to these mechanisms, any fluctuation on the spin interactions or g-tensor can relax the spin, as well as the hyperfine and exchange interactions. The total spin relaxation time is then calculated using the Overhauser relation: \cite{Overhauser1953}
\al{
\frac{1}{\tau_s}=\frac{1}{\tau_s^{EY}}+\frac{1}{\tau_s^{DP}} +\frac{1}{\tau_s^{BAP}}+...}
There have been many theoretical and experimental studies of the spin lifetime for both single spins and ensemble spin systems. It has been observed that spin centers of diamond \cite{Balasubramanian2009} and silicon carbide\cite{Koehl2011} possess long spin coherence times up to several milliseconds. Additionally, it has been theoretically calculated that spin lifetimes in III-V quantum wells are dominated by the Dyakanov-Perel relaxation mechanism. \cite{Lau2001}. Although spin dynamics in nonmagnetic and centrosymmetric materials have been receiving growing attention they have not been theoretically studied except for some studies of common group IV elements with a density functional theory \cite{Restrepo2012,Fedorov2013} or with simple effective models such as 8-band \kp Hamiltonians \cite{Li2011,Gmitra2013}. None of these include a full zone contribution from the electronic structure, and there has not been any study of spin lifetimes in \STO and \LAOSTO systems. 

\section{Elliott-Yafet Spin Relaxation Mechanism}
Since perovskite oxides have an inversion symmetric crystal structure, the relevant spin relaxation mechanism for \STO is Elliott-Yafet spin relaxation. This approach \cite{Elliott1954} is based on the fact that spin is not a good quantum number and, therefore, cannot be used to express the wavefunctions of the electrons in the conduction band. Without spin-orbit coupling, ($\v L\cdot \v S $ term), electrons can be depicted as Bloch functions multiplied by pure spin eigenstates of a direction. However, the spin-orbit coupling part of the Hamiltonian requires wave functions to be written in terms of the mixing of two spin states instead of a function which is a product of spatial and spin parts. In this picture (pseudo)spin up and (pseudo)spin down eigenstates are the following:\cite{Fabian1999}
\al{
\bs
&\psi_{\v k  n \uparrow} =[a_{\v k}(\v r)\ket \uparrow + b_{\v k}(\v r)\ket \downarrow]e^{i\mathbf k \cdot \mathbf r}\\
&\psi_{-\v k  n \downarrow} =[a^*_{-\v k}(\v r)\ket \downarrow - b^*_{-\v k}(\v r)\ket \uparrow]e^{i\mathbf k \cdot \mathbf r}
\es
}
where  $a_{\v k}(\v r)$ and $b_{\v k}(\v r)$ are Bloch type functions and have the same symmetry of the lattice and $\ket \uparrow$ and $\ket \downarrow$ are spin up and down states. In centrosymmetric crystals due to inversion symmetry and time reversal symmetry these two states have the same energy, therefore all of the bands are doubly degenerate.  As a result of time reversal symmetry \m{E_{\v k n \uparrow}=E_{-\v k n\downarrow}}. If there also exist inversion symmetry in addition to that then \m{E_{\v k n \uparrow}=E_{-\v k n \uparrow}}. So, it is concluded that \m{E_{\v k n \uparrow}=E_{\v k n \downarrow}}. In this perspective, a scattering process can deliver a (pseudo)spin-up state to a (pseudo)spin-down state elastically which is not possible for a pure spin state. By pseudo(spin) up and down I mean that a state consists of mostly up or down spins relative to a direction (usually the z-direction). Moreover, the magnitude of the $b_{\v k}(\v r)$ (also known as spin mixing 
coefficient) is usually small whereas $a_{\v k}(\v r)
$ is usually close to 1. This is the main reason for a scattering process resulting in a flip of the (pseudo)spin is quite seldom compared to the number of total scatterings. Because spin-orbit coupling is relatively small, it takes typically 10$^6$ scattering events before the spin is flipped.\cite{Simon2008}

\section{The Effective Dirac-Delta Type Impurity Potential}
We first model the scalar scattering potential of impurities for the electrons of the conduction band. Considering Bloch like wavefunctions the matrix elements of a potential between two states can be written using equations \ref{eq:effpot} and \ref{eq:so-tensor} from the previous section.
\al{
\bs
  \tilde{V}_{\v k' \alpha',\v k, \alpha} &= \int d\v r e^{-i\v k' \cdot \v r} \Big ( V(\uv{r})\delta_{\alpha' \alpha} 
 +\sum_{ij}\lambda_{ij}\nabla_i V(\uv{r}) \nabla_j \delta_{\alpha' \alpha} \\
&+i\sum_{ijk}\lambda_{ijk}\nabla_i V(\uv{r}) \nabla_j [\sigma_k]_{\alpha' \alpha} \Big )e^{i\v k \cdot \v r} 
\es
}
The first two terms in the parenthesis relax momentum but do not relax spin (note \m{\delta_{\alpha' \alpha}}), therefore, they don't play any role in the spin relaxation mechanism. However, they are necessary for momentum relaxation calculations.
For \STO with only six non-zero $\lambda$ values, the spin dependent part of the potential ($V^s$) which can flip the spin becomes:
\al{ \label{eq:effpotspin}
\bs
 \tilde{V}^s_{\v k' \alpha',\v k, \alpha}=&i \int d\v r e^{-i\v k' \cdot \v r} \Big[
  \lambda_{k}(\nabla_y V(\uv{r}) \nabla_x -\nabla_x V(\uv{r}) \nabla_y) \unit{\sigma}_z\\
&+ \lambda_{j}(\nabla_x V(\uv{r}) \nabla_z - \nabla_z V(\uv{r}) \nabla_x) \unit{\sigma}_y + \lambda_{i}( \nabla_z V(\uv{r}) \nabla_y -\nabla_y V(\uv{r}) \nabla_z ) \unit{\sigma }_x        
\Big] e^{i\v k \cdot \v r}
\es
}
where
$\unit{{\sigma}}_i = \matrixel{u_{c\v k \alpha'}}{\sigma_i}{u_{c\v k \alpha}}$ and i=x,y,z
We assume a Dirac-Delta type impurity potential \m{V(\v r)=V_0\delta(\v r)}, and define \m{\v q=\v {k'} -\v k} and substitute into Eq.~\ref{eq:effpotspin}.
\al{
\bs
  \tilde{V}^s_{\v k' \alpha',\v k, \alpha}=&iV_0 \int d\v r e^{-i\v q \cdot \v r}  \Big[ \lambda_k (\nabla_y \delta(\v{r}) (ik_x) -\nabla_x \delta(\v{r}) (ik_y)) \unit{\sigma}_z \\
  &+ \lambda_{j}(\nabla_x \delta(\v{r}) (ik_z) - \nabla_z \delta(\v{r}) (ik_x)) \unit{\sigma}_y+ \lambda_{i}( \nabla_z \delta(\v{r}) (ik_y) -\nabla_y \delta(\v{r}) (ik_z) ) \unit{\sigma }_x        
\Big]
\es
}
This result is basically the Fourier transform of the effective potential.
\al{\label{eq:Vscatt}
\bs
\tilde{V}^s_{\v q, \alpha', \alpha}=iV_0&[\lambda_{k}(-q_y k_x+q_x k_y)\unit{\sigma}_z+ \lambda_{j}(-q_x k_z+q_z k_x)\unit{\sigma}_y + \lambda_{i}(-q_z k_y+q_y k_z)\unit{\sigma}_x]
\es}
Thus the scattering potential, which describes the effect of a Dirac-Delta type impurity on the conduction band electrons, can be written in a compact form:
\al{\label{eq:compactscattering}
 V = i V_0 \epsilon_{ijk} q_i k_j \lambda_k = i V_0 (\v q \times \v k) \cdot  \lambda =i V_0 (\v k' \times \v k) \cdot  \lambda
 }
 where 
 \al{\label{eq:compactlambda}
{\lambda} =\lambda_i \sigma_x \v i + \lambda_j\sigma_y \v j+ \lambda_k \sigma_z \v k
}

\section{Spin Relaxation Times of \LAOSTO 2DEGs}
From the effective spin-orbit interaction and effective potential, \eqr{eq:compactscattering} and \eqr{eq:compactlambda} we calculate the spin relaxation times via the Elliott-Yafet mechanism  as a function of temperature. The relaxation time is defined as:
\al{ \label{eq:spinrelax}
  \frac{1}{\tau_s(\v k)}=\sum_{\v k' \neq \v k, \alpha' \neq  \alpha} P(\v k,\alpha ; \v k',\alpha')
}
where \m{P(\v k,\alpha ; \v k',\alpha')} is the probability of scattering from a state with wave vector \m{\v k} and spin $\alpha$ to \m{\v k'} and spin $\alpha'$ per unit time. Spin flips occur via mixing of different spin states into the wave functions of eigenstates of different momenta, which produces spin flips as the carriers scatter from interactions with impurities and phonons.  We model the scattering as being due to impurities, and calculate spin-flip scattering probabilities (transition rate) from Fermi's Golden rule:
\al{ \label{eq:goldenrule}
P(\v k,\alpha ; \v k',\alpha')=\frac{2\pi}{\hbar}|\matrixel{\psi_{\v k',\alpha'}}{V_{scattering}}{\psi_{\v k,\alpha}}|^2\delta(\epsilon_{\v k'}-\epsilon_{\v k})
}
The matrix element of a Dirac-Delta potential between scattered states ($V_{scattering}$ ) is already calculated in Eq.~\ref{eq:Vscatt}. Using this derivation we calculate the probability of spin flip for Dirac-Delta type potential:
\al{ \label{eq:PD}
\bs
P_D=V_0^2\frac{2\pi}{\hbar}&\Big| \lambda_{k}(-q_y k_x+q_x k_y)\unit{\sigma}_z +\lambda_{j}(-q_x k_z+q_z k_x)\unit{\sigma}_y\\&+\lambda_{i}(-q_z k_y+q_y k_z)\unit{\sigma}_x \Big|^2\delta(\epsilon_{\v k'}-\epsilon_{\v k})
\es
}
An overall spin relaxation time for a collection of electrons with different $\v k$ values can be calculated as:
\al{
  \label{eq:totalspinrelaxation}
\frac{1}{\tau_s}=\frac{\sum_{k}\tau_s(\mathbf k)^{-1}f_k(\epsilon)(1-f_{k'}(\epsilon))}{\sum_{k}f_k(\epsilon)(1-f_{k'}(\epsilon)}
}
where f is the Fermi-Dirac distribution function, and $\epsilon$ is the energy of the scattered electrons.

\begin{figure}[h]
\includegraphics[width=1\textwidth]{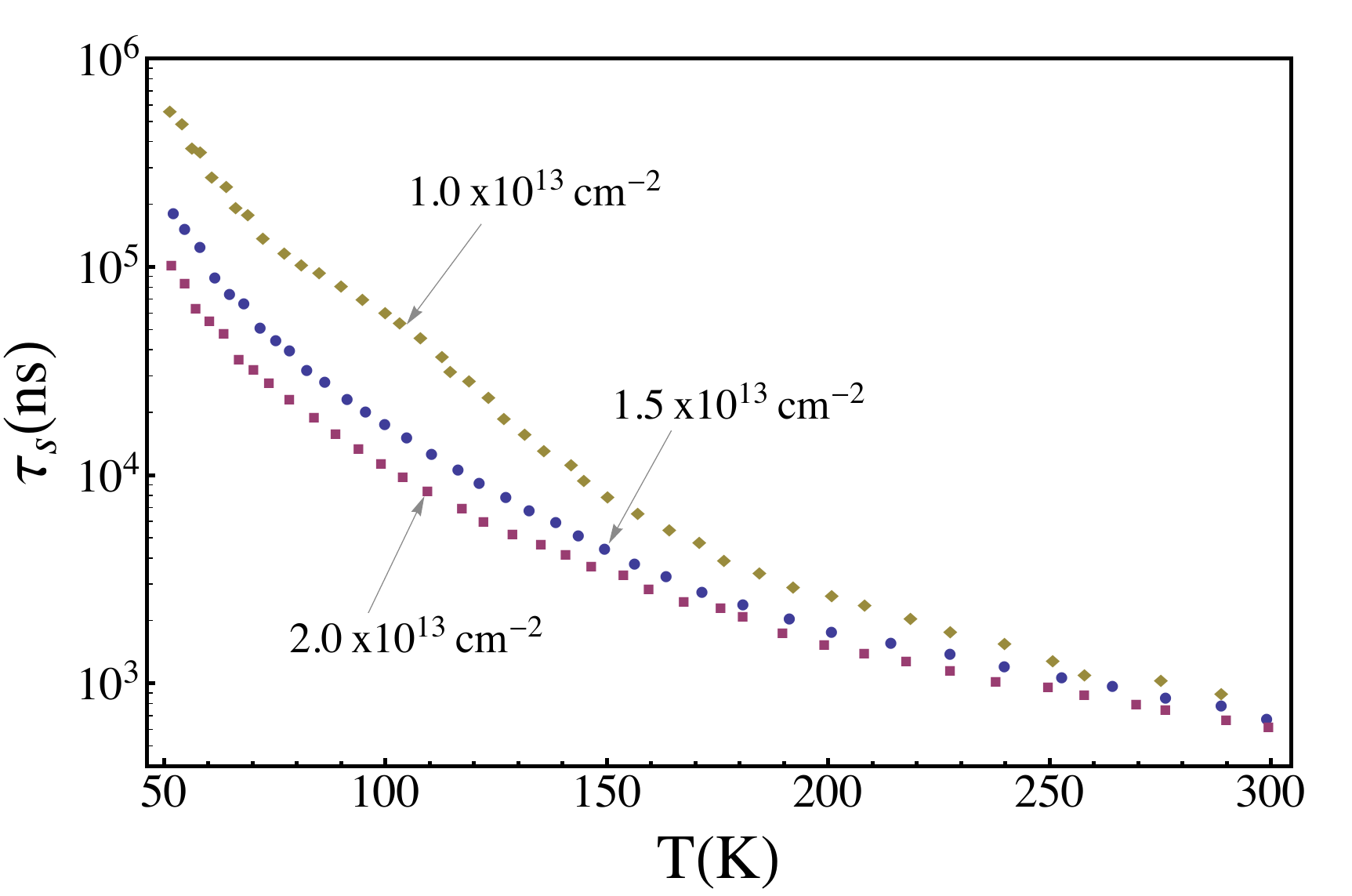}
\caption[Spin relaxation time as a function of temperature for various densities of carriers in the LAO/STO 2DEG]{ Spin relaxation time as a function of temperature for various densities of carriers in the LAO/STO 2DEG.  The graph with diamonds (yellow) is for a 2DEG grown at 10$^{-6}$ mbar partial oxygen pressure while others are grown at 10$^{-4}$ mbar partial oxygen pressure. Mobilities are between 10$^0$-10$^4$ cm$^2$V$^{-1}$s$^{-1}$ over the same temperature range. Using mobilities and densities measured by Kalabukhov et al.\cite{Kalabukhov2007} we find the strength of the scattering and, therefore, the relaxation time \cite{Sahin2014}.}
\label{fig:lao-sto-lifetime}
\end{figure}

As we model the scatterer as an impurity of Delta type, $V(\v r)=V_0 \delta(\v r)$, the strength of the impurity, namely $V_0$, must be calculated in order to find the spin relaxation times. This is done in two steps. First we calculate carrier densities as a function of $V_0$,  temperature and chemical potential,
\al{
n(\mu,T)=\frac{1}{8\pi^3}\int d\v k f_k(\epsilon)
}
where n is the electron carrier density in the conduction band and $\mu$ is the chemical potential. We should also note that the chemical potential at each carrier density and temperature is below the second conduction band suggesting that only the first conduction band is occupied by electrons. Then we calculate the mobility, which is a function of the momentum relaxation time and the effective mass, and compare to experimental Hall mobilities from Kalabukhov et al.\cite{Kalabukhov2007} and Ariando et al.\cite{Ariando2011} for different carrier densities and temperatures. $V_0$ is then deduced from this comparison. The mobility is defined as:
\al{
\mu_H=\frac{e}{m^*}<\tau_p >
}
where $<\tau_p >$ is average momentum relaxation time, e is electron charge, and $m^*$ is the effective mass. The average momentum relaxation time can be readily calculated the same way as the average spin relaxation time. The difference is that we use the full effective impurity scattering potential in momentum relaxation time calculations, while only the spin dependent part of the effective potential is used in spin relaxation calculations. Therefore $<\tau_p >$ can be written similarly to Eq. \ref{eq:totalspinrelaxation}
\al{
<\frac{1}{\tau_p}>=\frac{\sum_{k}\tau_p(\mathbf k)^{-1}f_k(\epsilon)(1-f_{k'}(\epsilon))}{\sum_{k}f_k(\epsilon)(1-f_{k'}(\epsilon))}
}
\al{
  \frac{1}{\tau_p(\v k)}=\sum_{\v k' \neq \v k} P(\v k,s ; \v k',s')
}

\begin{figure}[h]
\centering
\includegraphics[width=0.99\textwidth]{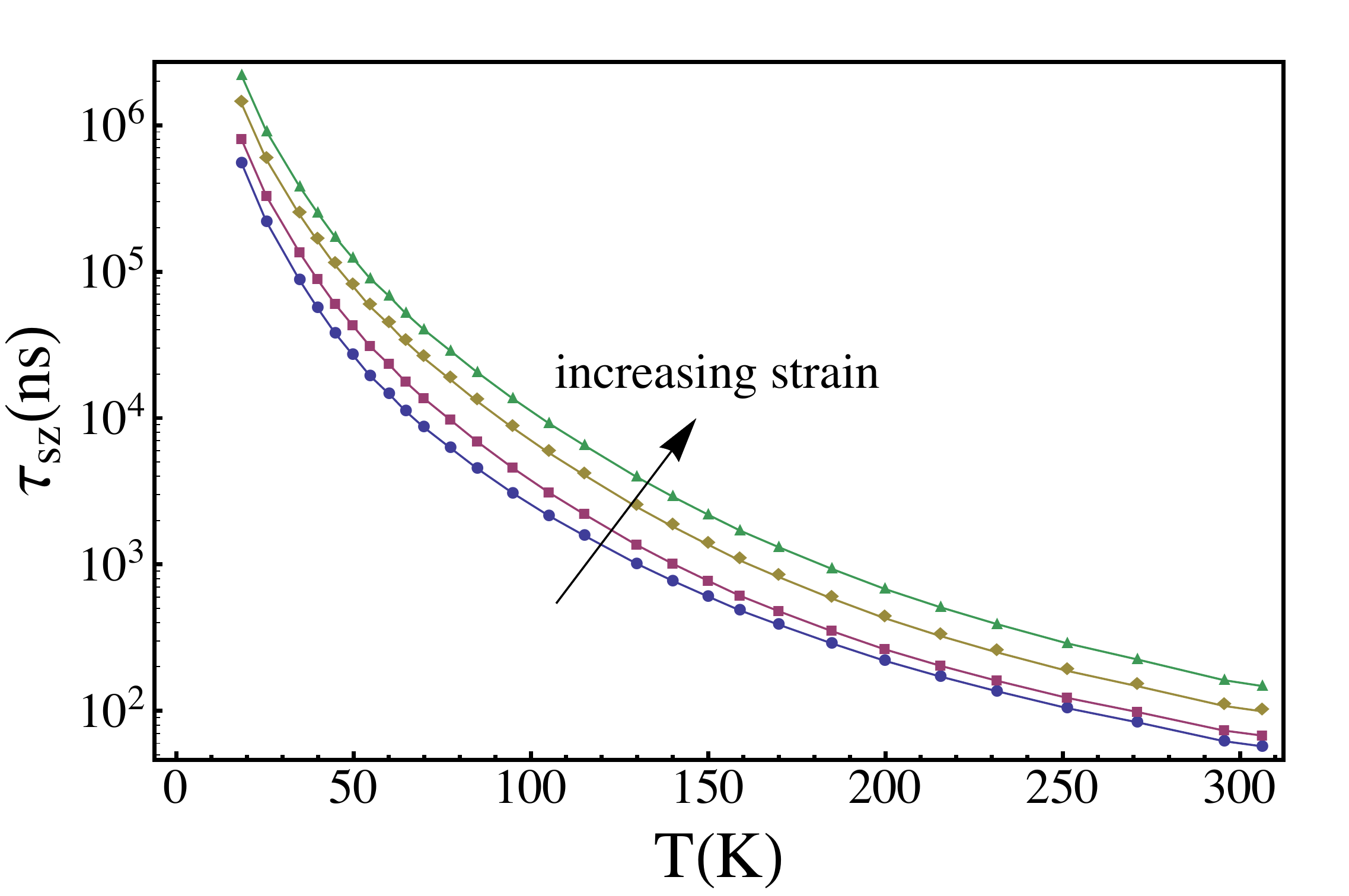}
\caption[Spin relaxation time of bulk strontium titanate as a function of temperature]{Spin relaxation time of bulk strontium titanate as a function of temperature. The carrier concentration of the strontium titanate is $1.0\times 10^{18}$cm$^{-3}$ and  mobilities vary from 5-7000 cm$^2$V$^{-1}$s$^{-1}$ over this temperature range\cite{Moos1995}. The energy splitting due to strain varies from 50 meV to 100 meV.(from blue to yellow curves) Increasing strain decreases the spin mixing ratio resulting in slightly higher spin lifetimes.}
\label{fig:sto-lifetime}
\end{figure}

The results of spin lifetime calculations are depicted in \figref{fig:lao-sto-lifetime} for the two-dimensional electron gas at  the LAO/STO interface for several experimentally achieved carrier densities (corresponding to several levels of oxygen partial pressure during the growth) \cite{Sahin2014}. The dominant source of the reduction of carrier spin lifetime with temperature is an increase in the scattering rate from phonons at higher temperatures. These spin lifetimes considerably exceed those of bulk III-V semiconductors at room temperature and are one to two orders of magnitude longer than room-temperature spin lifetimes in specially-designed GaAs quantum wells grown along the [110] direction\cite{Karimov2003}.  The spin lifetimes roughly follow a temperature-dependent power law T$^{-3.5}$, with most of the temperature dependence originating from the $T^{-2.7}$ dependence of the experimental carrier mobility\cite{Moos1995} used in the calculation. (\figref{fig:sto-lifetime})
Spins oriented along $\hat x$ or $\hat y$ exhibit the same lifetime dependence on temperature and strain, but are shorter by $\sim 15\%$ at low temperatures and $\sim 10\%$ at room temperature from $\tau_{sz}$.

Here the Rashba spin splittings induced by the effective confinement fields along the growth direction at the interface are ignored; these splittings further reduce the spin lifetimes, and do not  substantially change the spin-orbit structure of the wave functions of the conduction band\cite{Meier1984}, thus our results can be viewed as the long spin lifetimes obtainable if the confinement field that induces the Rashba spin splitting has been compensated by another field, such as a gate field\cite{Lau2005}. 
\section{Tuning Spin Lifetimes of \STO by Strain}
We have also studied spin lifetimes for bulk strained strontium titanate for spin parallel to $\hat z$ ($\tau_{sz}$ in Fig. \ref{fig:sto-lifetime} by evaluating from Eq. \ref{eq:totalspinrelaxation} using reported carrier mobilities and densities \cite{Moos1995} ). Spins oriented along $\hat x$ or $\hat y$ exhibit the same lifetime dependence on temperature and strain, but are shorter than $\tau_{sz}$ by 15\%. Strain splitting of the bands is increased uniformly from 50 meV to 110 meV which reduces the spin mixing of these bands, resulting in a longer spin lifetime.  The resulting spin lifetimes of \LAOSTO 2DEGs are of the same order as those of the strained STO at low temperatures, but one order of magnitude greater at room temperature. We also observe that spin lifetimes of the bulk strontium titanate have the same temperature dependence as the LAO/STO interface.

\section{Conclusions}
We introduce a systematic approach to the calculation of the effective spin-orbit interaction and the Elliot-Yafet spin relaxation rate in doubly-degenerate bands. This method is broadly applicable to centrosymmetric nonmagnetic materials. By using a low energy tight-binding Hamiltonian with spin-orbit coupling, we calculated the eigenenergies and eigenfunctions of bulk strontium titanate and the 2DEG at the LAO/STO interface. We derived a formula for the effective spin-orbit interaction in the conduction band of these systems and applied it to spin lifetimes. We calculated the spin lifetimes from the Elliott-Yafet mechanism as electrons scatter from an impurity. These results reproduce previous calculations via the \kp theory of spin lifetimes in III-V semiconductors. Our results also support the presence of robust, room-temperature spin dynamics in oxide materials such as STO and the \LAOSTO interfacial 2DEG. We have also shown that these spin lifetimes can be tuned by strain, which changes the strength of the spin-orbit coupling at the interface \cite{Sahin2014}. As centrosymmetric materials have recently taken  a more prominent role in spin-dependent phenomena, it is expected that this approach will apply to a broad range of materials and spin-dependent phenomena.

\chapter{Intrinsic Spin-Hall Effect}

\section{Introduction}
\begin{figure}[t]
\includegraphics[width=1\linewidth]{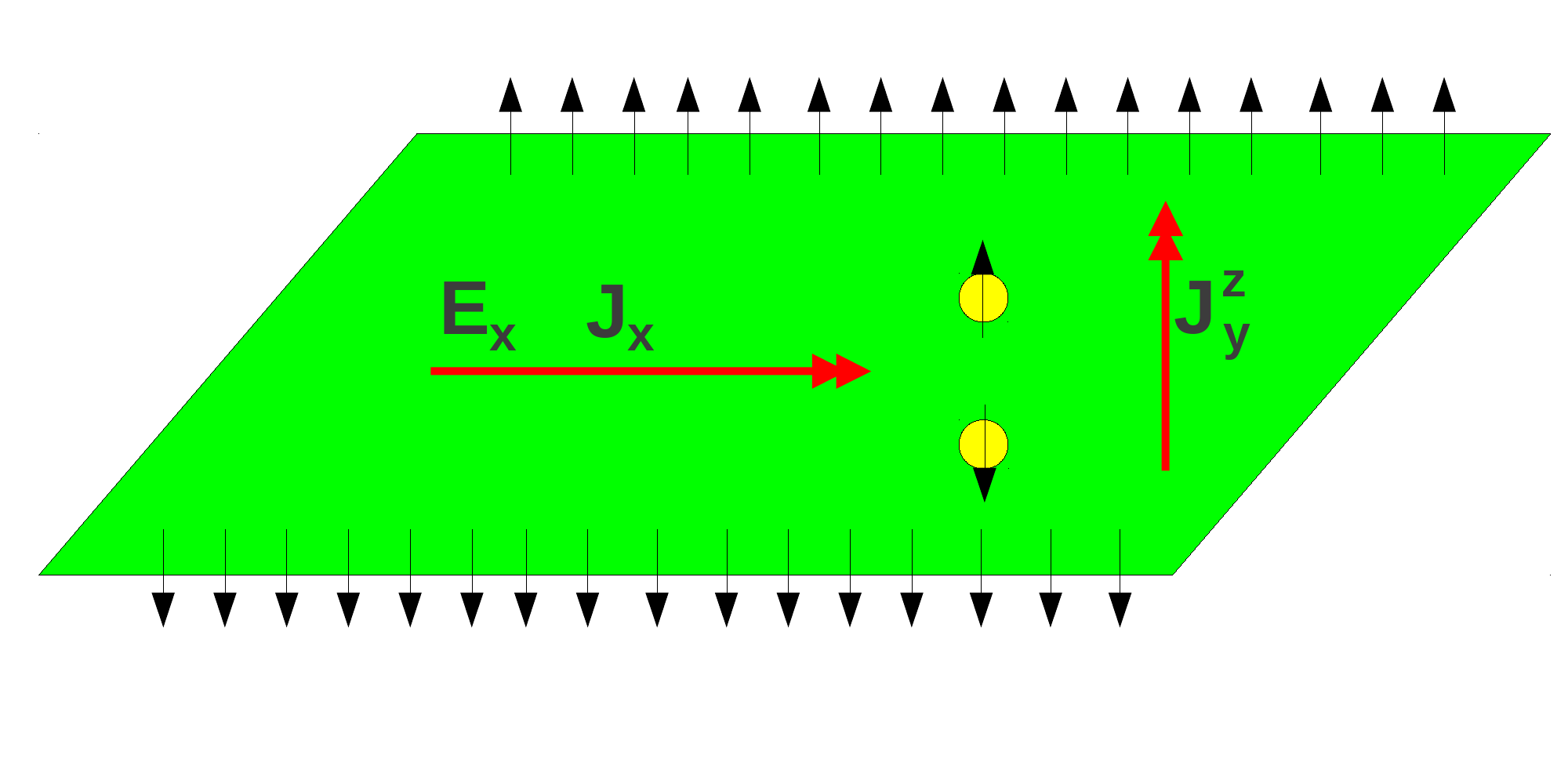}
\caption[Spin Hall effect]{Spin Hall effect. An electric current flowing in one direction results in a perpendicular spin current under the effect of spin-orbit interaction. When there exists a J$_x$ along the x direction under the effect of electric field E$_x$, electrons with opposite spins tend to move in opposite y directions and this results in spin accumulation at the opposite edges with different spins, and therefore a spin current perpendicular to the electrical current.(J$_y^z$). The spin Hall conductivity is defined as the ratio of the spin current to the electric field. ($\sigma_{yx}^z=\frac{J_y^z}{E_x}$)}
\label{fig:spinhalleffect}
\end{figure}

The Hall effect is defined as the deflection of the charge current perpendicular to its direction of flow due to an external magnetic field. This effect was observed by Edwin Hall 18 years before the discovery of the electron. Since then there have been found a few different members of the Hall effect family such as the ordinary Hall effect, the anomalous Hall effect, the quantum Hall effect, the fractional quantum Hall effect, and finally the spin Hall effect.\cite{Inoue2005} Among all of these phenomena, the spin Hall effect (SHE) is unique as it does not require an external magnetic field or magnetization to be observed. In other words, time-reversal symmetry is not broken, and there is no net magnetization as in the case of a similar effect, anomalous Hall effect. The SHE mainly describes the emergence of a perpendicular spin current as a result of an external electric field (Fig.~\ref{fig:spinhalleffect}). The spin Hall conductivity is the ratio of the spin current to the electric field and is a way of measuring the efficiency of a system to convert charge current to spin current. Materials with large SHC driven under the effect of spin-orbit interactions provide spin control with no external magnetic field and generate pure spin currents; therefore they are as significant candidates for future spintronics applications. 

There exist mainly three sources that contribute the strength of the spin Hall conductivity, which can be subcategorized into extrinsic and intrinsic effects:
\begin{enumerate}
     \item (extrinsic) Skew-scattering: This is due to spin-dependent scattering of electrons by impurities at a different rate depending on the polarization of their spins.
     \item (extrinsic) Side-jump: This effect arises from the non-canonical characteristics of the physical position and velocity operators. Under the effect of the spin-orbit interaction, the physical position operator is related to the canonical position operator:
     \al{
     \v r_{phys,i}=\v r_{canon,i} -\frac{\alpha}{\hbar}\v p_i \times \sigma_i}
      where $\alpha$ is the spin-orbit interaction in the system, index i denotes the ith electron. The physical velocity operator, which is the time derivative of the position operator, also has an extra term due to spin-orbit interaction, called as anomalous part of the velocity operator. As a result of that the center of a wave packet shifts in the perpendicular direction as a result of a momentum transfer (external electric field) to the wave packet, eventually causing a spin accumulation at the edges of the material.
     \item (intrinsic) Berry curvature: The intrinsic effect originates from the Berry curvature of the electronic band structure. The Berry curvature plays the role of a k-dependent effective magnetic field in momentum space and analogous to the Lorentz force in the ordinary Hall effect.
\end{enumerate}
In general, the spin Hall effect stems from a combination of the three mechanisms explained above. However, it is easy to distinguish materials where only one of the mechanisms is stronger than the other. All of the calculations in this chapter assume that the intrinsic effect is dominant and neglect extrinsic contribution to the spin Hall conductivity. 

The first experiment for the spin Hall effect was conducted by Kato \etal \cite{Kato2004} measuring optical Kerr rotation of incident linearly polarized light and thereby observing a spin accumulation at the edges of  GaAs and In$_x$Ga$_{1-x}$As films  in 2004. They concluded that crystal orientation and strain dependent pure spin current was the result of extrinsic spin Hall conductivity. Shortly after Wunderlich \etal \cite{Wunderlich2005} observed Kerr rotation of circularly polarized light in 2D hole systems at (Al,Ga)As/GaAs heterojunctions. However measurement of direct SHE by optical methods is only suitable for direct band gap materials. For semiconductors such as silicon, inverse spin Hall effect measurements are possible. Silicon has a small spin-to-charge current conversion efficiency. However, it can still be used as a spin-current detector.\ \cite{Ando2012}. One of the materials with the highest spin Hall conductivity is platinum, which was electrically detected to have a conductivity of 240 $\Omega$cm$^{-1}$  by Kimura \etal \cite{Kimura2007}. These results were verified by the theoretical study of Guo \etal \cite{Guo2008} using fist principle relativistic band structure calculations which concluded that platinum has an SHC of 200 $\Omega^{-1}$cm$^{-1}$ at room temperature.

\section{Linear Response Theory and Calculation of Spin Hall Conductivity from the Kubo Formula}
The Kubo formula expresses the linear response of a system, in other words, the linear response of an observable to a time-dependent source, force, or perturbation. There are other techniques to calculate the response of a system, such as semiclassical Boltzmann equations, which are useful for non-equilibrium situations, and the Landauer formalism for mesoscopic systems. The Kubo formula is a correlation/response function defining the susceptibility of the system. It may be referred as an x-y correlation function where x and y could be e.g. an electric current, a spin current. As an example spin Hall conductivity could be named as a spin current-electric current correlation function (or susceptibility). This formalism applies to many different physical problems, such as local density, charge current, spin current and magnetization.

Since the intrinsic spin Hall conductivity is an electrical current-spin current correlation function it can be described by a similar formula in the clean and static regime:
\al{\label{eq:spinhall}
Re[\sigma]^z_{yx}=\frac{e\hbar}{V}\sum_{\v kn}f_{\v k ,n}\Omega_n (\v k)}
which is the sum of all the Berry curvatures ($\Omega_n (\v k)$) multiplied by the Fermi function, where V is the total volume of the system and n is the band index. The Berry curvature is an intrinsic property of the system and directly originates from the band structure:
\al{\label{berrycurvature}\Omega_n (\v k)=\sum_{n \neq n'}Im \frac{\langle\psi_{n'\v k}|\hat J_y^z| \psi_{n\v k}\rangle\matrixel{\psi_{n\v k}}{\hat{\textit{v}}_x}{\psi_{n'\v k}}}{(E_{n\v k}-E_{n'\v k})^2}}
where $\psi_{n \v k}$ are eigenstates calculated from tight-binding Hamiltonian (Chapter 2). $\hat J_x^z$ and $\hat v_y$ are spin current and velocity operators which can be calculated as:
\al{ \label{eq:spincurrent}
J_y^z=\frac{\hbar}{4}(\hat v_y\sigma_z+\sigma_z\hat v_y) ~~\text{and}~~\hbar \hat v_i=\nabla_{k_i}\hat H
}
The Berry phase which was introduced by Michael Berry in 1984 \cite{Berry1984}, is the phase that a state gains after its Hamiltonian is adiabatically transformed along a parameter space (momentum space in our calculations). The simplest example of such a phase would be the vector at the North Pole. As the vector is moved on a curved space (sphere in this example) towards the initial position following a closed loop, the final direction of the vector is different from the initial vector by a phase. Here curvature of the sphere represents the curvature of the parameter space, and the path which a vector follows is the path is taken into account when adiabatic change is applied to the Hamiltonian. In quantum mechanics instead of vectors in real space, we deal with vectors in the Hilbert space, which are the wavefunctions of the Hamiltonian. Let us assume that $\psi (\v k)$ is a solution to the Schr\"{o}dinger equation with Hamiltonian H($\v k$). Here $\v k$ stands for a parameter on which the Hamiltonian depends and can be changed adiabatically. The Berry curvature (Eq.~\ref{berrycurvature}) is then simply the curl of the Berry phase and related to many physical observables, such as the spin Hall effect and the anomalous Hall conductivity. In the case of a non-degenerate band structure, n is just an index labeling bands. However, in the degenerate case there is more than one wavefunction associated with each n at each $\v  k$. Thus the Berry curvature $\Omega_n^z (\v k)$ becomes a NxN matrix where N denotes the number of the degeneracy. Berry curvature for a non-degenerate case is gauge invariant: 
\al{
\Omega_n (\v k) =\nabla \times \v{A_n}=\sum_{n \neq n'}Im \frac{\matrixel{u_{n \v k}}{ j_y} {u_{n' \v k}} \matrixel{u_{n' \v k}}{\hat v_x}{u_{n \v k}}}{(E_{n\v k} -E_{n' \v k})^2} 
}
On the other hand, the non-abelian Berry curvature of degenerate bands is gauge covariant and therefore is not an observable. However, the trace of this matrix is a gauge invariant quantity. Degenerate and, therefore, non-abelian Berry curvature is written as:
\al{
\Omega_n (\v k) &= D \times \v{A_n}= 
\nabla \times \v{A}_n - i \v{A}_n \times \v{A}_n \\
\Omega_{ij} (\v k)&= \bra{\nabla_{\v k} u_{ik}}\times \ket{\nabla_{\v k} u_{jk}}+i \sum_{l \in \Sigma} 
\braket{\nabla_{\v k} u_{i \v k}}{u_{l \v k}} \times \braket{u_{l \v k}}{\nabla_{\v k} u_{j \v k}}
}
where $D$ is the covariant derivative and $\Sigma$ is the sub-space spanned by degenerate wavefunctions. The second term in non-abelian curvature, $\v{A}_n \times \v{A}_n$ is zero in the abelian case. For a doubly degenerate band, the Berry curvature will be in the form of a 2x2 matrix,
\[
\Omega= 
  \begin{matrix}\\\mbox{}\end{matrix}
  \begin{pmatrix} \Omega_{11} & \Omega_{12} \\
          \Omega_{12}^* & \Omega_{22} \\
   \end{pmatrix}
\]
and the spin Hall conductivity would be calculated as:
\al{
\sigma_{xy}^z\approx Tr(S\Omega)
}
where S is the Dirac spin matrix. All of the calculations in this chapter deal with abelian Berry curvature, which is ensured by the crystal symmetry. Furthermore, computational time is significantly reduced due to the symmetry of the Berry curvatures at different k-points of the Brillouin zone, which may be summarized as:
\al{
\Omega_n^z(\v k)&=\Omega_n^z(-\v k)\\
\Omega_n^z(k_x,k_y,k_z)&=\Omega_n^z(-k_x,k_y,k_z)\\
\Omega_n^z(k_x,-k_y,k_z)&=\Omega_n^z(k_x,k_y,-k_z)
}

\section{\LAOSTO Two Dimensional Electron Gas}
\begin{figure}[h]
\includegraphics[width=1\textwidth]{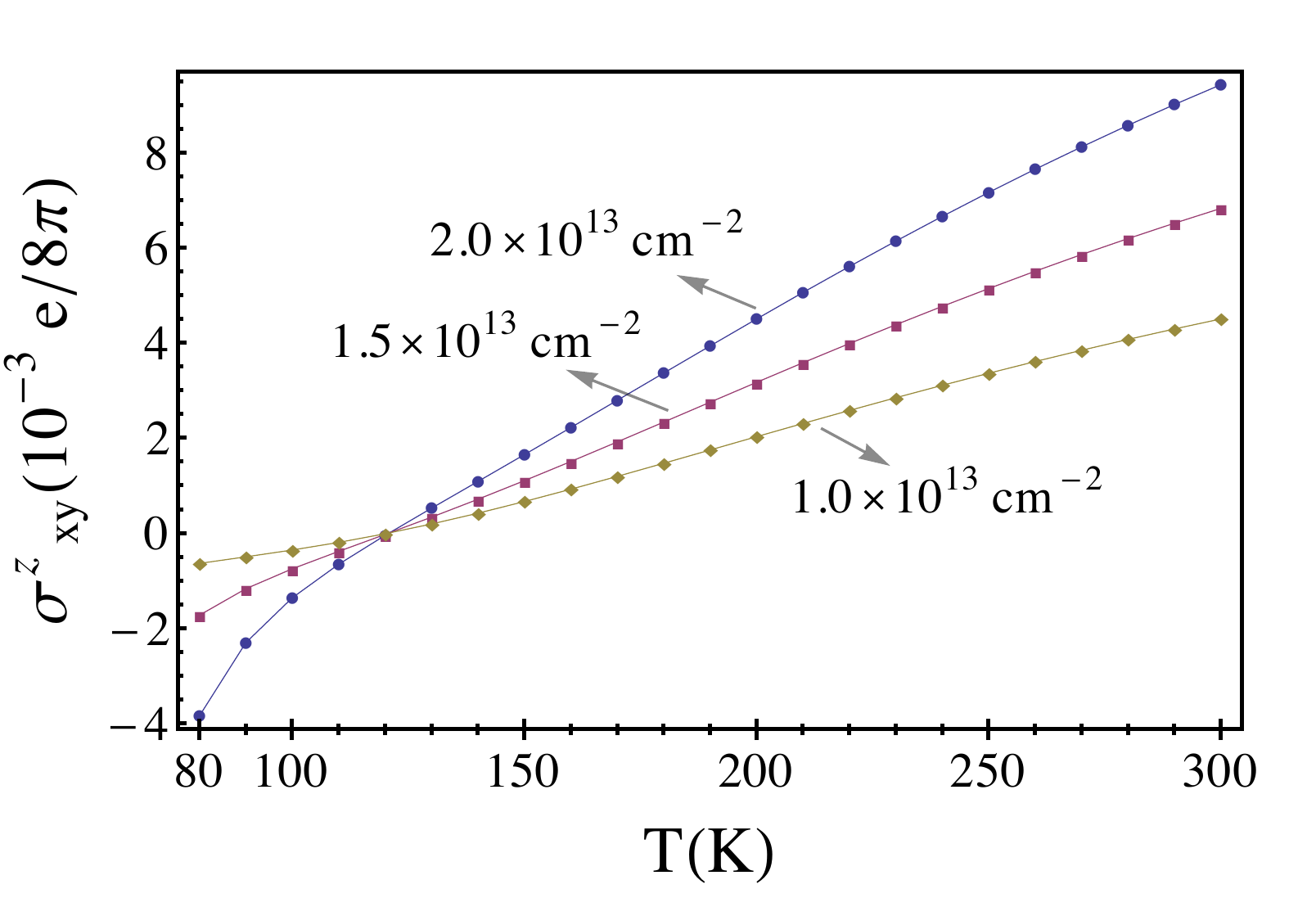}
\caption[Intrinsic spin Hall conductivity as a function of temperature for various densities of carriers in the \LAOSTO 2DEG]{Intrinsic spin Hall conductivity as a function of temperature for various densities of carriers in the \LAOSTO 2DEG. The unit of SHC in 2 dimensions is the electric charge e, or more commonly used one e/8$\pi$.}
\label{fig:SHC}
\end{figure}

The intrinsic spin Hall conductivity for several 2DEGs at the \LAOSTO interface has been calculated from the Kubo formula and is shown in  \figref{fig:SHC}. The total Hamiltonian for the \LAOSTO system consists of the tight-binding Hamiltonian, the spin-orbit Hamiltonian with spin-orbit couplings calculated from atomic spectra and a strain Hamiltonian which shifts bands in the growth direction [001] as a result of confinement effects. It is remarkable to observe that there exists a temperature where the spin Hall conductivity changes its sign. (from negative to positive) This sign change originates from the distribution of the Berry curvature within the Brillouin zone. The negative and positive Berry curvatures exist in energetically different k-points. Therefore, the overall contribution depends on the chemical potential. As the temperature is increased the chemical potential decreases and causes contributions from the edges of the Brillouin zone to be less significant, since their energies are higher than the energies around the zone center. These regions of the k-space have negative Berry curvature while the region around the zone center has a positive Berry curvature. (see lower figures in \figref{berrystraindependence} and \figref{berrysodependence})

\subsection{Strain Dependence of the Intrinsic Spin Hall Conductivity at the Interface of \LAOSTO}
\begin{figure}[t]
\centering
\includegraphics[width=1\textwidth]{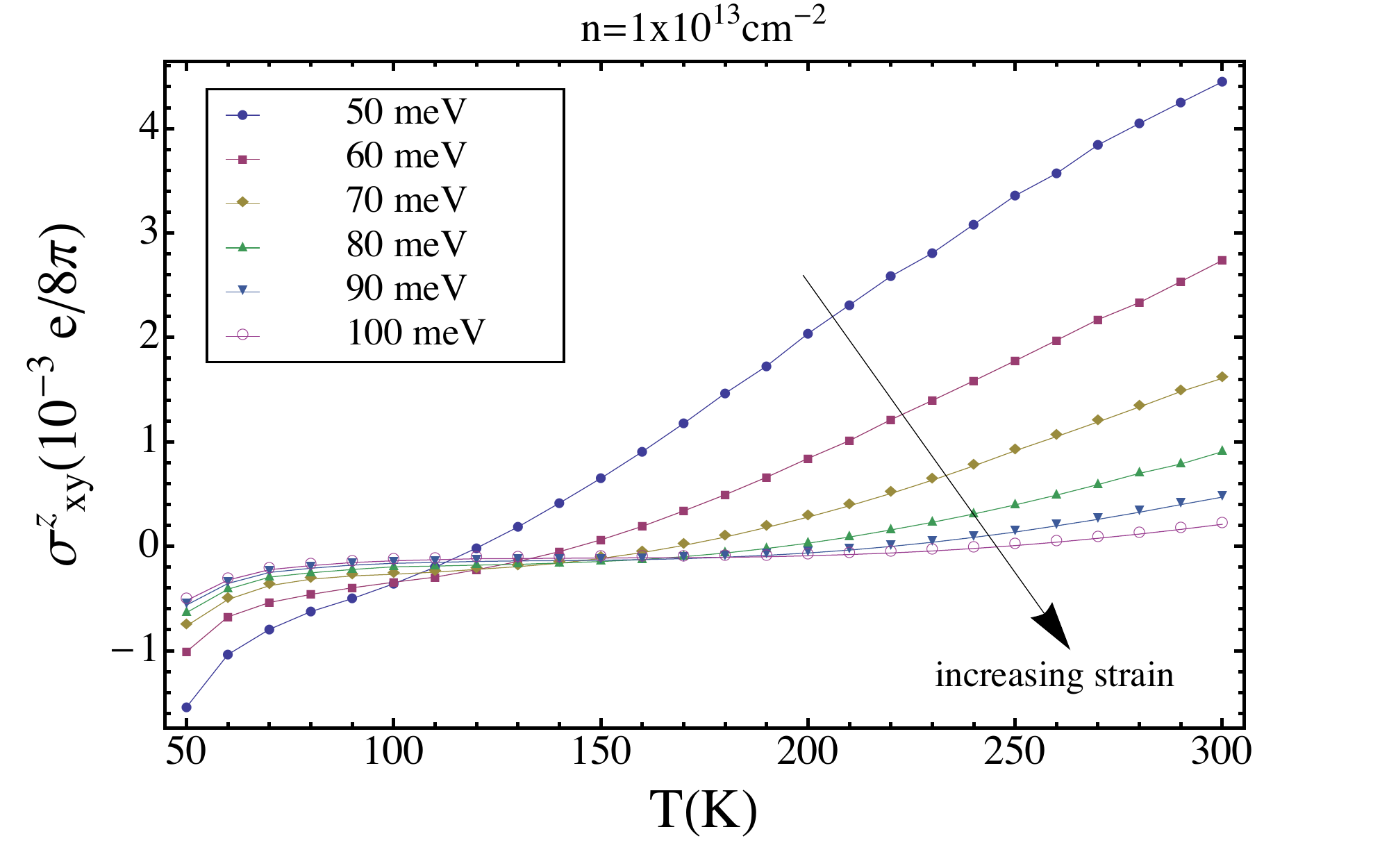}
\caption[Strain dependence of the intrinsic spin Hall conductivity]{Strain dependence of the intrinsic spin Hall conductivity. As the energy associated with strain increases from 50 meV to 100 meV, the conductivity drops dramatically. Strain causes shifting of the energy levels of conduction and valence bands. As higher conduction bands are shifted further, the spin Hall conductivity decreases since the Berry curvature (\eqr{berrycurvature}) is inversely related to the energy difference.}
\label{shcstraindependence}
\end{figure}

\begin{figure}[t]
\centering
\includegraphics[width=1\textwidth]{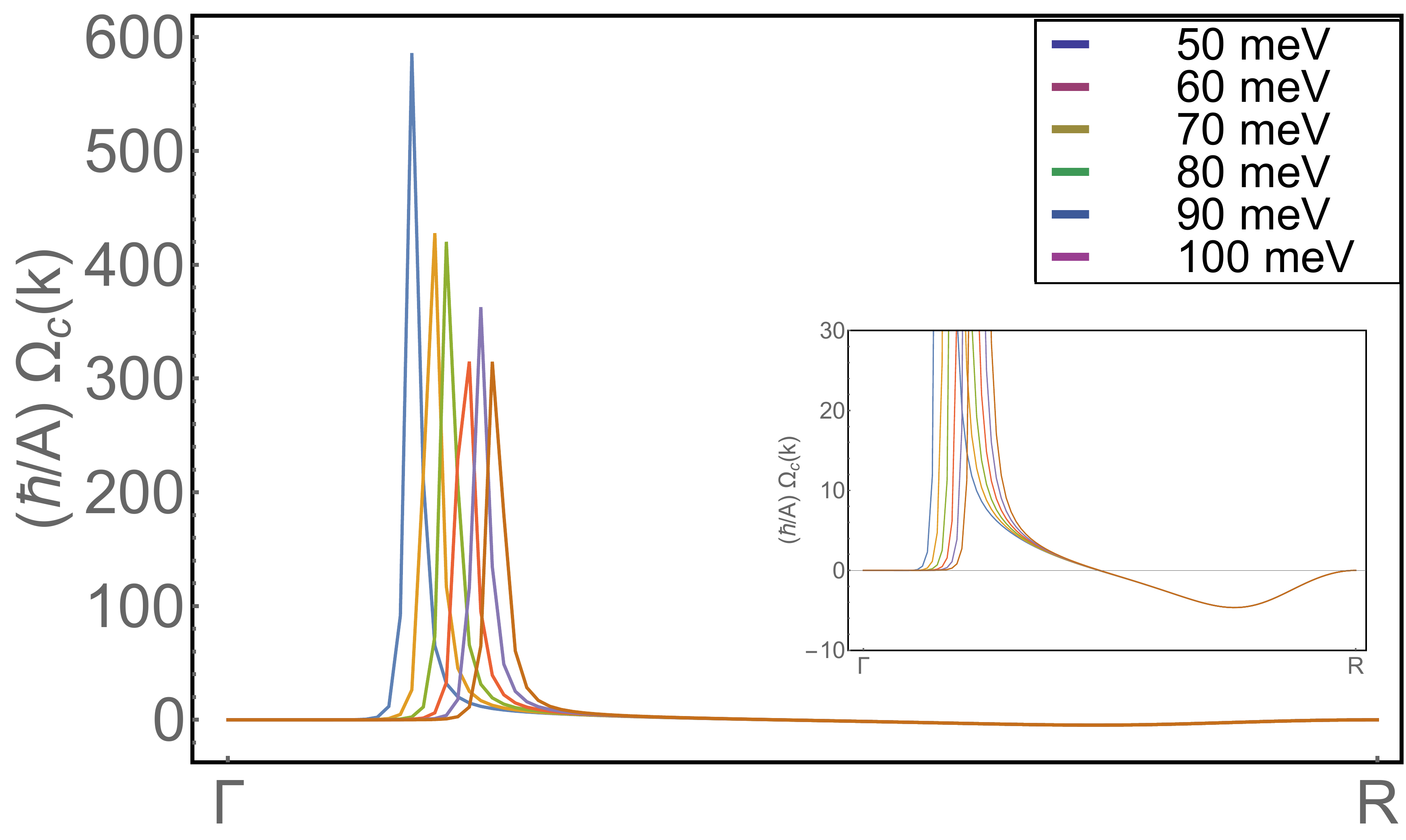}
\caption[Strain dependence of the Berry curvature]{Strain dependence of the Berry curvature. As the energy associated with strain increases from 50 meV to 100 meV the Berry curvature differs around the zone center, however this difference is negligibly small towards the Brillouin zone edges (inset).}
\label{berrystraindependence}
\end{figure}

From the same Hamiltonian, we have also calculated strain dependence of the intrinsic spin Hall conductivity and concluded that the magnitude of the intrinsic spin Hall conductivity dramatically depends on the strain. This relationship is shown in \figref{shcstraindependence}. Increasing strain results in vanishing spin Hall conductivity due to the fact that bands shift further from the lowest conduction band and the Kubo formula clearly shows that the ISHC is inversely related to the energy difference between calculated bands, that is the lowest conduction band of \LAOSTO 2DEG, and higher conduction bands. In addition to that the Berry curvature gets narrower in the positive regime while its negative values stay constant. (\figref{berrystraindependence}). Another important feature of the strain dependence of the ISHC is that the transition temperature where the ISHC changes its sign shifts to a higher temperature as the strain is increased.

\subsection{Spin-Orbit Dependence of the Intrinsic Spin Hall Conductivity in \LAOSTO}
\begin{figure}[h]
\includegraphics[width=1\textwidth]{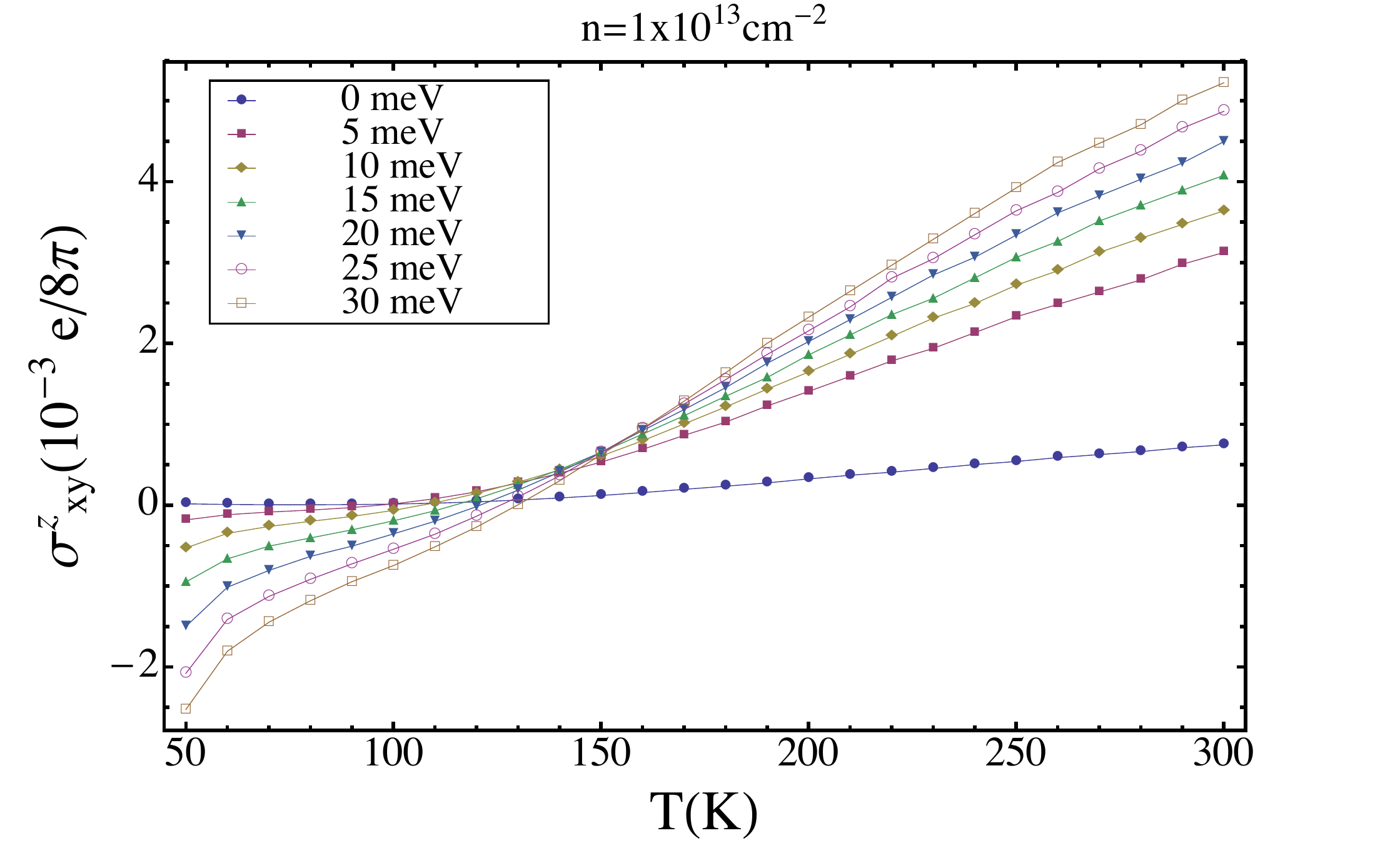}
\caption[Spin-orbit splitting dependence of the intrinsic spin Hall conductivity.]{Spin-orbit splitting dependence of the intrinsic spin Hall conductivity. As the spin-orbit splitting of conduction bands is increased from 0 meV to 30 meV, conductivity follows the strength of the splitting and grows steadily.}
\label{shcsodependence}
\end{figure}

\begin{figure}[h]
\includegraphics[width=1\textwidth]{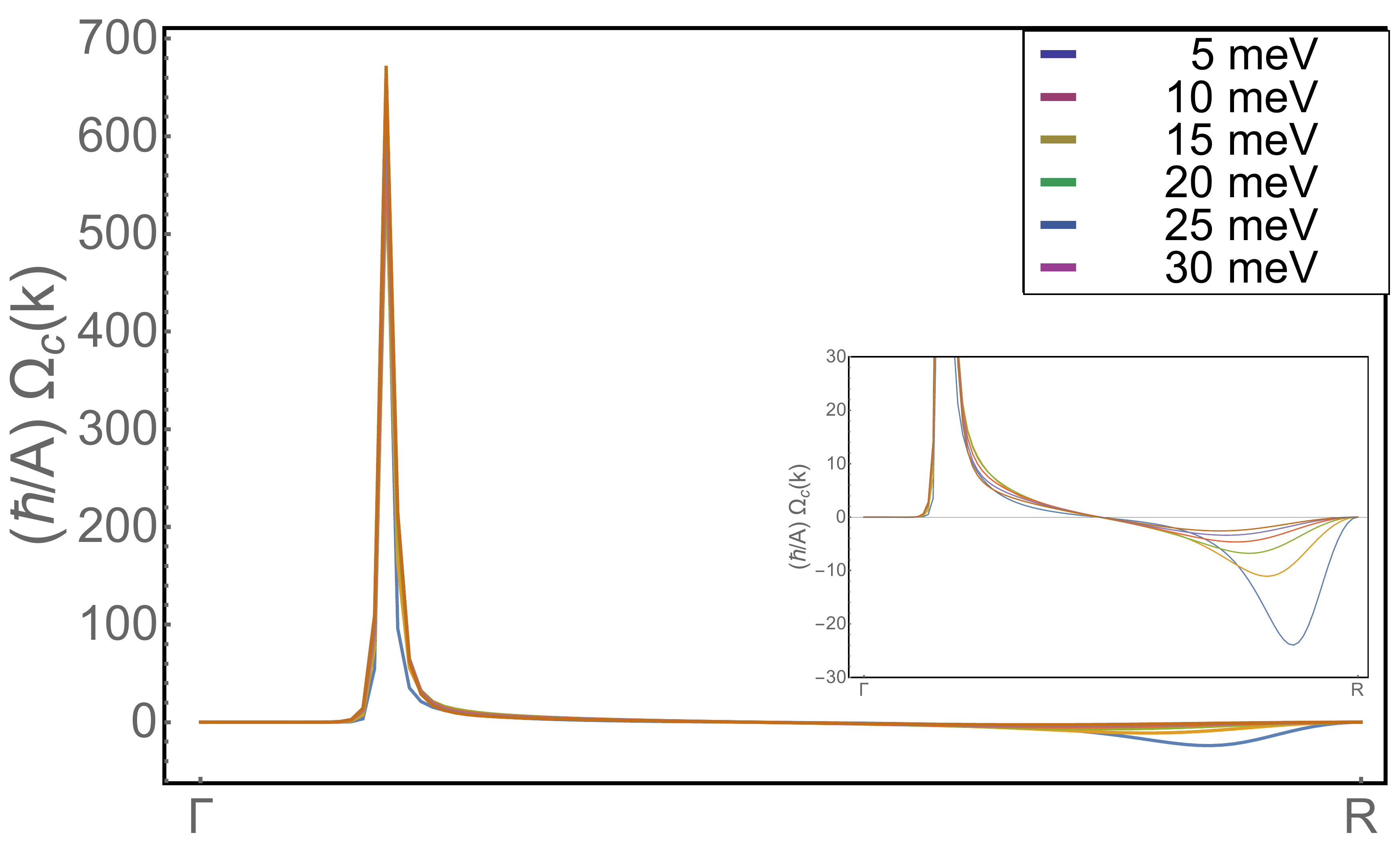}
\caption[Spin-orbit splitting dependence of the Berry curvature]{Spin-orbit splitting dependence of the Berry curvature. The Berry curvature doesn't exhibit much change around the zone center due to different splittings, but it gets modified by the splitting significantly in regions closer to the Brillouin zone edges (inset).}
\label{berrysodependence}
\end{figure}

We also studied how the spin-orbit splitting in the conduction band influences the ISHC. There have been different experiments showing a variety of spin-orbit splittings in the conduction band ranging from 0 meV to 30 meV. The carrier density is taken to be $1\times 10^{13}$cm$^{-2}$ and the spin-orbit splitting is changed in the range of experimental values, and then the ISHC has been calculated. (\figref{shcsodependence}). Since the spin Hall effect is a manifestation of the spin-orbit interaction, it is expected to get a lower conductivity for less spin-orbit splitting.  As expected, less spin-orbit splitting causes spin Hall conductivity to vanish while steadily increasing for larger splittings. Magnitudes of the ISHC don't change too much compared to the strain dependence of the ISHC (\figref{berrystraindependence}), however, the distribution of the Berry curvature over the Brillouin zone differs significantly. Unlike the case for strain dependence, the Berry curvature doesn't vary much around the zone center while exhibiting quite different levels of curvature close to the zone edges (inset of the \figref{berrysodependence}). That is due to the fact that the spin-orbit splitting increases its influence away from the zone center as a result of the higher spin mixing ratio in these regions. 

In addition to the variation of Berry curvature between strain dependent and spin splitting dependent ISHC, there exists an alteration of how the transition temperature behaves. In the strain dependent case, the transition temperature is shifted by more than 100 K. On the other hand, different spin splittings move the transition temperature by 25K, which reveals that the modification of the transition temperature depends more on the strain than the spin-orbit splitting.

\section{Bismuth-Antimony Alloys}
\subsection{Intrinsic Spin Hall Conductivity of \BiSb Alloys}
Bismuth and antimony are both semimetals with enormous spin-orbit couplings, 1.5 eV and 0.6 eV respectively\cite{Gonze1990}. Therefore, they are expected to exhibit very large spin Hall conductivities. Bismuth is not only a building block for many topological insulators but it also exhibits topologically insulating,and quantum Hall states when alloyed with antimony at 10\% as shown by angle-resolved photoemission spectroscopy (ARPES) measurements. \cite{Hsieh2008, Hsieh2009}. We calculated the intrinsic spin Hall conductivity  for these systems from the Kubo formula in the clean static limit, and investigated the behavior of the Berry curvature, which may rise to robust spin Hall conductivities as in the case of platinum \cite{Guo2008}. Energy and momentum resolved Berry curvatures are computed by a low-energy, three nearest neighbor tight-binding Hamiltonian \cite{Liu1995} and virtual crystal approximation in a full zone picture.

\begin{figure}[h]
    \includegraphics[width=.95\linewidth]{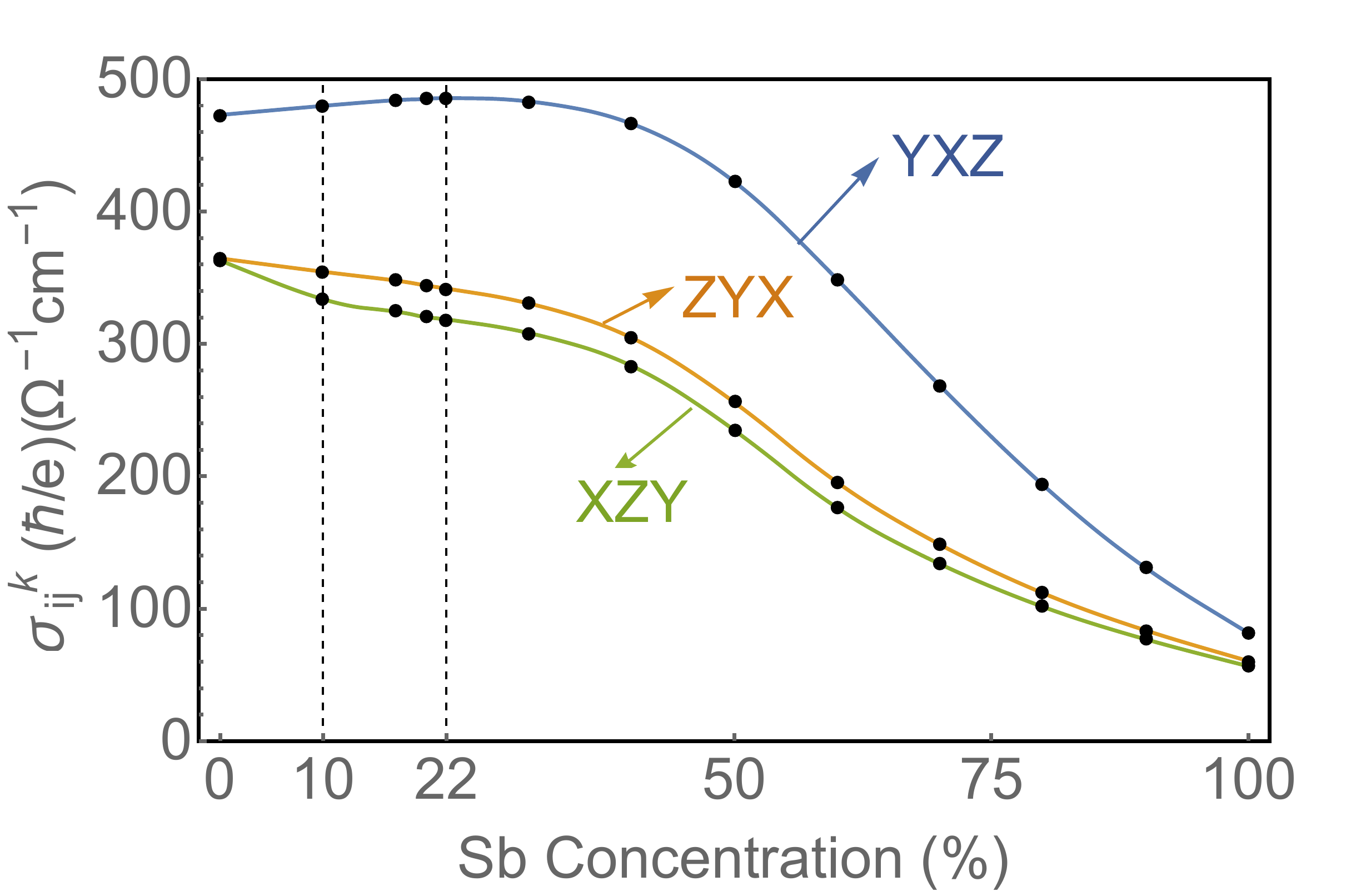}   
  \caption[Intrinsic spin Hall conductivity of Bi$_{1-x}$Sb$_{x}$ as a function of antimony concentration $x$]{Intrinsic spin Hall conductivity of Bi$_{1-x}$Sb$_{x}$ as a function of antimony concentration $x$. The largest spin Hall conductivity occurs near the semimetal-semiconductor transition at 22\% antimony. In all of the calculations for \BiSb alloys x, y and z-axes correspond to ($\bar 1$10), ($\bar 1 \bar 1 2$), and (111) in terms of the crystallographic axes of Ref.~\cite{Liu1995}. }
   \label{fig:sbdependence}
\end{figure}

Our results indicate very large intrinsic spin Hall conductivity in bismuth-based materials and topological insulators, such as Bi$_{0.9}$Sb$_{0.1}$ up to 474 ($\hbar$/e)($\Omega^{-1}$cm$^{-1}$) for bismuth and 96 ($\hbar$/e)($\Omega^{-1}$cm$^{-1}$) for antimony, larger than that of metals such as aluminium, comparable to platinum and stronger than conventional semiconductors. Our calculation in \figref{fig:sbdependence} is for $\sigma_{yx}^z$, where x, y, and z refers to  ($\bar 1$10), ($\bar 1 \bar 1 2$), and (111) in the crystallographic axes which are bisectrix, binary and trigonal axes of a typical rhombohedral unit cell. 

SHC calculations are carried out in every possible direction for the $\sigma_{ij}^k$ tensor. $\sigma_{yx}^z$ has the largest value for every antimony concentration. $\sigma_{zy}^x$ and $\sigma_{xz}^y$ have the exactly same values for pure bismuth and pure antimony, however, $\sigma_{zy}^x$ has slightly larger spin Hall conductivity for the case of an alloy. Bismuth shows a larger anisotropy compared to antimony between the three directions as shown in \figref{fig:sbdependence}.We observe that the intrinsic SHC, $\sigma_{yx}^z$, initially rises slightly as bismuth is alloyed with antimony from Fig.~\ref{fig:sbdependence}. However, it drops monotonically following the effective spin-orbit interaction in the system.  The initial increase of SHC can be explained by investigating the behavior of the Berry curvature in the system. We observe that Berry curvature has certain hot spots at certain symmetry points of the Brillouin zone when plotted along the band structure. These points are L and T points for bismuth which also correspond to the conduction and valence band edges respectively. The Brillouin zone region and k-points around the conduction band edge at L have enormous and negative curvatures while they are positive for the valence band edge at L. The same is also true for the symmetry point T. As antimony is introduced to pure bismuth and the curvature stays robust, the Fermi level begins to drop causing negative curvature contributions from L point to diminish. For a small concentration of antimony this effect dominates, however at larger concentrations the band structure completely changes and the Berry curvature decreases with increasing antimony concentration. Further details of this investigation can be found in the next section.

\subsection{Density of States and Density Berry Curvatures}
\begin{figure}[h]
  \begin{minipage}[b]{0.5\linewidth}
    \includegraphics[width=0.99\linewidth]{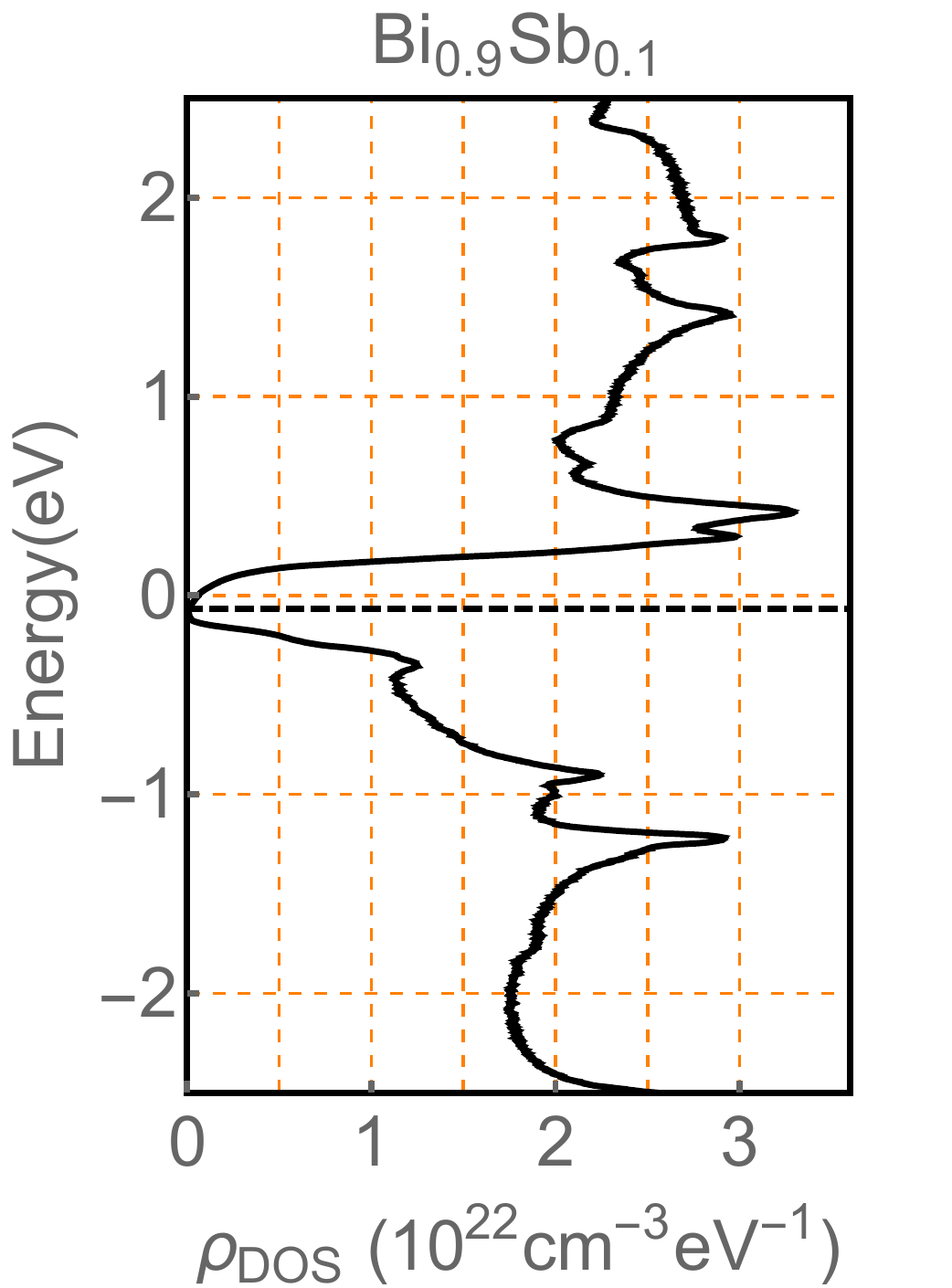} 
  \end{minipage} 
  \begin{minipage}[b]{0.5\linewidth}
    \includegraphics[width=0.99\linewidth]{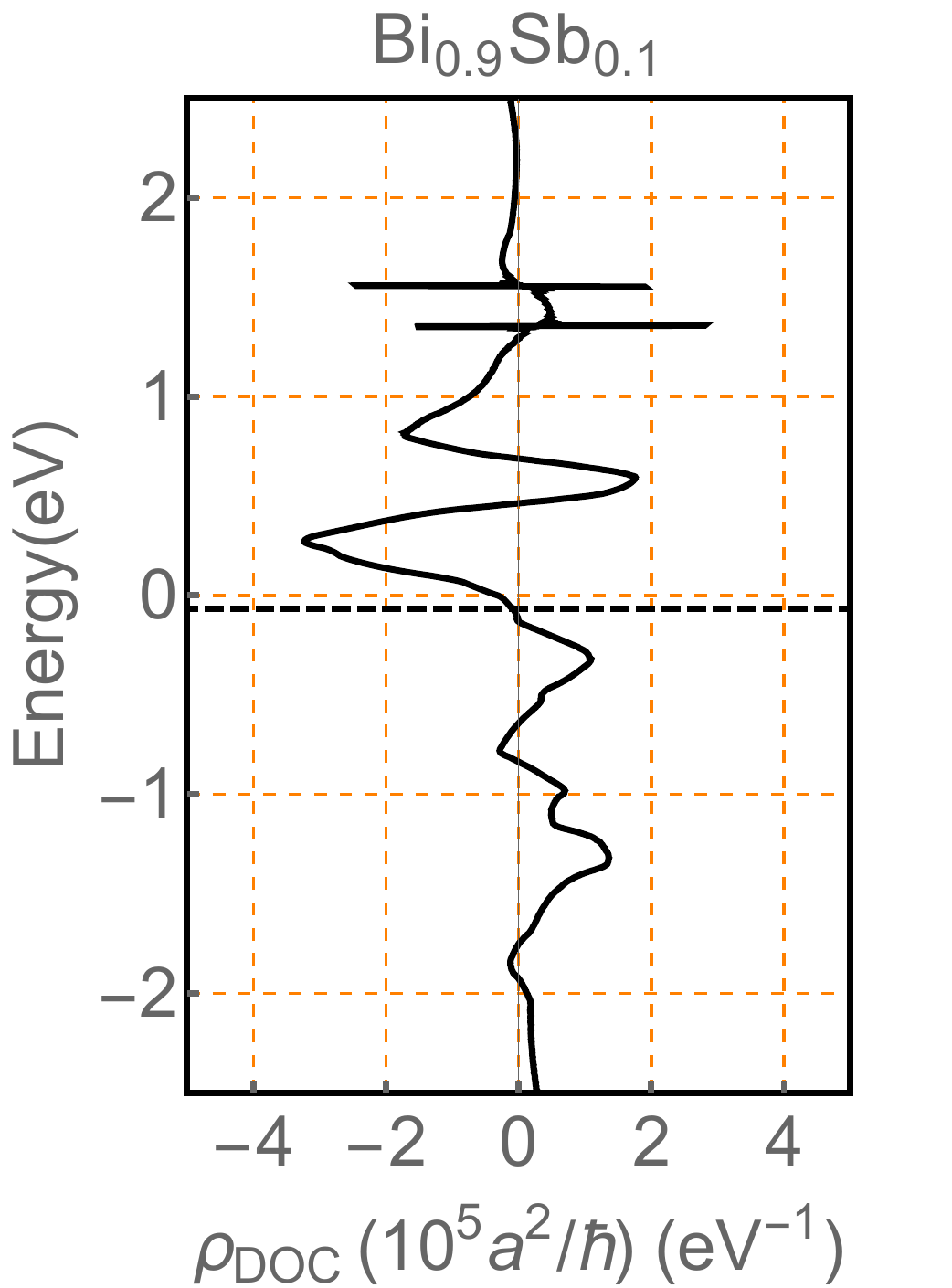} 
  \end{minipage} 
  \caption[Density of states and density of curvatures of Bi$_{0.9}$Sb$_{0.1}$]{Density of states and density of curvatures of Bi$_{0.9}$Sb$_{0.1}$. a) Density of states normalized per unit volume and unit energy and b) density of curvatures in the units of $\frac{a^2}{\hbar}$ $\frac{1}{eV}$ for Bi$_{0.9}$Sb$_{0.1}$ around the Fermi energy, where $a$ is the lattice constant. In these calculations, the Fermi level is located at -0.073 meV and indicated by a black dashed line.}
   \label{fig:densities}
\end{figure}

We have computed the density of states for all concentrations of antimony. There exists a band gap in the valence band for each of them between  $\Gamma$ and X points, where X always has higher energy. For bismuth this gap is approximately 2.433 eV between -8.087 eV ($\Gamma$) and -5.654 eV (X), while for Bi$_{0.9}$Sb$_{0.1}$ it is 2.238 eV and for antimony 0.567 eV. We have plotted the density of states around the Fermi level for Bi$_{0.9}$Sb$_{0.1}$. (topological insulator) in Fig.~\ref{fig:densities} part a). In the same figure part b) we have plotted the \textit{density of curvatures} (DOC). We define the DOC as the amount of Berry curvature, Eq.~\ref{berrycurvature}, per unit energy, such that the spin Hall conductivity becomes:
\al{\label{eq:doc}
\sigma_{yx}^z=\frac{e\hbar}{V} \int_{-\infty}^\mu d\epsilon \rho_{doc}(\epsilon)f(\epsilon),
}
where $\mu$ is the chemical potential. The density of curvatures in Fig.~\ref{fig:densities} is plotted between -2.5 and 2.5 eV from where most of the contributions come. Energies below and above these limits do not show particularly interesting behavior At lower energies the curvatures cancel each other and we don't observe much contributions up to around -5 eV, which is the first energy level after the band gap in the valence band. For the purpose of this work states below -5 eV can be omitted. Since spin Hall conductivity can be expressed as an integral of a density, some features of the SHC can clearly be seen by comparing the energy dependence of the Berry curvature originating from the electronic structure. For instance, one can compute how spin Hall conductivity depends on chemical potential by changing the limits of the integral in Eq.~\ref{eq:doc}. From knowledge of the DOC, one could choose how to change Fermi level, either by doping or gate voltage, in order to obtain largest SHC, since positive curvature increases SHC and negative curvature decreases.

The change in sign in $\rho_{\rm DOC}$ near the Fermi energy is an additional remarkable feature that originates from the nature of the topological insulator state. The formation of a topological insulator state corresponds to the opening of a gap between strongly spin-orbit correlated states. The composition of the states at the conduction edge and the valence edge are very similar, but with opposite sign matrix elements in Eq.~(\ref{berrycurvature}). As the Fermi energy is brought closer to the conduction edge or the valence edge, that contribution begins to dominate due to the energy denominator in Eq.~(\ref{berrycurvature}). Thus, this behavior of $\rho_{\rm DOC}$, changing sign across the Fermi energy, appears to be a generic feature of topological insulators.

\begin{figure}[h]
     \begin{minipage}[b]{0.49\linewidth}
    \includegraphics[width=1\linewidth]{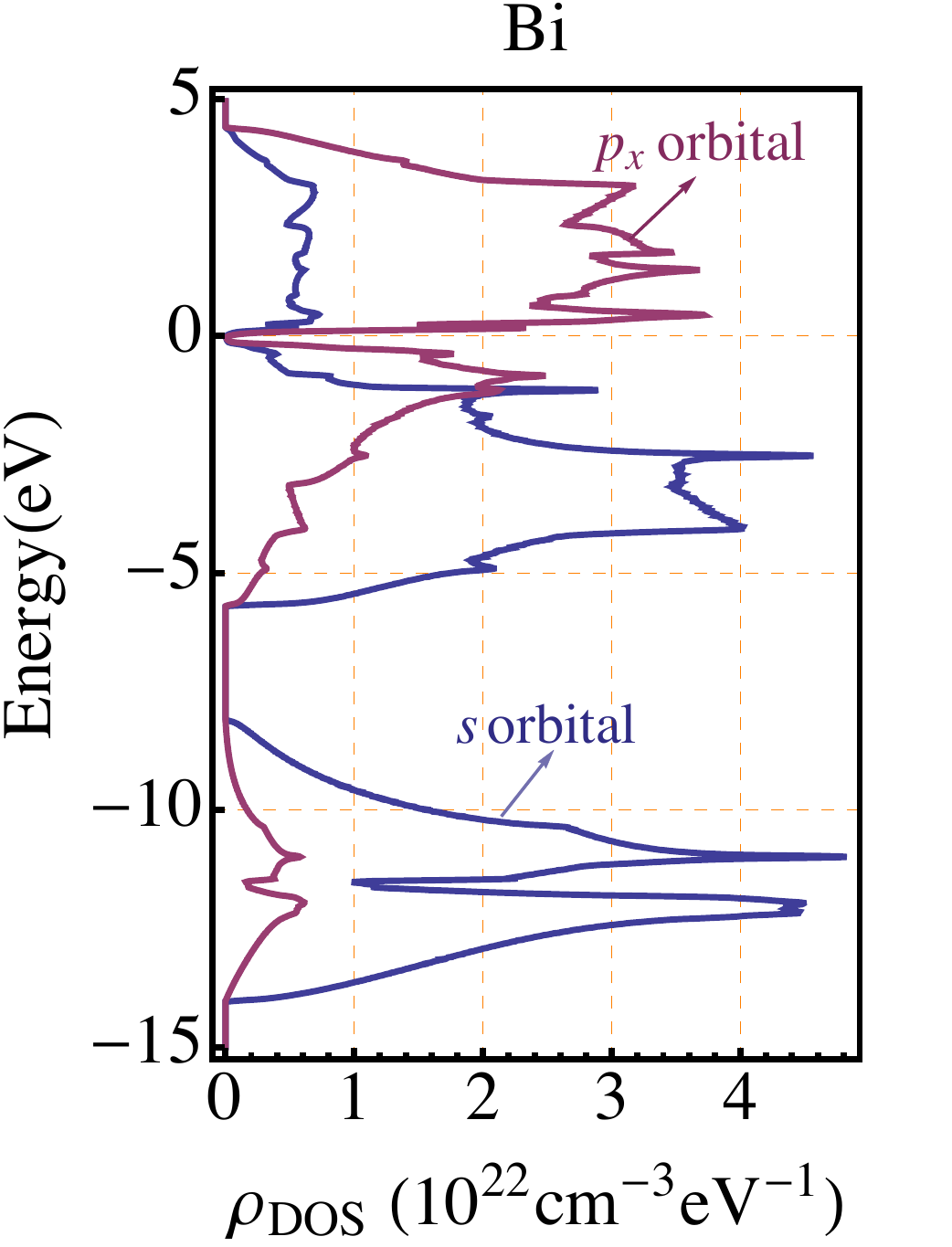} 
  \end{minipage} 
  \begin{minipage}[b]{0.49\linewidth}
    \includegraphics[width=1\linewidth]{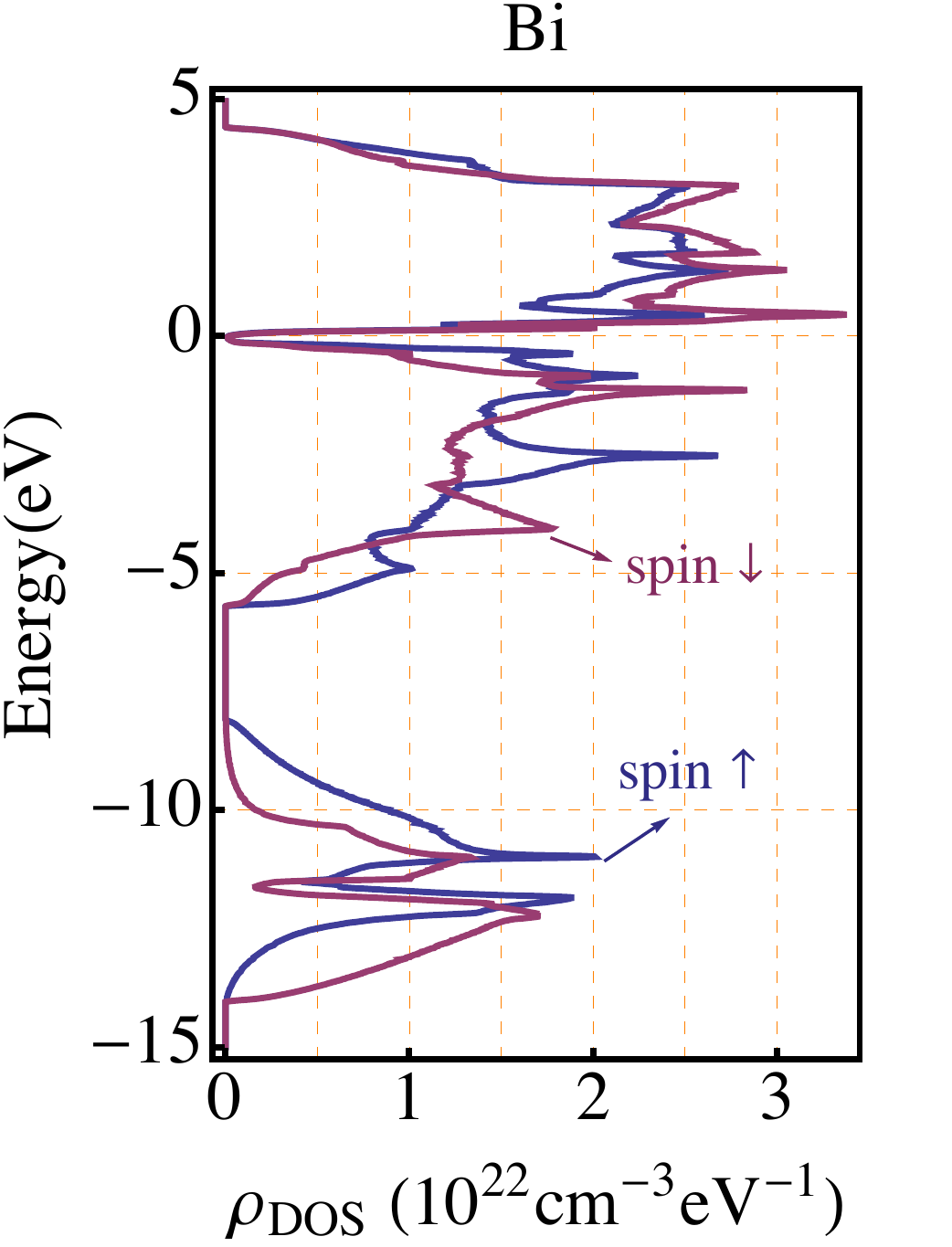} 
  \end{minipage} 
  \caption[ Orbital resolved density of states for bismuth]{Orbital resolved density of states for bismuth. a) Blue curve indicates s-orbital DOS while the red curve is for p$_x$ orbitals. Other p orbitals, p$_y$, and p$_z$, have a similar behavior as p$_x$, and therefore, are not shown in the figure. b) Spin resolved DOS for bismuth. Here we plot spin-up density with (blue curve) and spin-down density (red curve) including all orbitals.}
   \label{fig:spinorbitalDOS}
\end{figure}

We have also calculated spin and orbital resolved density of states. Orbital resolved DOS clearly indicates that states around Fermi level and above consist of dominantly p-type orbitals while states at lower energies are formed from s-like orbitals.(Fig.~\ref{fig:spinorbitalDOS} part a) We observe the emergence of p-type orbitals above the band gap in the valence band while s orbitals decrease dramatically after an energy of -2.5 eV. On the other hand, the spin-resolved DOS (Fig.~\ref{fig:spinorbitalDOS} part b) doesn't exhibit a similar behavior. The spin-up and spin-down DOS are not separated by energy. 

\subsection{Tuning of Spin Hall Conductivity and Temperature Dependence}
\begin{figure}[h]
    \includegraphics[width=1\linewidth]{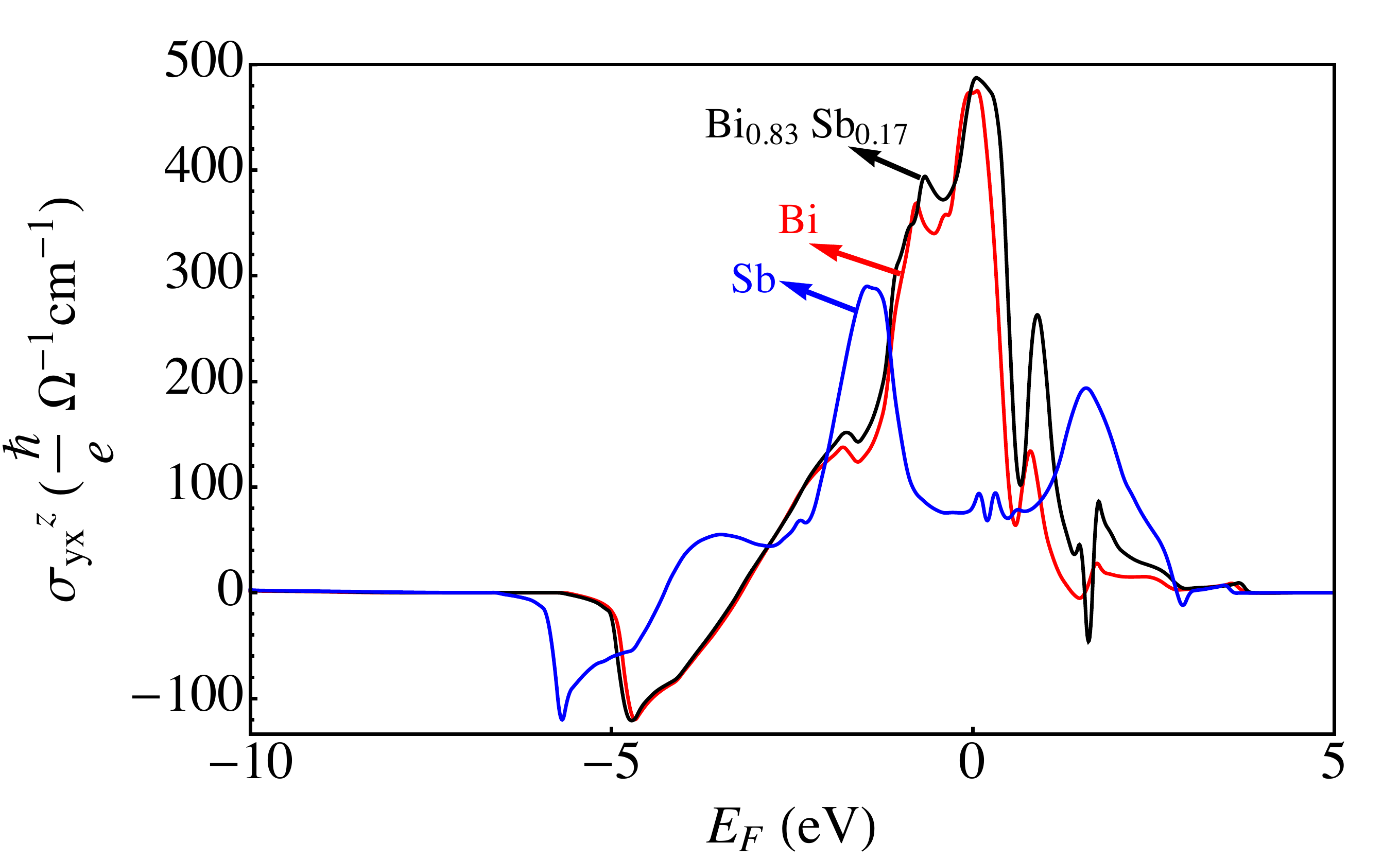} 
  \caption[Intrinsic spin Hall conductivity as a function of Fermi level]{Intrinsic spin Hall conductivity as a function of Fermi level. Intrinsic spin Hall conductivity for bismuth, antimony and the Bi$_{0.83}$Sb$_{0.17}$ alloy which has the largest band gap. The Fermi levels are at 0 eV for Bi and Sb while it is at -0.107 eV for Bi$_{0.83}$Sb$_{0.17}$. }
   \label{fig:fermidependence}
\end{figure}

We now consider the effects on the spin Hall conductivity that would come from varying the carrier concentration and Fermi energy by doping. As expected from Fig.~\ref{fig:densities}, we find a sensitive dependence of the spin Hall conductivity on the Fermi energy for both  bismuth, antimony, and Bi$_{0.83}$Sb$_{0.17}$, which has the largest semiconducting band gap, shown in  Fig.~\ref{fig:fermidependence}. For each material, there exists an optimum range for the Fermi energies so that the intrinsic SHC is a maximum. This range approximately corresponds from -20 meV to +40 meV for bismuth. In the case of antimony, we observe that there exist several Fermi energies that exceed the intrinsic SHC of antimony at 0 eV. For example, a Fermi energy of -1.5 eV produces a spin Hall conductivity four times that of undoped antimony, and more than half that of bismuth (288  $\hbar$/e)($\Omega^{-1}$cm$^{-1}$). By comparison the topological insulator material Bi$_{0.83}$Sb$_{0.17}$ does not possess a larger spin Hall conductivity than bismuth, and in fact its spin Hall conductivity as a function of Fermi energy is very similar to that of bismuth. We thus note that the dominant contribution to the spin Hall conductivity comes from the large spin-orbit interaction in the materials, rather than the topological character of the band structures. There also exists a region where the conductivity drops to 0 from about -10 to -5 eV of Fermi levels. This gap in the SHC is associated with the fact that the Berry curvatures at these energies are negligibly small creating a curvature gap in the system similar to a band gap in the density of the states plots. The presence of this band gap deep within the valence structure of bismuth or antimony is a property of the electronic structure model Hamiltonian for these systems (Ref.~\cite{Liu1995}). By increasing the Fermi level over this curvature gap we observe a negative spin Hall conductivity around  -5 eV.

\begin{figure}[h]
     \begin{minipage}[b]{0.49\linewidth}
    \includegraphics[width=1\linewidth]{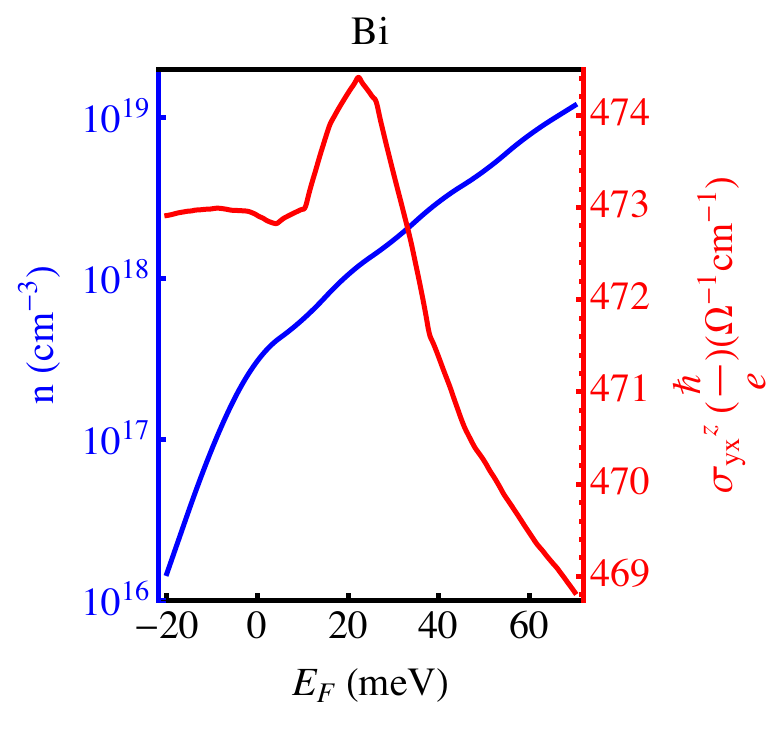} 
  \end{minipage} 
  \begin{minipage}[b]{0.49\linewidth}
    \includegraphics[width=1\linewidth]{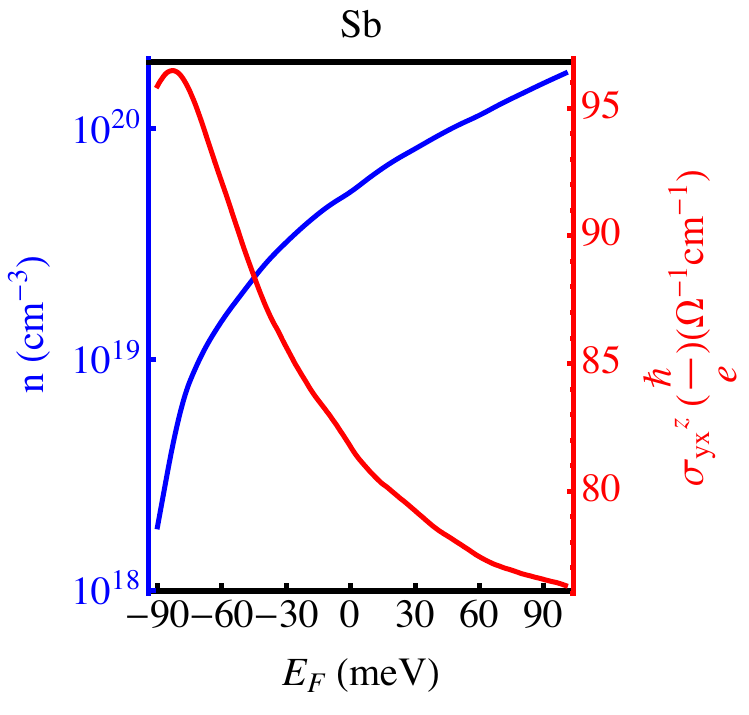} 
  \end{minipage} 
  \caption[Carrier densities and corresponding intrinsic spin Hall conductivities as a function of Fermi level for bismuth and antimony]{Carrier densities and corresponding intrinsic spin Hall conductivities as a function of Fermi level for bismuth and antimony. 100 meV change in the Fermi level results in several orders of magnitude change in the carrier density. However, SHC is not affected much by the position of the Fermi energy indicating that there exists no strong dependence on the temperature as well.}
   \label{fig:carrierdensity}
\end{figure}

From the same band structure we computed the change in the density of the charge carriers as the Fermi level is shifted from 0 eV, which is shown in Fig.~\ref{fig:carrierdensity}. Carrier densities for bismuth and antimony can be altered by a gate voltage. Initially, electron and hole carrier densities are equal to each other when the Fermi level is at 0 eV. The carrier density of semimetal bismuth is 3.1x10$^{17}$cm$^{-3}$ while it is 5.3x10$^{19}$cm$^{-3}$ for antimony; lower than that of typical metals. As the Fermi level is shifted to higher energies by a gate voltage, the materials exhibit more metallic behavior. Electron densities increase up to 10$^{19}$cm$^{-3}$ and 10$^{20}$cm$^{-3}$ for bismuth and antimony respectively with increasing Fermi level. However, several orders of magnitude change in the carrier densities is not reflected in the spin Hall conductivities. Bismuth changes its SHC by about 1\%, and antimony changes its by about 20\% within the range shown in Fig.~\ref{fig:carrierdensity}. 

\begin{figure*}[ht!]
\begin{tabular}{ccc}
\includegraphics[width=0.3\columnwidth]{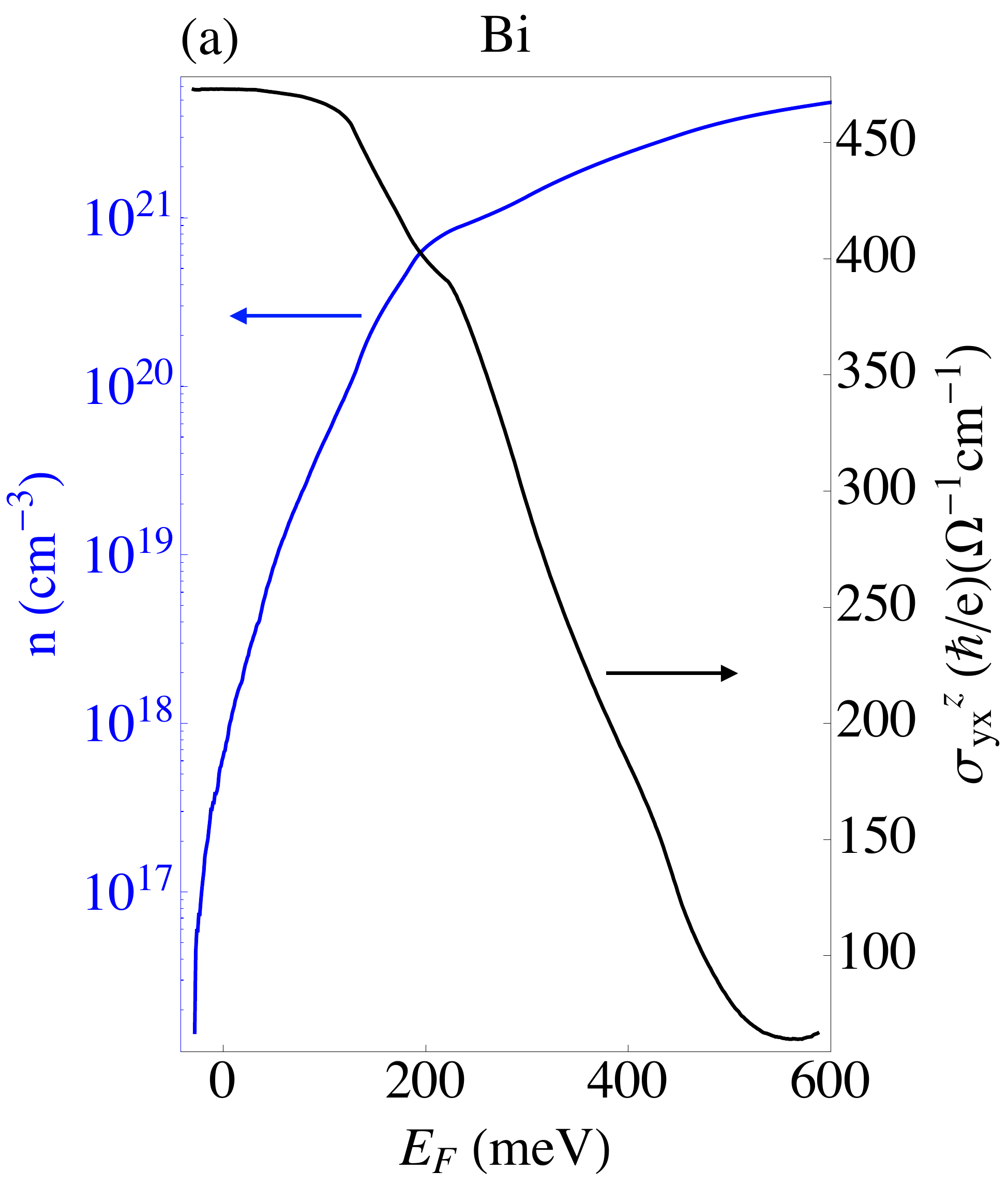}&
\includegraphics[width=0.3\columnwidth]{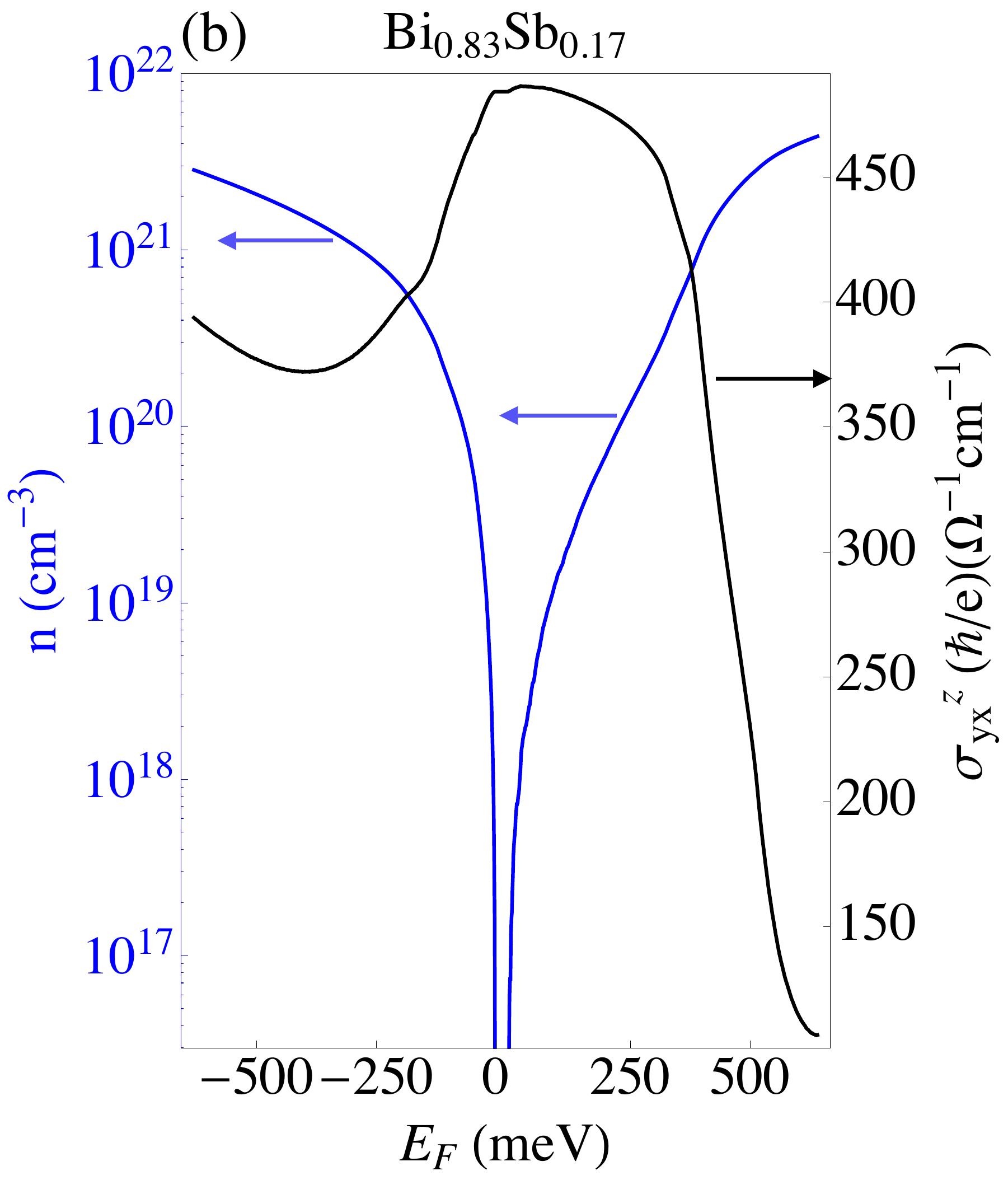}&
\includegraphics[width=0.3\columnwidth]{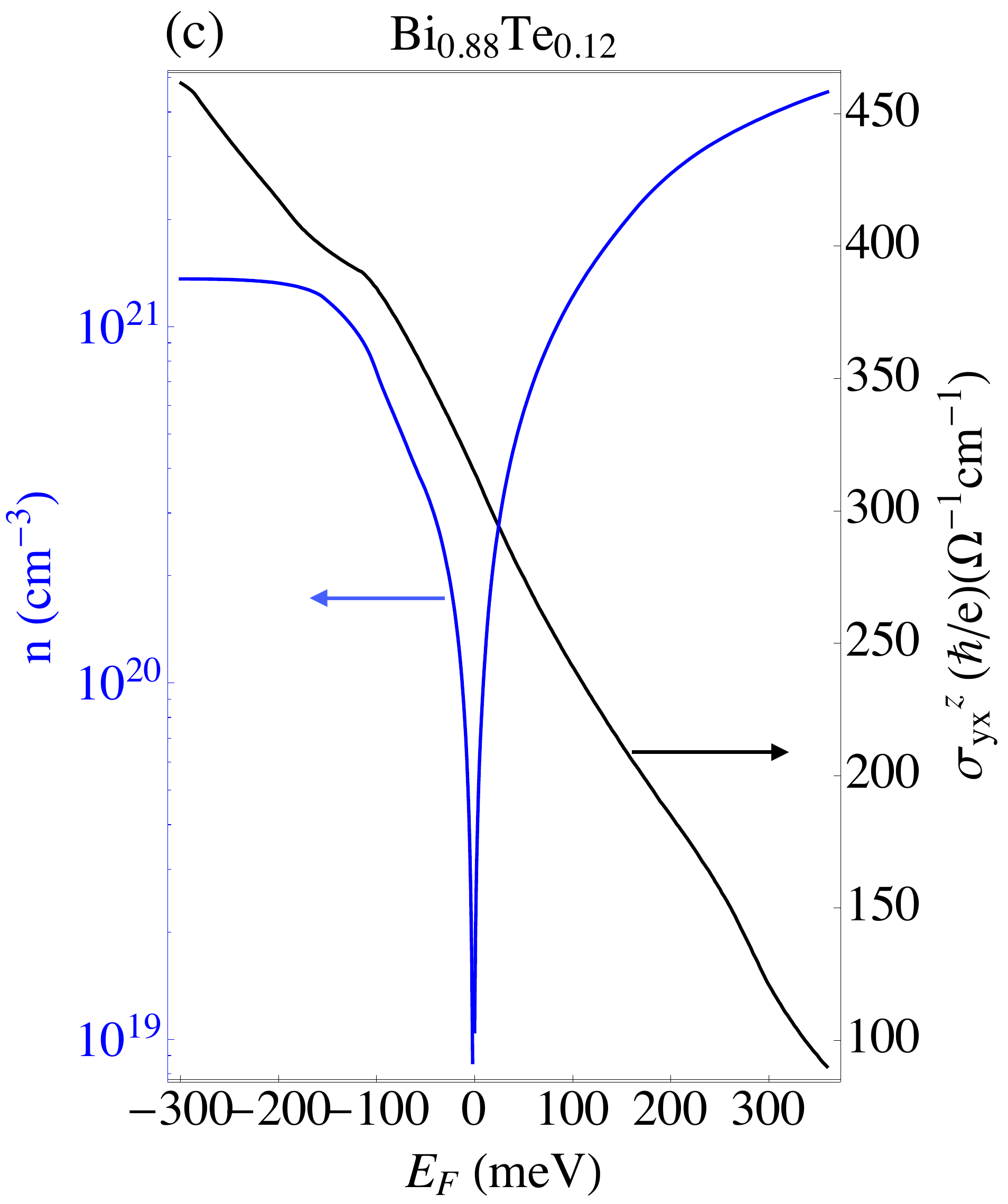}
\end{tabular}
  \caption[Gate-induced carrier densities and corresponding intrinsic spin Hall conductivities as a function of Fermi level]{Gate-induced carrier densities and corresponding intrinsic spin Hall conductivities as a function of Fermi level. (a) bismuth, (b) Bi$_{0.83}$Sb$_{0.17}$ and (c) Bi$_{0.88}$Te$_{0.12}$. A range of spin Hall conductivities varying by a factor of five is achievable by doping, either via a gate or through the introduction of dopants such as Te.}
    \label{fig:carrierdensity2}
\end{figure*}

Now we focus on bismuth, Bi$_{0.83}$Sb$_{0.17}$, and Bi$_{\rm 0.88}$Te$_{\rm 0.12}$ with a larger change in the Fermi energy. We expect that the change in the Fermi level would be achieved through accumulation or depletion via an electrical gate in a field-effect transistor device. The change in carrier density is plotted as a function of the change in carrier density (electron or hole).  For Bi$_{\rm 0.83}$Sb$_{\rm 0.17}$ the equilibrium bulk carrier concentration vanishes at low temperature. Changes in the carrier concentration modify the spin Hall conductivity by approximately a factor of five, suggesting that gate-tuning the spin Hall conductivity of such materials is possible. For bismuth, there is little change in the spin Hall conductivity for an initial change of the Fermi energy by 150 meV. Instead of gate-tuning to this point, it should be possible to dope the material with a group-VI dopant such as Te. For a Te concentration of 12\% the spin Hall conductivity lies in between the upper and lower extremes, producing the largest tuning range with voltage. Thus, we present in Fig.~\ref{fig:carrierdensity2}(c) the carrier-dependence of the spin Hall conductivity for Bi$_{\rm 0.88}$Te$_{\rm 0.12}$. We note that this doping consists of adding Te to the crystal structure of Bi, not shifting to the crystal structure of Bi$_2$Te$_3$.  As the longitudinal conductivity of these materials will change as well with a change in the Fermi energy we expect that the spin Hall angle, defined as the ratio of the spin Hall conductivity to the longitudinal conductivity, could be substantially varied as well.

\begin{figure}{\label{fig:bitempdep}}
  \centering
    \includegraphics[width=1\textwidth]{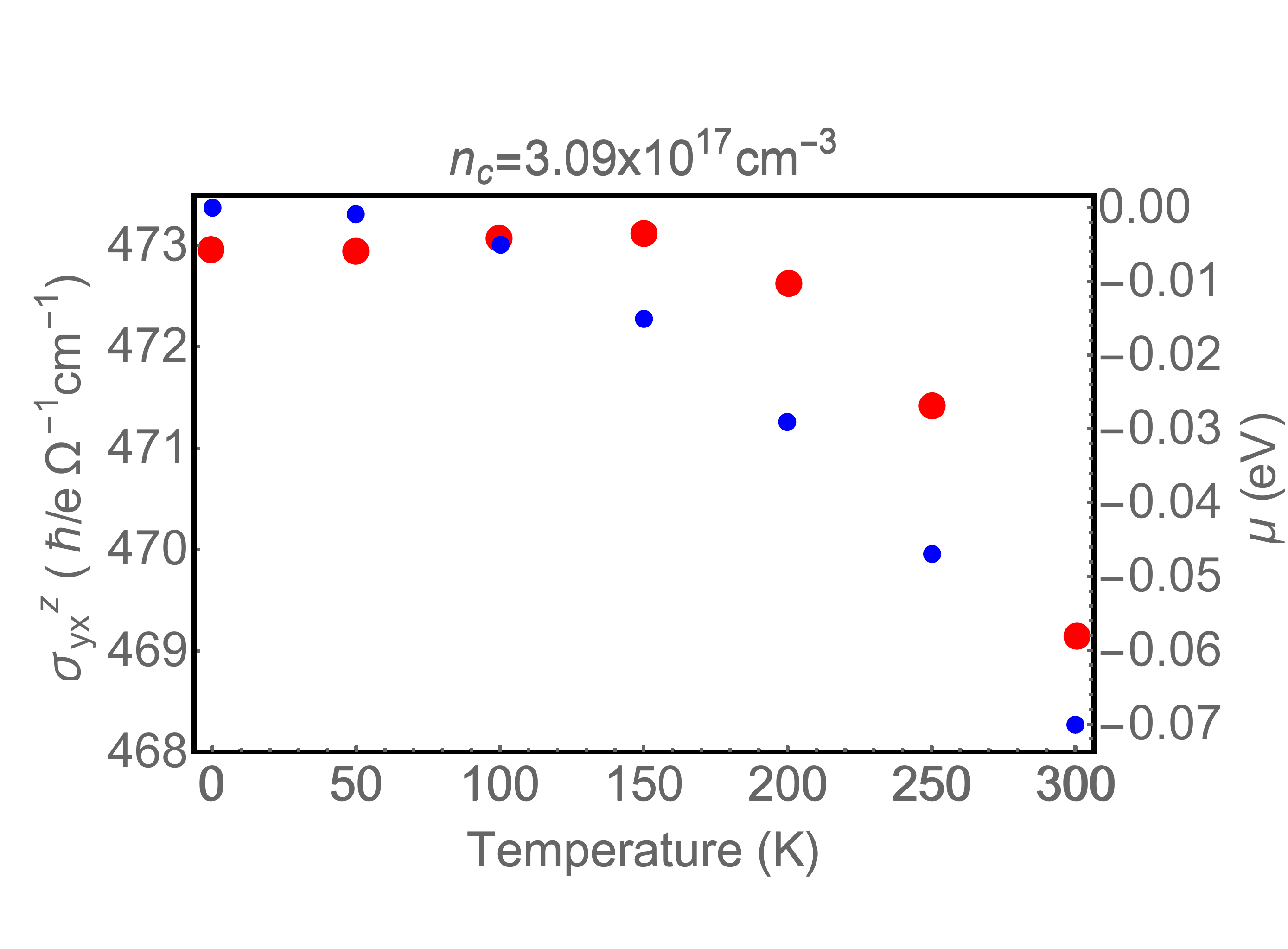}
   \caption[Temperature dependence of the spin Hall conductivity for bismuth]{Temperature dependence of the spin Hall conductivity for bismuth. Calculation of the dependence of the SHC on the temperature shows only a 1\% change in the SHC at room temperature compared to low temperatures. Blue dots represent the location of the chemical potential while red dots represent calculated SHC as a function of temperature.}
  \end{figure}

We have also investigated the temperature dependence of SHC for bismuth as shown in \figref{fig:bitempdep}. We don't observe more than a 1\% change in the SHC as the temperature is increased from 0K to room temperature. In this calculation, the carrier density of the bismuth is taken as constant at 3.09$\times$10$^{17}$cm$^{-3}$ and chemical potentials for each temperature are calculated. Blue dots in \figref{fig:bitempdep} indicates the location of the chemical potential, which moves down to -70 meV with respect to the top of the valence band. The top of the valence band consists of states with relatively large and positive Berry curvatures. Therefore, lowering the chemical potential leads to a slight decrease in the SHC as states with positive curvatures are not included in the SHC calculations. 
  
\subsection{An Alternative Virtual Crystal Approximation} \label{sec:alternativeVCA}
\begin{figure}
  \begin{minipage}[c]{0.55\textwidth}
    \includegraphics[width=0.99\textwidth]{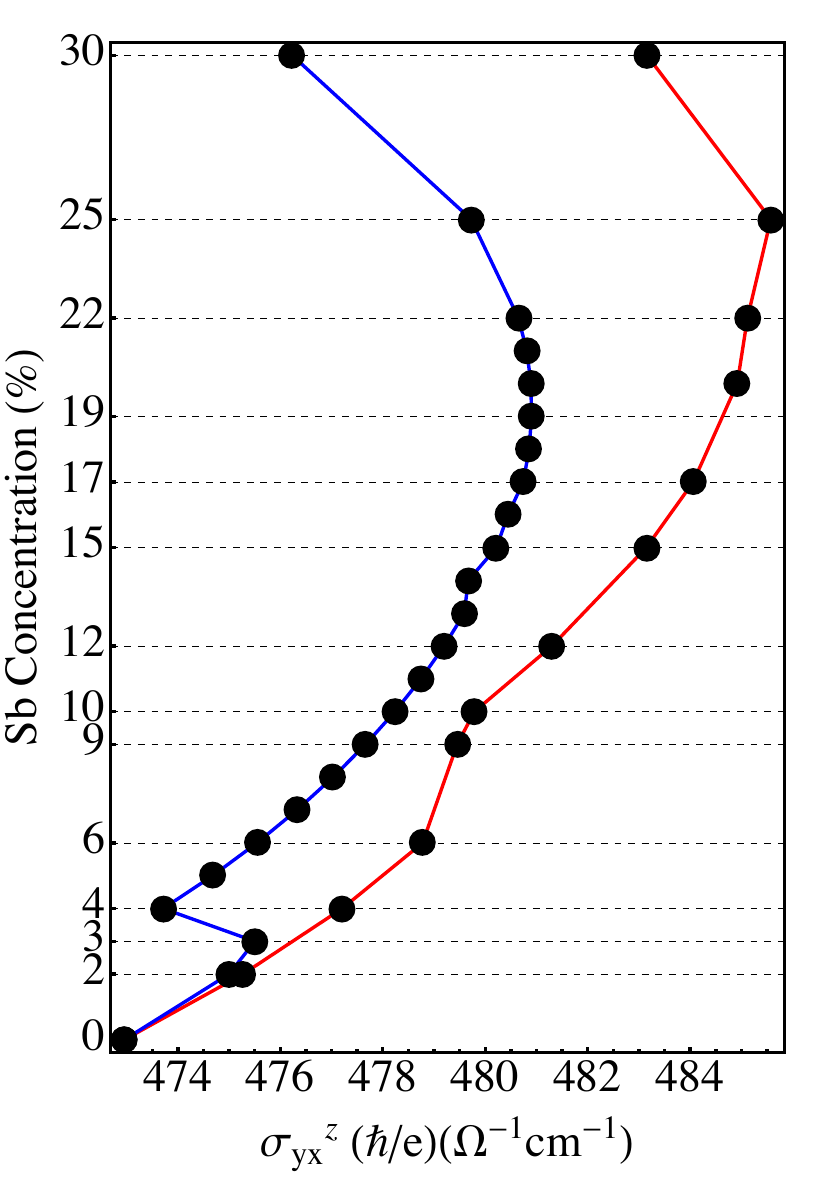}
  \end{minipage}\hfill
  \begin{minipage}[c]{0.45\textwidth}
 \caption[Intrinsic spin Hall conductivity calculated from an alternative virtual crystal approximation]{{\label{fig:fukane}}Intrinsic spin Hall conductivity calculated from an alternative virtual crystal approximation. ISHC of \BiSb as a function of antimony concentration for the band structure using the tight-binding parameters of Ref.~ \cite{Liu1995} and the virtual crystal approximation proposed by Ref.  \cite{Teo2008}. The red line shows the SHC results of a linearly interpolated VCA while the blue curve is for the alternative VCA as introduced in Eq.~\ref{eq:newVCA}. The abrupt change at 3\% for the alternate VCA is due to the inversion of conduction and valence bands states at the L point of the Brillouin zone, which have Berry curvatures with opposite signs. Both parametrizations behave similarly to each other except at this point \cite{Sahin2015}.}
  \end{minipage}
  \end{figure}

For bismuth and antimony alloys, a different version of the virtual crystal approach was proposed by Teo, Fu, and Kane \cite{Teo2008}. The primary goal of this parametrization is to fix the sign of the g-factor and symmetry of the bands at the L point. The L$_s$ and L$_a$ states should be inverted at around x=3\%. This could be achieved by interpolating the overlap integrals as:
\al{\label{eq:newVCA}
V_{Bi_{1-x}Sb_x}(sp\sigma)=(1-x^2)V_{\text{Bi}}(sp\sigma)+xV_{\text{Sb}}(sp\sigma),
}
while keeping the on-site energies and other parameters unchanged. This scheme is only valid for small antimony concentrations (x$\approx$0.1). However, it improves the sign of the g-factor extensively \cite{Teo2008}. We have also considered this VCA approximation by using the tight-binding parameters of Ref.~\cite{Liu1995} and haven't observed much difference in the spin Hall conductivity from the new parametrization. (\figref{fig:fukane}) The difference is less than 1\% and the newer band structure is only valid for antimony concentration x less than approximately 20\%. 

\section{Bismuth Chalcogenides: \BiSe and \BiTe}
\subsection{Spin Hall Conductivity of \BiSe and \BiTe}
\begin{figure}[t]
\centering
\begin{minipage}[t]{0.49\linewidth}
\includegraphics[width=1\textwidth]{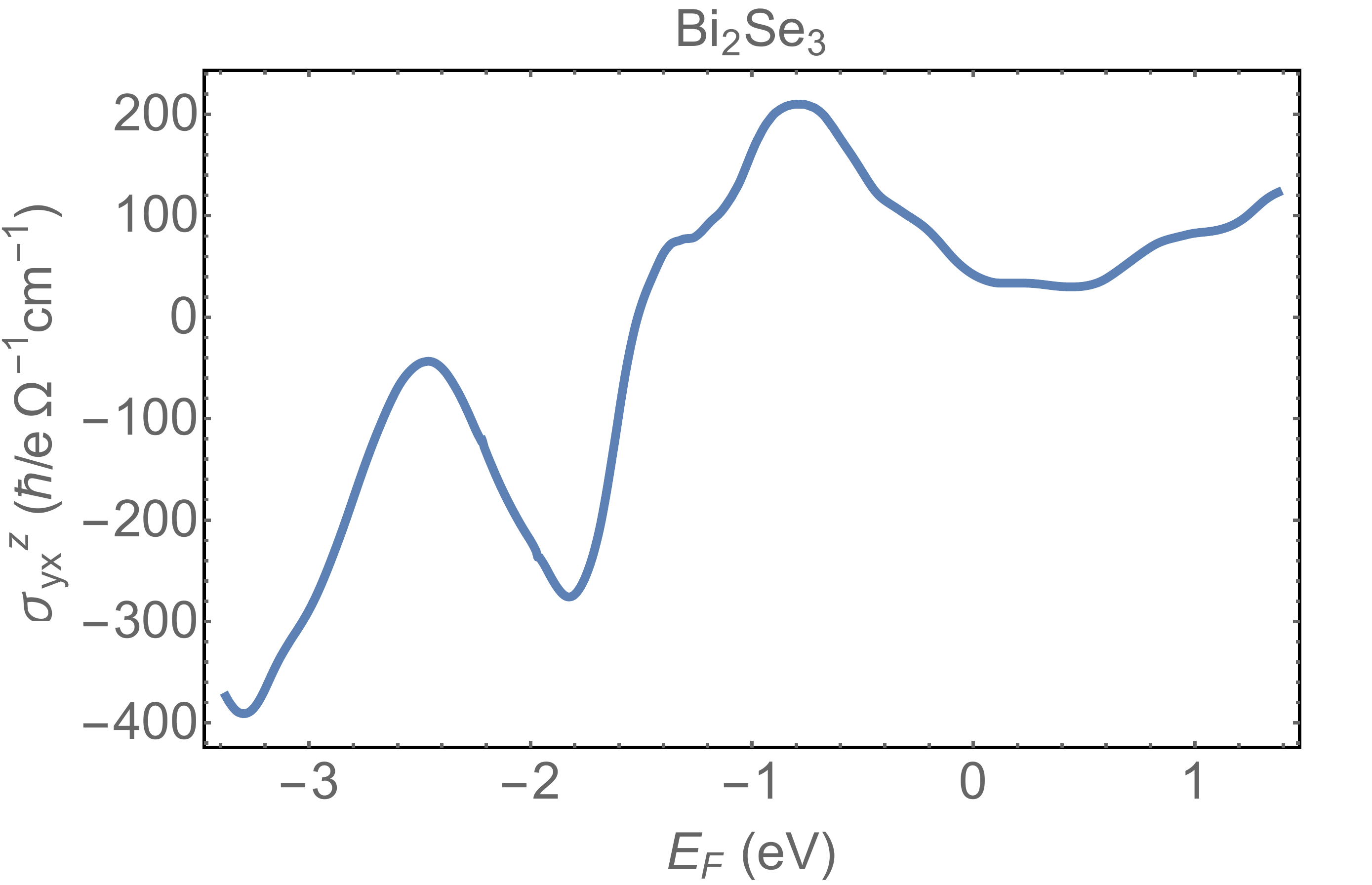}
  \end{minipage} 
  \begin{minipage}[t]{0.49\linewidth}
\includegraphics[width=1\textwidth]{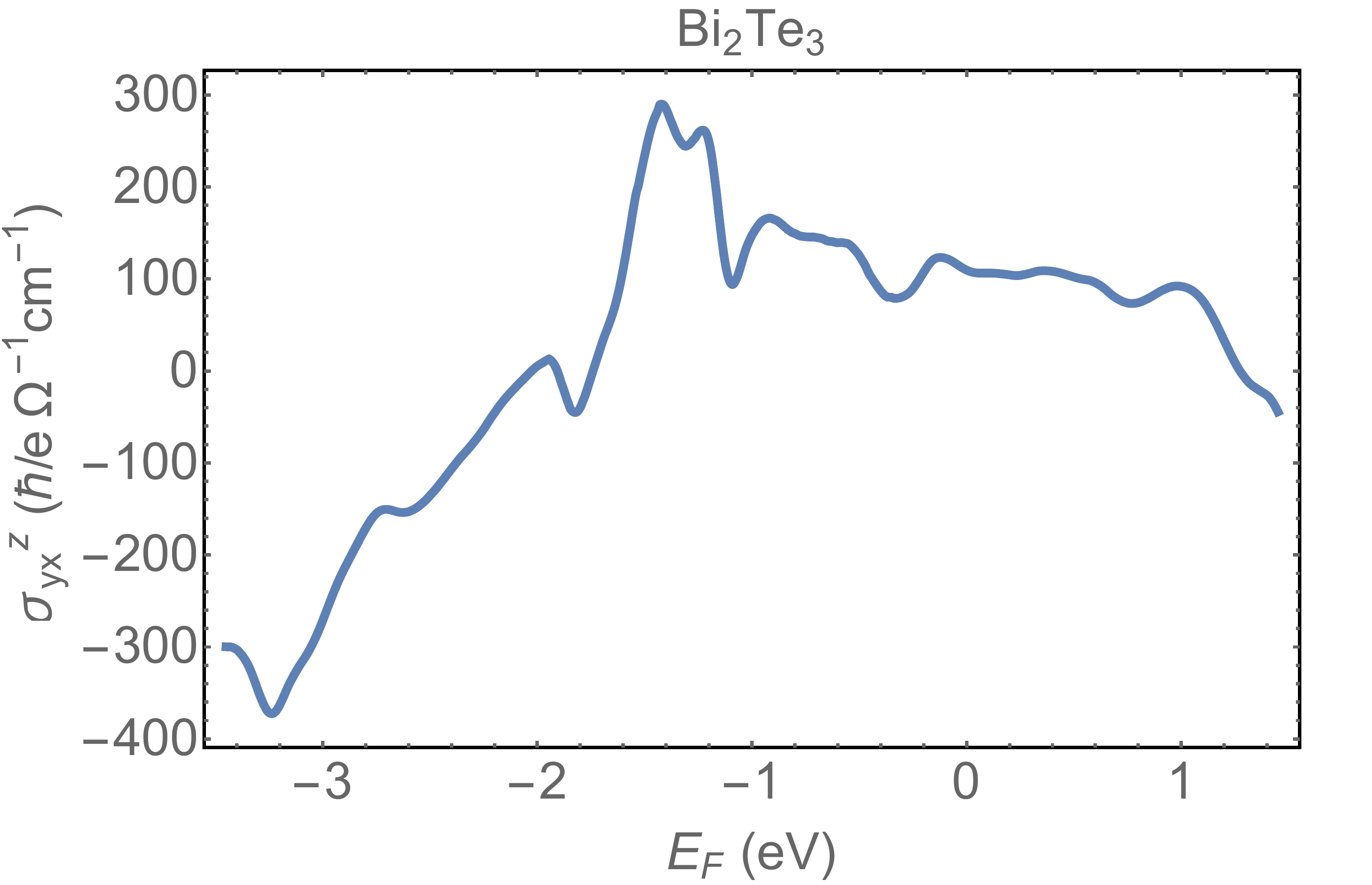}
\end{minipage}
\caption[Fermi energy dependence of spin Hall conductivities of \BiSe and \BiTe ]{Fermi energy dependence of spin Hall conductivities of \BiSe and \TeBi. Dirac points are located within the band gap at 0 eV. In all of the calculations for \BiSe and \BiTe the x, y and z-axes correspond to (10$\bar 1$), ($\bar 1 2 \bar 1 $), and (111) in terms of crystallographic axes of Ref.~\cite{Kobayashi2011a}}
\label{fig:bisete-shc}
\end{figure}

Bismuth based materials with small band gaps and strong spin-orbit interactions such as bismuth selenide, \BiSe \cite{Xia2009, Zhang2009} and bismuth telluride, \BiTe \cite{Chen2009} exhibit three-dimensional topological insulator states  with single Dirac cones, as well as novel phenomena such as the quantum spin Hall effect. Recently it has been shown by spin-transfer torque measurements that \BiSe exhibits a gigantic spin Hall conductivity \cite{Mellnik2014}. It is expected that \BiTe, a topological insulator with similar crystal structure as well as large spin-orbit coupling, would also exhibit a giant SHC. In this part of the study, we have utilized a tight-binding Hamiltonian and calculated the intrinsic spin Hall conductivity emerging from the bulk band structure of these materials. Both results show that bismuth chalcogenides \BiSe and \BiTe exhibit very large, tunable spin Hall conductivities as shown in \figref{fig:bisete-shc}. The dependence of the SHC on the chemical potential is also depicted in the same figure. At a Fermi energy of 0 eV, \BiSe exhibits an intrinsic SHC of 48 $(\hbar/e) \Omega^{-1}cm^{-1}$ which can be tuned up to 200 $(\hbar/e) \Omega^{-1}cm^{-1}$ by reducing the chemical potential. Similarly \BiTe exhibits a large SHC at 0 eV which is around 100 $(\hbar/e) \Omega^{-1}cm^{-1}$ and can be increased to 300 in the same units. However in the vicinity of a Fermi energy of 0 eV these results don't change much, thus indicating that there is no significant temperature dependence as in the case of platinum. \cite{Guo2008}
Therefore, these materials, as well as bismuth-antimony alloys, are promising candidates for transverse spin current generation and spintronic applications.  The Fermi level dependence of these conductivities suggests that both materials can be tuned by either doping or gate voltage such that their SHC can be changed substantially, but not as large as bismuth-antimony alloys. 

 \begin{figure}[h]\label{otherdirections}
\includegraphics[width=1\textwidth]{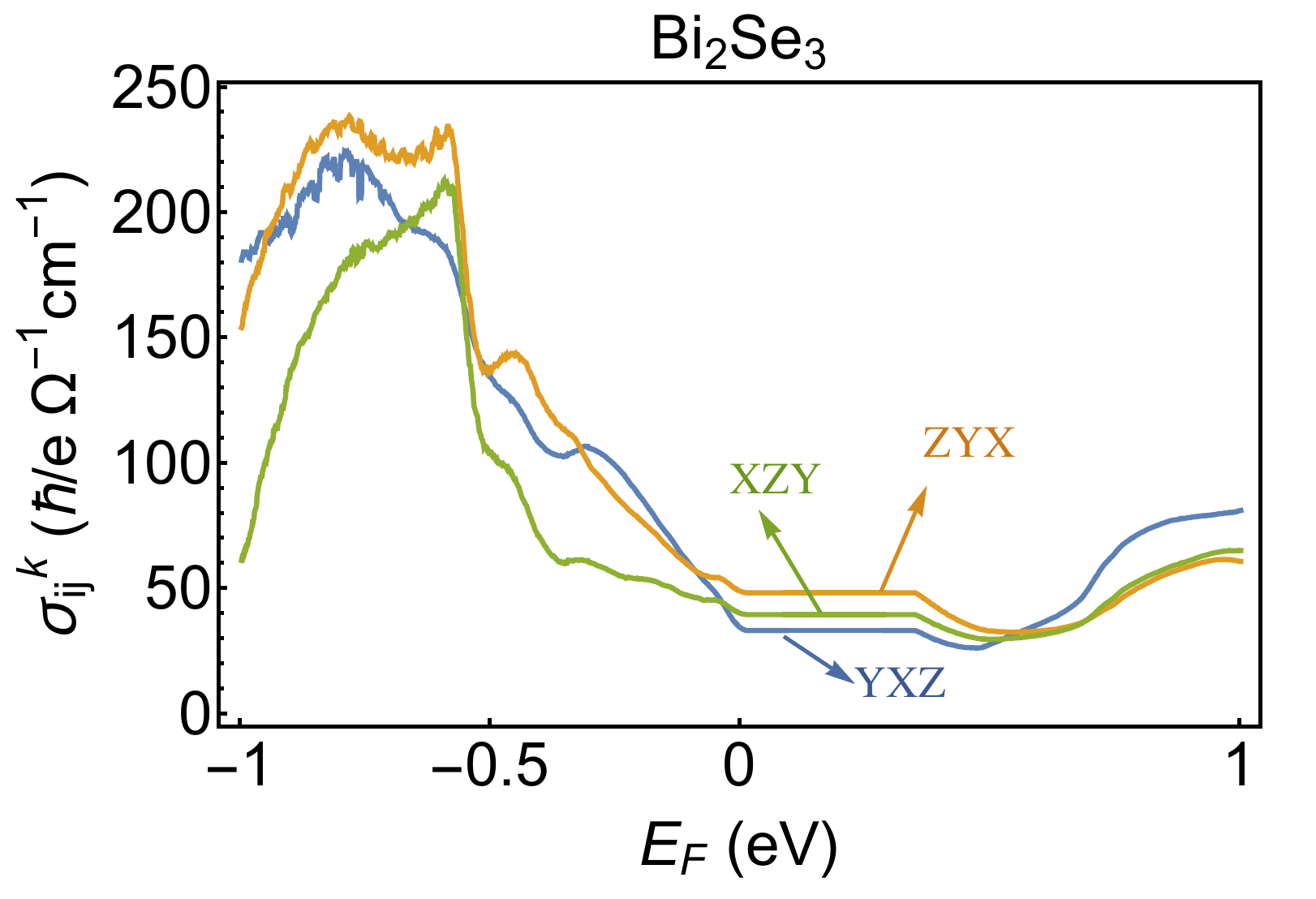}
\caption[Spin Hall conductivity calculated in other directions]{Spin Hall conductivity calculated in other directions. Here IJK stands for the $\sigma_{ij}^k$ element of the SHC tensor. Axes x, y and z  correspond to (10$\bar 1$), ($\bar 1 2 \bar 1 $), and (111) in terms of the crystallographic axes of Ref.~\cite{Kobayashi2011a}}
\end{figure}
 
Furthermore, we have not observed a large anisotropy in the SHC calculated in different directions for \BiSe as shown in \figref{otherdirections}. This is due to the fact that the crystal structure of bismuth-based materials is rhombohedral, meaning a cubic structure that is slightly distorted in the body diagonal. Figure~\ref{otherdirections} shows the Fermi level dependence of a general SHC tensor $\sigma_{ij}^k$. The index $i$ is the direction of the spin current and $j$ is the direction of electric field, while $k$ stands for the direction of the spin polarization. Recent experiments such as Ref.~\cite{Mellnik2014} chose to measure the spin signal in an orientation with charge current out of plane while spin current and magnetization are in the plane. The SHC of \BiSe with a Fermi energy at 0 eV is around 48-52 ($\hbar/e) \Omega^{-1}cm^{-1}$. On the other hand we observe a large anisotropy when the same calculation is done for \BiSe in three possible directions. Spin Hall conductivity with an out-of-plane spin polarization is almost twice larger than the two other directions that are in the plane. The values of the SHC at 0 eV Fermi level read as 108, 77, and 62 in the units of ($\hbar /e$)($\Omega^{-1} cm^{-1}$) for $\sigma_{yx}^z$, $\sigma_{zy}^x$, and $\sigma_{xz}^y$ respectively.

 \begin{figure}[h]\label{biteotherdirections}
\includegraphics[width=0.99\textwidth]{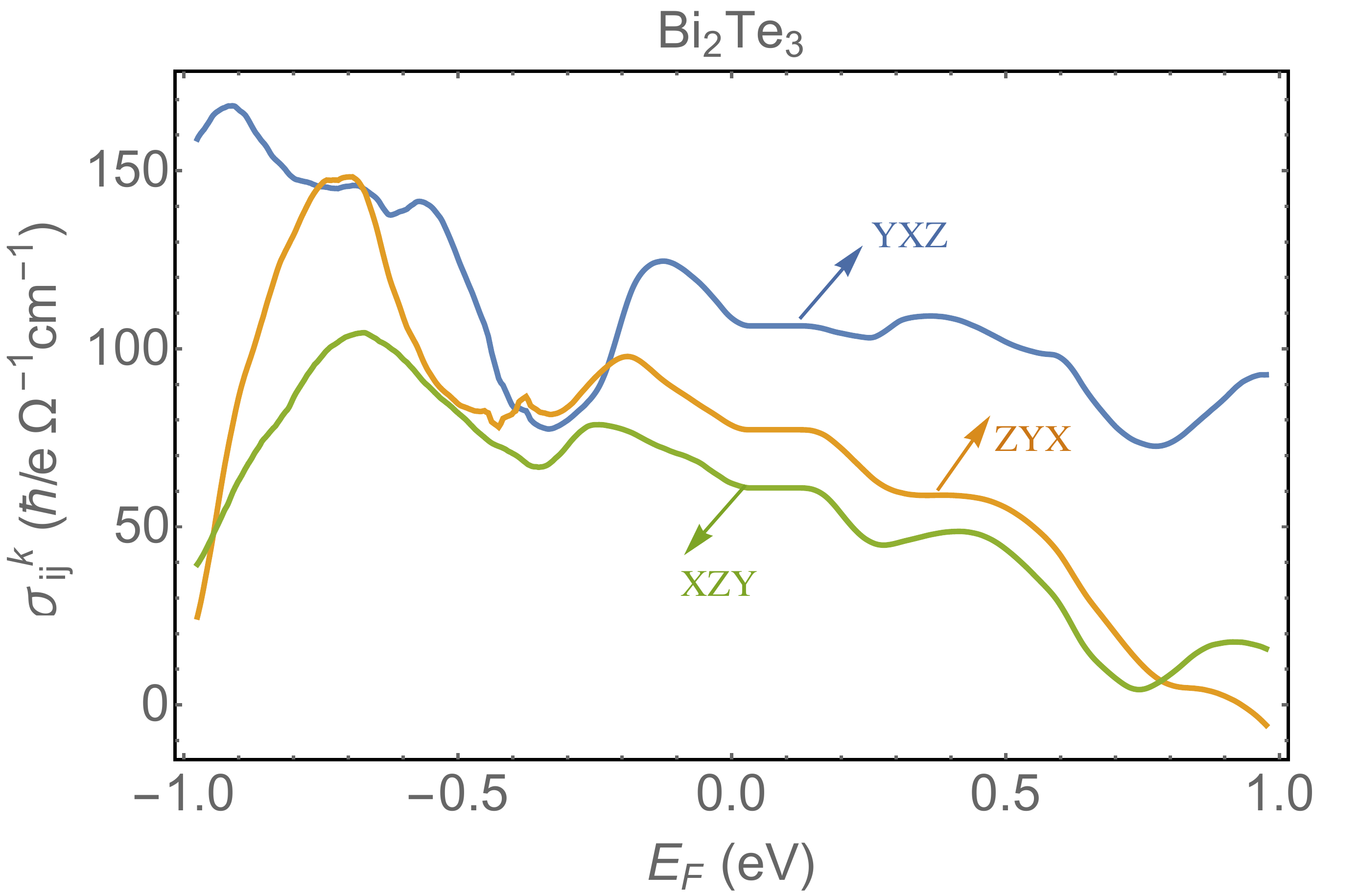}
\caption[Comparison of the spin Hall conductivity of \BiTe calculated in three directions]{Comparison of the spin Hall conductivity of \BiTe calculated in three directions. Spin Hall conductivity calculated in other directions shows a larger directionality for \BiSe crystals. Here IJK stands for the $\sigma_{ij}^k$ element of SHC tensor. The axes x, y and z  correspond to (10$\bar 1$), ($\bar 1 2 \bar 1 $), and (111) in terms of the crystallographic axes of Ref.~\cite{Kobayashi2011a}}
\end{figure}

\subsection{Band Resolved Spin Hall Conductivity Contributions}
\begin{figure}[h]\label{fig:bisebandberry}
\centering
\includegraphics[width=0.99\textwidth]{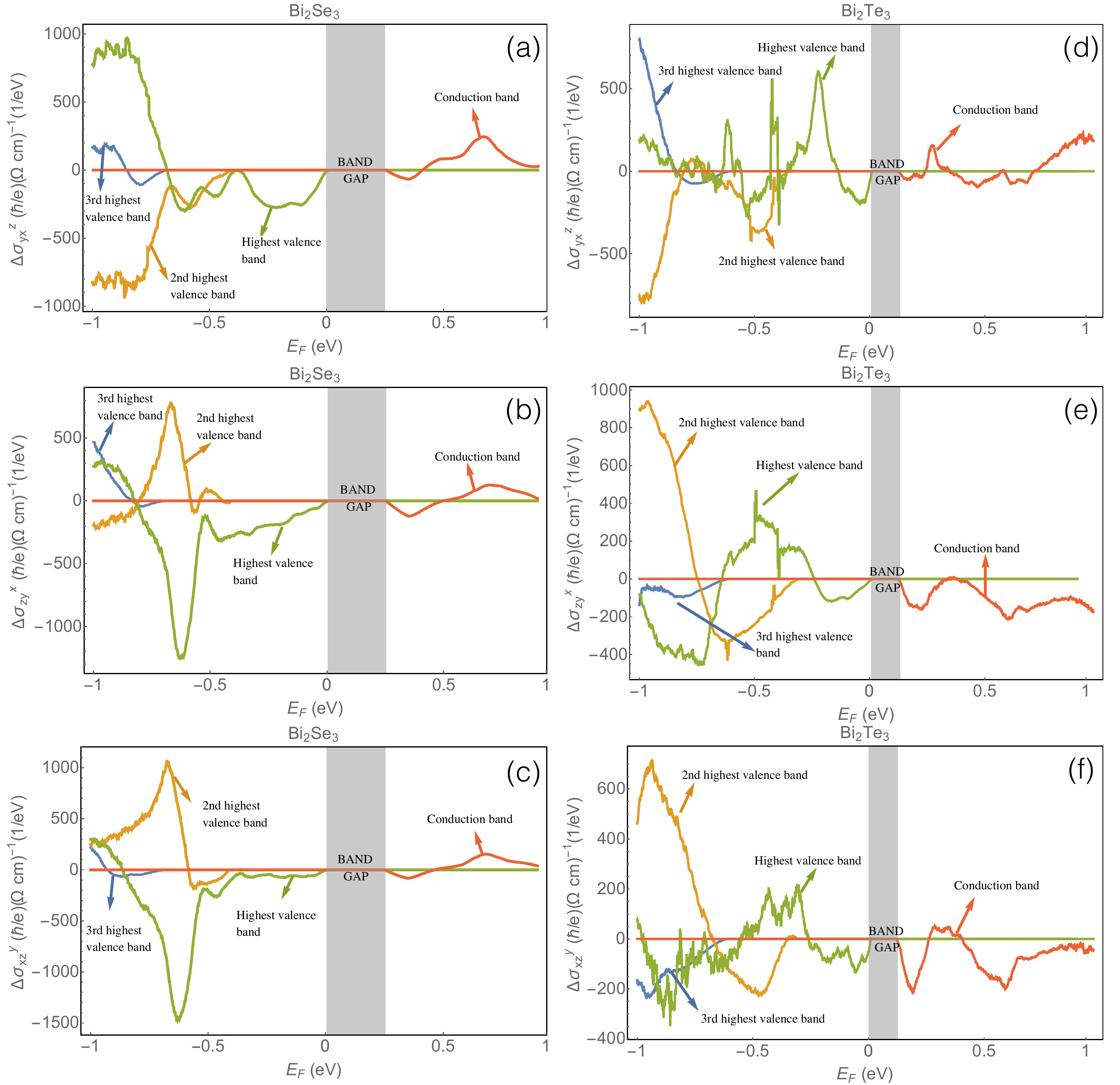}
\caption[Contribution of different bands to the spin Hall conductivity in \BiSe and \BiTe crystals]{Contribution of different bands to the spin Hall conductivity in \BiSe and \BiTe crystals. Shifting the Fermi energy below the valence band would increase the total SHC since the highest valence band has a very large negative Berry curvature.}
\end{figure}

It is possible to investigate contributions of individual bands around the Fermi energy to the SHC (\figref{fig:bisebandberry}). This allows us to predict how much the SHC would change as the Fermi level is shifted by an external perturbation, and it also shows the relative contribution of bulk bands to the SHC. The surface state contribution would be in the gray area in \figref{fig:bisebandberry}, however, they are not included in own bulk calculations. We report that the highest valence band and lowest conduction band have a negative Berry curvature around the band gap for both \BiSe and \BiTe in every direction. This indicates that the shifting chemical potential to lower than the valence band edge would result in an increase of the SHC up to some energy level, which is different for each case. On the other hand positioning the Fermi energy higher than the conduction band edge has a reverse effect. Furthermore, the negative Berry curvatures of the conduction bands change to positive curvatures at several hundreds of milielectronvolts above the conduction band edge and stay positive for \BiSe in all directions. This is also supported by \figref{otherdirections} in which SHC increases steadily following a decrease as a function of Fermi energy above the conduction band. 

While band resolved contributions exhibit similar features for \BiSe in three different directions, they are more distinctive in the case of the \BiTe crystal. This also explains the dependence of the SHC on the directions of the $\sigma_{ij}^k$ tensor for \BiTe and therefore a larger anisotropy in the SHC that is plotted in \figref{biteotherdirections}. The conduction band contributions of \BiTe for $\sigma_{zy}^x$ never become positive, indicating that any change in the Fermi level in the direction towards higher energies will destroy the SHC eventually. (Shown in \figref{fig:bisebandberry} part e.) 

\section{Conclusions and Comparison with Other Materials}
We list some experimental and theoretical values for SHC of different materials in Table~\ref{tab:SOIcomparison}. By comparing these values with the ones that we calculated, we conclude that bismuth-based materials have a robust SHC compared to most of the semiconductors such as silicon, gallium arsenide, and zinc selenide. Comparison of bismuth and antimony alloys with platinum and aluminum reveals that \BiSb alloys have larger SHC than the metals which are known to exhibit giant SHC. There exist a significant difference between the reported experimental SHC of \BiSe in Ref.~\cite{Mellnik2014} and our calculations. However, these experimental results will be reduced by a factor of 3 as we learnt through private communication. Such a modification suggest that our calculations may be similar to this experiment. The agreement between theory and experiment could be enhanced by including of the surface states into the SHC calculations. Our calculations demonstrate that \BiTe also exhibits a very high SHC as in \BiSe and \BiSb alloys. One observation that might be drawn from this table is that strength of the spin-orbit coupling is directly related to the magnitude of the spin Hall conductivity in a material.

 \begin{table}[h]\caption{Comparison of the Spin Hall Conductivities for Different Materials} \label{tab:comparison}
\begin{center}
\begin{tabular}{|l|l|l|l|}
\hline
Material & SHC ($(\hbar/e)\Omega^{-1}cm^{-1}$)& Type of Study & Reference\\
\hline
Silicon (p-type)& 0.02 &Experiment &\cite{Ando2012}\\
GaAs & 0.01 - 1 & Experiment &\cite{Matsuzaka2009}, \cite{Tse2006}\\
GaAs & 0.009 & Theory  &\cite{Engel2005a}\\
Platinum & 240 & Experiment& \cite{Vila2007}, \cite{Kimura2007}\\
Platinum & 200 (at room T) & Theory & \cite{Guo2008}\\
ZnSe & 0.01 & Experiment & \cite{Stern2006}\\
Aluminum & 10-36 & Experiment & \cite{Valenzuela2006}, \cite{Valenzuela2007}\\
Bismuth Selenide & 550-1000 & Experiment & \cite{Mellnik2014}\\
Bismuth-Antimony Alloys & 96-494 & Theory & This work \cite{Sahin2015}\\
Bismuth Selenide & 48-200 & Theory & This work\\
Bismuth Telluride & 104-300 & Theory & This work\\
\hline
\end{tabular}
\end{center}
\end{table}

In conclusion, we use a low energy effective spin-orbit Hamiltonian within a tight-binding approach for Bi and Sb as well as \BiSb alloys. Beginning with this low-energy Hamiltonian and band structure we calculate the intrinsic spin Hall conductivity using a Berry curvature technique in the clean static limit. We have also investigated the behavior of the Berry curvature in a full zone picture in terms of the density of Berry curvature \cite{Sahin2015}. We have observed that SHC of \BiSb alloys can be altered by a factor of 5 by either gate voltage or doping. We have also calculated the spin Hall conductivities of other bismuth-based materials, bismuth chalcogenides such as \BiSe and \BiTe topological insulators from a tight-binding Hamiltonian including two nearest-neighbor interactions. We have concluded that bismuth, antimony, \BiSb alloys as well as \BiSe and \BiTe exhibit giant spin Hall conductivities consistent with the results of Ref. \cite{Mellnik2014}. Robust spin-orbit couplings and Berry curvatures in these materials result in spin Hall conductivities which are comparable to other materials with giant SHC and considerably larger than conventional semiconductors and metals. Thus, we conclude that bismuth-antimony alloys and bismuth chalcogenides are very promising materials for the generation of fully polarized spin current.

\chapter{Conclusions and Future Directions}
\section{Conclusions}
We performed spin calculations for several different systems, each of which is a prominent candidate material for future spintronic applications. As a two dimensional system, \LAOSTO 2DEGs exhibit large spin lifetimes even at room temperature, which is a significant advantage for spin manipulation. In order to calculate spin lifetimes, a rigorous prescription for the construction of an effective spin-orbit Hamiltonian near the conduction band edge is provided and applied to materials that are spatially inversion symmetric with doubly degenerate bands. This Hamiltonian produces an effective spin-orbit interaction tensor, $\lambda_{ijk}$, and knowledge of the $\lambda$'s allows us to construct the effective spin-orbit interaction between electrons in a particular  band and scattering from a scalar spin-independent potential $V(\v r)$ ({\it e.g.} impurity scattering or phonon scattering in a quasi-elastic approximation\cite{Yu3ed}). Furthermore, we verify this prescription by testing it at the Brillouin zone center of some direct-gap III-V semiconductors with the double degeneracy. We compare the Hamiltonian and spin lifetimes from a tight-binding band structure to those from a \kp model describing the single conduction valley. We then extract from this formalism an effective spin-orbit Hamiltonian for \STO and use it to predict spin lifetimes for conduction electrons in strained \STO and a \LAOSTO 2DEG. We find exceptionally long spin lifetimes in both systems, suggesting that STO-based materials should have robust room-temperature spintronic properties. This prescription to construct an effective spin-orbit Hamiltonian is also useful in calculating a broad assortment of spin-related properties, including spin diffusion lengths, spin Hall conductivities, $g$-tensors, and spin precession lengths, when valid electronic structure calculations are available. In addition, this systematic approach to the calculation of the effective spin-orbit interaction and the Elliot-Yafet spin relaxation rate is broadly applicable to other centrosymmetric nonmagnetic materials. As centrosymmetric materials have recently started to play a more significant role in spin-dependent physics (e.g. large spin Hall effects in cubic metals, spin lifetimes in diamond-based materials) it is expected that our approach will apply to a broad range of materials and spin-dependent phenomena.  

We have calculated the intrinsic spin Hall conductivity for LAO/STO 2DEGs,  bismuth and antimony alloys and bismuth chalcogenides, using the Kubo formula in the clean and static regime. The oxide interface shows unusual behavior as a function of the temperature, strain and spin-orbit coupling in the system. For example, there exists a transition temperature where the SHC changes sign, and this critical temperature can be tuned by strain or by altering the strength of the spin-orbit coupling at the interface. On the other hand, the electronic structures of \BiSb alloys are described by a three-nearest-neighbor tight-binding Hamiltonian while we used a TB Hamiltonian with two nearest neighbor hopping integrals in \BiSe and \TeBi. \BiSb alloys are treated in two different virtual crystal approximations, both of which give rise to very similar spin Hall conductivities. The virtual crystal technique provides a qualitatively good description of the semi-metal semiconductor transition regime, where the \BiSb alloy opens a band gap and exhibits topologically protected states. We have not observed a substantial difference in the SHC of semimetal and topological insulator phases, however due to the difference in the charge conductivity we expect a very large spin Hall angle in the topological insulator regime compared to the semimetallic one. We have also computed the density of Berry curvature in order to investigate the Fermi energy dependence of the SHC in these systems. Calculations of the Fermi level dependence of the spin Hall conductivity suggest that substantial gate tuning of the spin Hall conductivity is possible in bismuth-based materials. Bismuth, antimony and \BiSb alloys with large spin-orbit couplings exhibit robust intrinsic spin Hall conductivities, larger than conventional semiconductors and metals with large spin Hall conductivity. Bismuth, antimony, and bismuth-antimony alloys are thus promising candidates for transverse spin current generation and spintronic applications.

\section{Future Directions}
\subsection{Relation between Spin-Orbit Interaction and Intrinsic SHC}
Comparison of the effective spin-orbit interaction, \eqr{eq:so-tensor}, which is derived in Chapter 3 reveals the fact that it is in the form of a Berry curvature which is evaluated at a k-point in the Brillouin zone (zone center in this case):
\al{
\lambda_{ijk}=\frac{1}{2}Im\sum_{\alpha \alpha'}[\sigma_k]_{\alpha \alpha'} \sum_{n\neq c, \beta} \frac{\matrixel{u_{c \tilde{\mathbf k} \alpha'}}{\nabla_{\tilde{k}_i} \unit H_{\tilde{\mathbf k}}} {u_{n \tilde{\mathbf k} \beta}} \matrixel{u_{n \tilde{\mathbf k} \beta}}{\nabla_{\tilde{k}_j} \unit H_{\tilde{\mathbf k}}}{u_{c \tilde{\mathbf k} \alpha}}}{(E_{c \tilde{\mathbf k}} -E_{n \tilde{\mathbf k}})^2} \Bigg |_{\tilde{\mathbf k}=0} 
} 
On the other hand, the intrinsic spin Hall conductivity of Chapter 4 (\eqr{eq:spinhall}) is also in the form of a Berry curvature. When we substitute all the quantities and rearrange the matrix elements, we get:
\al{
\sigma^z_{ji}=\frac{e}{V}\sum_{\v kn}f_{\v k ,n} \frac{1}{2}Im\sum_{n \neq n'}\frac{
\matrixel{\psi_{n \v k}}{\nabla_{k_i}  H_{\v k}}{\psi_{n' \v k }}\matrixel{\psi_{n'\v k}}{\nabla_{k_j}   H_{\v k}\hat{\sigma}_z+\hat{\sigma}_z \nabla_{k_j}   H_{\v k}}{ \psi_{n\v k}} } {(E_{n\v k}-E_{n'\v k})^2}
}
The two expressions differ by some coefficients, the Fermi-Dirac distribution function and the form of the second matrix element of an operator. Both the effective spin-orbit interaction tensor($\lambda_{ijk}$) and the intrinsic spin Hall conductivity ($\sigma_{xy}^z$) are calculated from the same band structure. In addition, these both quantities are intrinsic to the system. One of the directions this research may take is connecting these two intrinsic quantities and relating intrinsic SHC to the effective spin-orbit interaction in the system. This may give rise to a systematic calculation of several important phenomena in spin dynamics such as spin lifetimes, effective masses, g-factors, and spin Hall angles from the same physical quantity. 

\subsection{Inclusion of the Surface States into SHC Calculation of Bismuth Chalcogenides}
All of the results reported in this work for \BiSe and \BiTe stem from the bulk band structure of these materials. While bulk bands are dominant for most of the properties of these materials, surface states also are a result of characteristics of these materials and exhibit topologically protected behavior, and so contribute to the spin Hall conductivity significantly. Theoretical attempts to include these states by using an effective two-dimensional Hamiltonian result in either very small or zero conductivity. This is due to the fact that an effective Hamiltonian cannot express correct curvatures of these bands although it can be used for the purpose of electronic band calculations. 

One possible solution to this problem is calculating surface states from the bulk states of the material. There have been several studies aiming to construct a calculation based on the tight-binding Hamiltonian. Chadi and Cohen \cite{Chadi1975a} calculated such surface band structure by taking a finite number of layers (16) for Ge, GaAs and ZnSe and changing interatomic interaction parameters by moving surface layers such that:
\al{H_{ij}=H_{ij}^0e^{-\beta \Delta R}}
where $\Delta R$ is the change in the nearest neighbor distance, $H_{ij}$ and $H_{ij}^0$ are the modified and the original tight-binding matrix elements respectively. 

On the other hand, Lee and Joannopoulos \cite{Lee1981} proposed to use a technique called the transfer-matrix approach, which matches the wavefunctions of neighboring layers. The transfer matrix includes all the information needed for solving eigenvalues and eigenfunctions of the surface states. This method is extremely fast and efficient since bulk projected bands, surface energies, and decay lengths can be obtained by only diagonalizing the transfer matrix without performing time-consuming Green's function calculations.

Extending this work so that the contribution of the surface states into SHC is included may be interesting from many perspective. The topologically protected surface states of bismuth chalcogenides offer new physics to be discovered and utilized. The connection between topology and spintronics is also one of the main reasons to extend this work to other materials with similar properties.

\appendix \label{sec:Appendix}

\chapter{Tight-Binding Hamiltonian for Strontium Titanate}\label{app:strontiumhamiltonian}
In this section I list the matrix elements of the tight-binding Hamiltonian for strontium titanate that I constructed using parameters of Kahn and Leyendecker \cite{Kahn1964} in terms of Slater-Koster parameters \cite{Slater1954}. These parameters are already listed in Chapter 2. First, I introduce some shorthand notation for the atomic orbitals used throughout this section. (Table~\ref{shorthand})

\begin{table}[h]\caption[Notation for Atomic Orbitals]{Notation for Atomic Orbitals}
\label{shorthand}
\begin{center}
\begin{tabular}{|c|c|}

Atomic Orbital & Shorthand Notation\\ 
\hline
& \\
s & $s$ \\
& \\
$p_x$, $p_y$, and $p_z$ &$x$, $y$, and $z$ \\
& \\
$d_{xy}$, $d_{yz}$, and $d_{zx}$ & $d_1$, $d_2$, and $d_3$\\
& \\
$d_{x^2-y^2}$, and $d_{3z^2-r^2}$  & $d_4$ and $d_5$\\
& \\
\hline
\end{tabular}
\end{center}
\end{table}
For complex oxides, a state that is written as $x_i$ denotes an orbital of thpe $p_x$ for the oxygen atom which is located at {\it{i}}th position where i runs from 1 to 3. Furthermore S stands for strain element for the bulk strontium titanate (or confinement energy for the \LAOSTO 2DEG) which shifts the energy of the orbital which has a z-component. Tight-binding Hamiltonian for strontium titanate is tabulated in Table~\ref{sto1} and Table~\ref{sto2}.

\begin{table}[h]\caption{Matrix Elements of the Hamiltonian for \STO I}
\label{sto1}
\begin{center}
\begin{tabular}{|l|l|}
\hline
&\\
 $\matrixel{d_1}{H}{d_1}=$ E$_d$-M$_{Ti}$ &  $\matrixel{d_2}{H}{d_2}=$ E$_d$-M$_{Ti}+S$ \\
&\\
\hline
&\\
 $\matrixel{d_3}{H}{d_3}=$ E$_d$-M$_{Ti}+S$ &  $\matrixel{d_4}{H}{d_4}=$ E$_d$-M$_{Ti}$+d$_{el}$\\
 &\\
 \hline
&\\
 $\matrixel{d_5}{H}{d_5}=$ E$_d$-M$_{Ti}$+d$_{el}$ & $\matrixel{z_1}{H}{z_1}=E_p+p_{el}-M_O+S$ \\
 &\\
 \hline
&\\
  $\matrixel{z_2}{H}{z_2}=E_p+p_{el}-M_O+S$ & $\matrixel{z_3}{H}{z_3}=E_p-M_O+S$ \\
&\\
\hline
&\\
$\matrixel{x_1}{H}{x_1}=E_p-M_O$ & $\matrixel{x_2}{H}{x_2}=E_p+p_{el}-M_O$ \\
&\\
\hline
&\\
 $\matrixel{x_3}{H}{x_3}=E_p+p_{el}-M_O$& $\matrixel{y_1}{H}{y_1}=E_p+p_{el}-M_O$
 \\
 &\\
 \hline
&\\
 $\matrixel{y_2}{H}{y_2}=E_p-M_O$  & $\matrixel{y_3}{H}{y_3}=E_p+p_{el}-M_O$  \\
 &\\
 \hline
&\\
  $\matrixel{x_1}{H}{x_2}=$ 4C$_1$cos($\frac{1}{2}$\kx)cos ($\frac{1}{2}$\ky)     & $\matrixel{x_1}{H}{x_3}= $ 4C$_1$cos($\frac{1}{2}$\kx)cos ($\frac{1}{2}$\kz)  \\ 
 &\\ 
 \hline
&\\
 $\matrixel{y_1}{H}{y_3}= $ 4C$_4$cos($\frac{1}{2}$ \kx) cos($\frac{1}{2}$ \kz)  & $\matrixel{z_1}{H}{z_2}= $ 4C$_4$cos($\frac{1}{2}$ \kx) cos($\frac{1}{2}$ \ky) \\ 
 &\\
 \hline
&\\
$\matrixel{x_2}{H}{x_3}= $4C$_4$cos($\frac{1}{2}$ \ky) cos($\frac{1}{2}$ \kz) & $\matrixel{y_2}{H}{y_3}= $ 4C$_1$cos($\frac{1}{2}$ \ky) cos($\frac{1}{2}$ \kz)\\
&\\
\hline
&\\
$\matrixel{z_2}{H}{z_3}= $ 4C$_1$cos($\frac{1}{2}$ \ky) cos($\frac{1}{2}$ \kz)& $\matrixel{x_1}{H}{y_2}= $ -4C$_2$sin($\frac{1}{2}$ \kx)sin ($\frac{1}{2}$\ky) \\
&\\
\hline
&\\
$\matrixel{x_1}{H}{z_3}= $ -4C$_2$sin[($\frac{1}{2}$ \kx)sin ($\frac{1}{2}$\kz)& $\matrixel{y_1}{H}{x_2}= $-4C$_3$sin($\frac{1}{2}$ \kx)sin ($\frac{1}{2}$\ky) \\
&\\
\hline
&\\
$\matrixel{z_1}{H}{x_3}= $ -4C$_3$sin($\frac{1}{2}$ \kx)sin ($\frac{1}{2}$\kz) & $\matrixel{y_2}{H}{z_3}= $-4C$_2$sin($\frac{1}{2}$ \ky)sin($\frac{1}{2}$\kz) \\
&\\
\hline
&\\
$\matrixel{z_1}{H}{z_3}= \matrixel{x_1}{H}{x_3} $ & $\matrixel{y_1}{H}{y_2}= \matrixel{x_1}{H}{x_2}$\\
&\\
\hline
\end{tabular}
\end{center}
\end{table}

\begin{table}[h]\caption{Matrix Elements of the Hamiltonian for \STO II}
\label{sto2}
\begin{center}
\begin{tabular}{|l|l|}
\hline
&\\
$\matrixel{z_2}{H}{y_3}= $-4C$_3$sin($\frac{1}{2}$ \ky)sin($\frac{1}{2}$\kz) & $\matrixel{x_1}{H}{d_4}= $-i(pd$\sigma$)sin($\frac{1}{2}$\kx) \\
&\\
\hline
&\\
$\matrixel{x_1}{H}{d_5}= $ $\sqrt 3$i(pd$\sigma$)sin($\frac{1}{2}$\kx) & $\matrixel{y_1}{H}{d_1}= $2i(pd$\pi$)sin($\frac{1}{2}$\kx) \\ 
&\\
\hline
&\\
$\matrixel{z_1}{H}{d_3}= $ 2i(pd$\pi$)sin($\frac{1}{2}$\kx) & $\matrixel{x_2}{H}{d_1}= $2i(pd$\pi$)sin($\frac{1}{2}$\ky)  \\
&\\
\hline
&\\
$\matrixel{y_2}{H}{d_4}= $ -i(pd$\sigma$)sin($\frac{1}{2}$\ky) & $\matrixel{y_2}{H}{d_5}= $ -$\sqrt 3$i(pd$\sigma$)sin($\frac{1}{2}$\ky)\\
&\\
\hline
&\\
$\matrixel{z_2}{H}{d_2}= $ 2i(pd$\pi$)sin($\frac{1}{2}$\ky) & $\matrixel{x_3}{H}{d_3}= $ 2i(pd$\pi$)sin($\frac{1}{2}$\kz) \\ 
&\\
\hline
&\\
$\matrixel{y_3}{H}{d_2}= $ 2i(pd$\pi$)sin($\frac{1}{2}$\kz) & $\matrixel{z_3}{H}{d_4}= $2i(pd$\sigma$)sin($\frac{1}{2}$\kz) \\ 
&\\
\hline
\end{tabular}
\end{center}
\end{table}

\newpage{}
\chapter{Tight-Binding Hamiltonian for Strained Strontium Titanate}\label{sec:strainhamiltonian}
This section will be added soon.

\newpage{}

\chapter{spds* Tight-Binding Hamiltonian for Zincblende Crystals}\label{appendixzincblende}

In this section I list the Hamiltonian in the spds* basis for a generic zincblende structure which can be used to write Hamiltonians for both group IV elements and III-V compounds. First I introduce the four parameters as discussed in Chapter 2 of Yu and Cardona. \cite{YuCardona}. This parameters come from the relative positions of atoms in the crystal.
\al{\bs
&g_0 = +cos\left (\frac{k_1}{2}\right )cos\left (\frac{k_2}{2}\right )cos\left (\frac{k_3}{2}\right ) - isin\left (\frac{k_1}{2}\right )sin\left (\frac{k_2}{2}\right )sin\left (\frac{k_3}{2}\right )\\
&g_1 = -cos\left (\frac{k_1}{2}\right )sin\left (\frac{k_2}{2}\right )sin\left (\frac{k_3}{2}\right ) +  isin\left (\frac{k_1}{2}\right )cos\left (\frac{k_2}{2}\right )cos\left (\frac{k_3}{2}\right )\\
&g_2 = -sin\left (\frac{k_1}{2}\right )cos\left (\frac{k_2}{2}\right ) sin\left (\frac{k_3}{2}\right ) +  icos\left (\frac{k_1}{2}\right ) sin \left (\frac{k_2}{2}\right )cos \left (\frac{k_3}{2}\right )\\
&g_3 = -sin\left (\frac{k_1}{2}\right )sin\left (\frac{k_2}{2}\right )cos\left (\frac{k_3}{2}\right ) + icos\left (\frac{k_1}{2}\right )cos\left (\frac{k_2}{2}\right )sin\left (\frac{k_3}{2}\right )
\es}

The subscript "a" denotes an anion of the III-V compound where "c" is used for cation. To get Hamiltonian for a group IV element such as carbon in diamond in the same basis, one should make all cation and anion matrix elements equal to each other in the generic zinc-blende Hamiltonian. The tight-binding Hamiltonian consists of atomic orbitals such as s, p, d, and s$^*$, in total 20 states. Therefore TB Hamiltonian is a 20$\times$20 matrix. First we redefine some overlap integrals to shorten the notations as shown in Table~\ref{redefined}.
\begin{table}[h]\caption{Overlap Integrals Redefined } \label{redefined}
\begin{center}
\begin{tabular}{|c|c|}
\hline
& \\
$V_{s^*d}=\frac{4}{\sqrt 3}s^*d\sigma$ & $V_{sd}=\frac{4}{\sqrt 3}sd\sigma$\\
 & \\
\hline 
& \\
$V_{pd}=\frac{4}{3\sqrt 3}(\sqrt 3 pd\sigma +pd\pi)$ &
$V'_{pd}=\frac{4}{3\sqrt 3}(\sqrt 3 pd\sigma -2 pd\pi)$\\& \\
\hline
& \\
$V_{dd}=\frac{4}{9} (3 dd\sigma +2 dd\pi +4 dd\delta)$ &
$V'_{dd}=\frac{4}{9}(3 dd\sigma - dd\pi -2 dd\delta)$\\
& \\
\hline
& \\
$V''_{dd} =\frac{4}{3\sqrt 3} (dd\pi -dd\delta)$ & \\
& \\
\hline
\end{tabular}
\end{center}
\end{table}
These are the only non-zero matrix elements of tight-binding Hamiltonian. One should also keep in mind that Hamiltonian is Hermitian and other elements can be easily found by complex conjugation. Furthermore when spin-orbit Hamiltonian is added, the Hamiltonian doubles its size and becomes a 40$\times$40 matrix.
\begin{table}[h]\caption{Diagonal Matrix Elements of the Hamiltonian}
\begin{center}
\begin{tabular}{|c|c|c|}
\hline
& & \\
  $\matrixel{s_a}{H}{s_a}=E_{sa}$ & $\matrixel{d_{1a}}{H}{d_{1a}}=E_{da}$ & $\matrixel{s_c}{H}{s_c}=E_{sc} $ \\
  & & \\
  \hline
  & & \\
  $\matrixel{d_{2a}}{H}{d_{2a}}=E_{da}$  &
  $\matrixel{x_a}{H}{x_a}=E_{pa} $ & $\matrixel{d_{3a}}{H}{d_{3a}}=E_{da} $  \\
  & & \\
  \hline
  & & \\
  $\matrixel{y_a}{H}{y_a}=E_{pa} $ & $\matrixel{d_{4a}}{H}{d_{4a}}=E_{da} $  &
  $\matrixel{z_a}{H}{z_a}=E_{pa} $ \\
  & & \\
  \hline
  & & \\
   $\matrixel{d_{5a}}{H}{d_{5a}}=E_{da} $ &
  $\matrixel{x_c}{H}{x_c}=E_{pc} $ & $\matrixel{d_{1c}}{H}{d_{1c}}=E_{dc} $  \\
  & & \\
  \hline
  & & \\
  $\matrixel{y_c}{H}{y_c}=E_{pc} $ & $\matrixel{d_{2c}}{H}{d_{2c}}=E_{dc} $  &
  $\matrixel{z_c}{H}{z_c}=E_{pc} $ \\
  & & \\
  \hline
  & & \\
   $\matrixel{d_{3c}}{H}{d_{3c}}=E_{dc} $  &
  $\matrixel{s^*_a}{H}{s^*_a}=E_{s^*a} $ & $\matrixel{d_{4c}}{H}{d_{4c}}=E_{dc} $  \\
  & & \\
  \hline
  & & \\
  
  $\matrixel{s^*_c}{H}{s^*_c}=E_{s^*c} $ & $\matrixel{d_{5c}}{H}{d_{5c}}=E_{dc} $ & \\
  & & \\
  \hline
\end{tabular}
\end{center}
\end{table}

\begin{table}[h]\caption{Matrix Elements of the Hamiltonian for $s$ and $s^*$ States }
\begin{center}
\begin{tabular}{|c|c|c|}
\hline
& & \\
 $\matrixel{s_a}{H}{s_c}=V_{ss} g_0 $     & $\matrixel{s^*_a}{H}{s_c}=V_{s^*s} g_0 $ &
 $\matrixel{s_a}{H}{s^*_c}=V_{ss^*} g_0 $ \\
 & & \\
 \hline
 & & \\
  $\matrixel{s^*_a}{H}{s^*_c}=V_{s^*s^*} g_0 $ &
 $\matrixel{s_a}{H}{x_c}=V_{sp} g_1 $     & $\matrixel{s^*_a}{H}{x_c}=V_{s^*p} g_1 $ \\
 & & \\
 \hline
 & & \\
 $\matrixel{s_a}{H}{y_c}=V_{sp} g_2 $     & $\matrixel{s^*_a}{H}{y_c}=V_{s^*p} g_2 $ &
 $\matrixel{s_a}{H}{z_c}=V_{sp} g_3 $     \\
 & & \\
 \hline
 & & \\
  $\matrixel{s^*_a}{H}{z_c}=V_{s^*p} g_3 $ &
 $\matrixel{s_a}{H}{d_{1c}}=V_{sd} g_3$  & $\matrixel{s^*_a}{H}{d_{1c}}=V_{s^*d} g_3$ \\
 & & \\
 \hline
 & & \\
 $\matrixel{s_a}{H}{d_{2c}}=V_{sd} g_1 $ & $\matrixel{s^*_a}{H}{d_{2c}}=V_{s^*d} g_1 $ &
 $\matrixel{s_a}{H}{d_{3c}}=V_{sd} g_2 $ \\
 & & \\
 \hline
  & & \\
  $\matrixel{s^*_a}{H}{d_{3c}}=V_{s^*d} g_2 $ & &\\ 
  & & \\
  \hline
\end{tabular}
\end{center}
\end{table}

\begin{table}[h]\caption{Matrix Elements of the Hamiltonian for $d_4$ and $d_5$ States }
\begin{center}
\begin{tabular}{|c|c|}
\hline
 & \\
$\matrixel{d_{4a}}{H}{x_c}=-\matrixel{d_{5a}}{H}{x_c}=\frac{-4}{\sqrt 3} pd\pi g_1$ & $\matrixel{d_{4a}}{H}{d_{2c}}=-\sqrt 3 V''_{dd} g_1$ \\
  & \\
  \hline
  & \\
$\matrixel{d_{4a}}{H}{y_c}=-\matrixel{d_{5a}}{H}{y_c}=\frac{-4}{\sqrt 3} pd\pi g_2 $ & $\matrixel{d_{5a}}{H}{z_c}=\frac{-8}{3}pd\pi g_3$ \\
  & \\
   \hline
  & \\
$\matrixel{d_{4a}}{H}{d_{3c}}=\sqrt 3 V''_{dd} g_2$ & $\matrixel{d_{5a}}{H}{d_{1c}}=-2V''_{dd} g_3$ \\
 & \\
  \hline
  & \\
$\matrixel{d_{4a}}{H}{d_{4c}}=\frac{4}{3}(2dd\pi+dd\delta) g_0 $ & $\matrixel{d_{5a}}{H}{d_{2c}}=V''_{dd} g_1$ \\
 & \\
  \hline
  & \\
  $\matrixel{d_{5a}}{H}{d_{3c}}=V''_{dd} g_2$ & $\matrixel{d_{5a}}{H}{d_{5c}}=\frac{4}{3}(2dd\pi+dd\delta) g_0 $ \\
 & \\
\hline
\end{tabular}
\end{center}
\end{table}

\begin{table}[h]\caption{Matrix Elements of the Hamiltonian for $x$, $y$ and $z$ States }
\begin{center}
\begin{tabular}{|c|c|c|}
\hline
& & \\
 $\matrixel{x_a}{H}{s_c}=-V_{sp} g_1  $ & $\matrixel{y_a}{H}{s_c}=-V_{sp} g_2  $ & $\matrixel{z_a}{H}{s_c}=-V_{sp} g_3  $\\
 & & \\
 \hline
 & & \\
 $\matrixel{x_a}{H}{x_c}= V_{xx} g_0 $ &  $\matrixel{y_a}{H}{x_c}= V_{xy} g_3 $ & $\matrixel{z_a}{H}{x_c}= V_{xy} g_2 $ \\
  & & \\
 \hline
 & & \\
 $\matrixel{x_a}{H}{y_c}= V_{xx} g_3  $ & $\matrixel{y_a}{H}{y_c}= V_{xx} g_0  $ & $\matrixel{z_a}{H}{y_c}= V_{xx} g_1  $\\
  & & \\
 \hline
 & & \\
 $\matrixel{x_a}{H}{z_c}= V_{xx} g_2 $ &  $\matrixel{y_a}{H}{z_c}= V_{xy} g_1 $ & $\matrixel{z_a}{H}{z_c}= V_{xx} g_0 $\\
  & & \\
 \hline
 & & \\
 $\matrixel{x_a}{H}{s^*_c}= -V_{s^*p} g_1  $ &  $\matrixel{y_a}{H}{s^*_c}= -V_{s^*p} g_2$ &  $\matrixel{z_a}{H}{s^*_c}= -V_{s^*p} g_3$ \\
  & & \\
 \hline
 & & \\
 $\matrixel{x_a}{H}{d_{1c}}=V_{pd} g_2  $ &  $\matrixel{y_a}{H}{d_{1c}}=V_{pd} g_1  $ & $\matrixel{z_a}{H}{d_{1c}}=V'_{pd} g_0  $\\
  & & \\
 \hline
 & & \\
 $\matrixel{x_a}{H}{d_{2c}}=V'_{pd} g_0  $ & $\matrixel{y_a}{H}{d_{2c}}=V_{pd} g_3  $ &$\matrixel{z_a}{H}{d_{2c}}=V_{pd} g_2  $ \\
  & & \\
 \hline
 & & \\
 $\matrixel{x_a}{H}{d_{3c}}=V_{pd} g_3  $ &  $\matrixel{y_a}{H}{d_{3c}}=V'_{pd} g_0  $ & $\matrixel{z_a}{H}{d_{3c}}=V_{pd} g_3  $\\
  & & \\
 \hline
 & & \\
 $\matrixel{x_a}{H}{d_{4c}}=\frac{4}{\sqrt 3} (pd\pi) g_1  $ &  $\matrixel{y_a}{H}{d_{4c}}=-\frac{4}{\sqrt 3} (pd\pi) g_2  $ &$\matrixel{z_a}{H}{d_{4c}}=0$ \\
  & & \\
 \hline
 & & \\
 $\matrixel{x_a}{H}{d_{5c}}=-\frac{4}{3} (pd\pi) g_1  $ &  $\matrixel{y_a}{H}{d_{5c}}=-\frac{4}{3} (pd\pi) g_2  $ &  $\matrixel{z_a}{H}{d_{5c}}=-\frac{8}{3} (pd\pi) g_3 $ \\
 & & \\
\hline
\end{tabular}
\end{center}
\end{table}

\begin{table}[h]\caption{Matrix Elements of the Hamiltonian for $d_1$, $d_2$ and $d_3$ States }
\begin{center}
\begin{tabular}{|c|c|c|}
\hline
& & \\
$\matrixel{d_{1a}}{H}{s_c}=V_{sd} g_3$ & $\matrixel{d_{2a}}{H}{s_c}= V_{sd} g_1$ & $\matrixel{d_{3a}}{H}{s_c}=V_{sd} g_2$ \\
 & & \\
 \hline
 & & \\
$\matrixel{d_{1a}}{H}{s^*_c}=V_{s^*d} g_3$ & $\matrixel{d_{2a}}{H}{s^*_c}= V_{s^*d} g_1$ & $\matrixel{d_{3a}}{H}{s^*_c}=V_{s^*d} g_2$ \\
 & & \\
 \hline
 & & \\
$\matrixel{d_{1a}}{H}{x_c}= -V_{pd} g_2 $ & $\matrixel{d_{2a}}{H}{x_c}= -V'_{pd} g_0$ & $\matrixel{d_{3a}}{H}{x_c}=-V_{pd} g_3$ \\
 & & \\
 \hline
 & & \\
$\matrixel{d_{1a}}{H}{y_c}= -V_{pd} g_1 $ & $\matrixel{d_{2a}}{H}{y_c}= V_{pd} g_3$ & $\matrixel{d_{3a}}{H}{y_c}=-V'_{pd} g_0$\\
 & & \\
 \hline
 & & \\
$\matrixel{d_{1a}}{H}{z_c}= -V'_{pd} g_0 $ & $\matrixel{d_{2a}}{H}{z_c}= = V_{pd} g_2$ & $\matrixel{d_{3a}}{H}{z_c}=-V_{pd} g_1$ \\
 & & \\
 \hline
 & & \\
$\matrixel{d_{1a}}{H}{d_{1c}}=V_{dd} g_0$ & $\matrixel{d_{2a}}{H}{d_{1c}}=V'_{dd} g_2$ & $\matrixel{d_{3a}}{H}{d_{1c}}=V'_{dd} g_1$ \\ 
 & & \\
 \hline
 & & \\
$\matrixel{d_{1a}}{H}{d_{2c}}=V'_{dd} g_2$ & $\matrixel{d_{2a}}{H}{d_{2c}}= V_{dd} g_0$ & $\matrixel{d_{3a}}{H}{d_{2c}}=V'_{dd} g_3$ \\
 & & \\
 \hline
 & & \\
$\matrixel{d_{1a}}{H}{d_{3c}}=V'_{dd} g_1$ & $\matrixel{d_{2a}}{H}{d_{3c}}= V'_{dd} g_3$ & $\matrixel{d_{3a}}{H}{d_{3c}}=V_{dd} g_0$ \\
 & & \\
 \hline
 & & \\
$\matrixel{d_{1a}}{H}{d_{4c}}=0$ &  $\matrixel{d_{2a}}{H}{d_{4c}}=-\sqrt{3} V''_{dd} g_1$ & $\matrixel{d_{3a}}{H}{d_{4c}}=\sqrt 3 V''_{dd} g_2$ \\ 
 & & \\
 \hline
 & & \\
$\matrixel{d_{1a}}{H}{d_{5c}}=-2V''_{dd} g_3$ & $\matrixel{d_{2a}}{H}{d_{5c}}= V''_{dd} g_1$ & $\matrixel{d_{3a}}{H}{d_{5c}}= V''_{dd} g_2$ \\
& &\\
\hline
\end{tabular}
\end{center}
\end{table}

\pagebreak{}

\biblio{central-bibliography.bib} 

\end{document}